\documentclass[fleqn,usenatbib,useAMS]{mnras}

\usepackage{graphicx}	
\usepackage{amsmath}	
\usepackage{amssymb}	
\usepackage{multicol}        
\usepackage{bm}		
\usepackage{pdflscape}	
\usepackage{subfiles}
\usepackage{multirow}
\usepackage{xr-hyper}
\usepackage[normalem]{ulem}

\usepackage{courier}
\usepackage[dvipsnames]{xcolor}

\usepackage[T1]{fontenc}
\usepackage{ae,aecompl}

\usepackage{newtxtext,newtxmath}
\usepackage{comment}

\title[Compact binary merger hosts and environments]{Exploring compact binary merger host galaxies and environments with \texttt{zELDA}}

\author[S. Mandhai et al.]{S. Mandhai$^{1,2}$\thanks{Contact e-mail: \href{mailto:soheb.mandhai@manchester.ac.uk}{soheb.mandhai@manchester.ac.uk}},
G. P. Lamb$^{1}$\thanks{Contact e-mail:\href{mailto:gpl6@leicester.ac.uk}{gpl6@leicester.ac.uk}},
N. R. Tanvir$^{1}$\thanks{Contact e-mail: \href{mailto:nrt3@leicester.ac.uk}{nrt3@leicester.ac.uk}},
J. Bray$^{3, 4}$,
C. J. Nixon$^{1}$,
\newauthor 
R. A. J. Eyles-Ferris$^{1}$,
A. J. Levan$^{5,6}$,
B. P. Gompertz$^{7}$.
\\
$^{1}$ School of Physics and Astronomy, University of Leicester, University Road, Leicester, LE1 7RH, United Kingdom\\
$^{2}$ Jodrell Bank Centre for Astrophysics, Department of Physics and Astronomy, The University of Manchester, Manchester, M13 9PY,  United Kingdom\\
$^{3}$ Department of Physical Sciences, Robert Hooke Building, Kents Hill, Milton Keynes MK7 6AA, United Kingdom\\
$^{4}$ Department of Physics, The University of Auckland, Private Bag 92019, Auckland, New Zealand \\
$^{5}$ Department of Astrophysics/IMAPP, Radboud University, Nijmegen, Netherlands \\
$^{6}$ Department of Physics, University of Warwick, Coventry, CV4 7AL, UK \\
$^{7}$ School of Physics and Astronomy, University of Birmingham, Birmingham, B15 2TT, UK
}

\date{Accepted XXX. Received YYY; in original form ZZZ}

\pubyear{2021}

\begin{document}
\label{firstpage}
\pagerange{\pageref{firstpage}--\pageref{lastpage}}
\maketitle

\begin{abstract}
Compact binaries such as double neutron stars or a neutron star paired with a black-hole, 
are strong sources of gravitational waves during coalescence
and also the likely progenitors of various electromagnetic phenomena, notably short-duration gamma-ray bursts (SGRBs), and kilonovae. 
In this work, we generate populations of synthetic binaries and place them in galaxies from the large-scale hydrodynamical galaxy evolution simulation \texttt{EAGLE}. With our \texttt{zELDA} code, binaries are seeded in proportion to star formation rate, and we follow their evolution to merger using both the \texttt{BPASS} and \texttt{COSMIC} binary stellar evolution codes. We track their dynamical evolution within their host galaxy potential, to estimate the galactocentric distance at the time of the merger.
Finally, we apply observational selection criteria to allow comparison of this model population with the legacy sample of SGRBs.
We find a reasonable agreement with the redshift distribution (peaking at $0.5<z<1$), host morphologies and projected galactocentric offsets (modal impact parameter $\lesssim10$\,kpc). Depending on the binary simulation used, we predict $\sim16-35\%$ of SGRB events 
would appear ``host-less", i.e. sources that merge with high impact parameters or have hosts fainter than the detection limit ($H>26$).

\end{abstract}




\begingroup
\let\clearpage\relax
\endgroup
\newpage

\section{Introduction}

The orbits of compact binaries (containing neutron stars or stellar mass black holes) decay via gravitational wave emission, with inspiral times ranging from a few Myr to many Hubble times.
The final mergers have become of increasing interest following the first gravitational wave (GW) detections \citep{LIGO2016}.
Where one or both of the components are a neutron star, such a coalescence can also produce detectable electromagnetic counterparts, opening the route to multi-messenger astrophysics \citep{Branchesi2016,abbott2017MMA,Murase2019,Huerta2019,Meszaros2019}. This was exemplified by the historic discovery of GW170817, and the accompanying emission from both a kilonova and relativistic jetted material \citep{GW170817LIGO,Tanaka2017,Levan2017,Myungshin2017,Tanvir2017,Evans2017,Pian2017,Alexander2017,Arcavi2017,Chornock2017,Coulter2017,Cowperthwaite2017,Haggard2017,Hallinan2017,Kasliwal2017,McCully2017,Margutti2017,Nicholl2017,Shappee2017,Smartt2017,Soares-Santos2017,Troja2017,Villar2017,Corsi2018,Lyman2018,Resmi2018,Mooley2018,Nynka2018,Hajela2019,Piro2019,Troja2019,Lamb2019,Troja2020}.

The kilonova is powered by the radioactive decay of newly synthesised heavy elements from ejected neutron star material, and may be the dominant site of r-process nucleosynthesis in the universe.
The kilonova accompanying GW170817 (AT2017gfo) was extensively studied from ultraviolet to infrared wavelengths, thanks to its relative proximity at $\approx40$\,Mpc \citep{GW170817LIGO}.
However, more typically, the current generation of GW detectors \,\footnote{Third Observing Run (O3)}\,\footnote{https://gracedb.ligo.org/superevents/public/O3/} finds neutron-star containing mergers at $\sim200$\,Mpc \citep{GWProspectsv6}, which combined with the large (10s or 100s of sq-deg) error regions, presents a significant challenge to counterpart searches \citep{mandhai2018rate,dichiara2019short}. 

The production of 
relativistic jets
during mergers is less well understood, but 
is believed to be responsible for the short-duration gamma-ray burst (SGRB) emission; these events have typical durations, $T_{90}<2\,\rm{s}$ \citep{Mazets81,Kouveliotou93}. 
This emission is much brighter and can be seen at cosmological distances, indeed, apart  from GRB\,170817A accompanying GW170817, the nearest SGRB with  a secure  redshift is GRB\,080905A at $z=0.1218$ \citep{rowlinson2010discovery} whilst the most distant SGRBs have been found at $z\gtrsim 2$ \citep[e.g.][]{levan06b,fong2017electromagnetic,selsing18}. 
The association of SGRBs with compact binary mergers confirmed both by analysis of the non-thermal emission from AT2017gfo \citep{Perego_2017,Wang2017}, and the evidence of kilonova emission following other SGRBs \citep[e.g.][]{Tanvir2013,Jin2016,Gompertz2018,lamb2019short,Troja2019GRB160821B,Jin2020}.

SGRBs were first localised to arcsec precision in the {\it Neil Gehrels Swift Observatory} (hereafter {\it Swift}) era, which resulted in the identifications of host galaxies and hence redshifts  \citep[e.g.][]{Gehrels2005,Bloom2006,Fox2005,Hjorth2005}. 
The typical redshifts for SGRBs at the the depth reached by {\it Swift}/BAT is $z<1$ \citep{lien2016}. 
These events originate from a wide range of host galaxies and environments, including some with little or no ongoing star formation \citep[e.g.][]{fong2013demographics, fong2013locations, berger2014short}. 
The {\it Swift} rate of SGRB discovery is $\sim10$ yr$^{-1}$, in contrast, detectors with larger fields-of-view such as the \textit{Fermi}/GBM (Gamma-ray Burst Monitor) have a rate of $\sim40$ yr$^{-1}$, although the {\it Fermi} localisations are 
much worse (typically a $\sim \rm{few}\times10\,\rm{deg^2}$)
, so  redshifts are rarely obtained. Since the afterglows of SGRBs are also typically fainter than those from long-GRBs ($T_{90}>2\,\rm{s}$)
, the net result is that the total sample of 
SGRBs with better constrained ($\sim$arcsec--arcmin) sky-positions is $\sim140$, and only 25--30\% of these have confident redshifts \citep[e.g.][]{Fong_2015, lien2016}.
It is worth cautioning that the assignation of a given GRB to the short-hard class is also uncertain in some cases \citep{Bromberg2013}, and also that a small proportion of the candidate host galaxies may be chance alignments rather than the real host \citep{levan07}. The existing sample should be understood to be incomplete and inhomogeneous, but still sufficient to show general trends. 

Based on this observed population of SGRBs, a non-negligible fraction  \citep[$\sim 15\%$;][]{fong2013demographics} appear to have no obvious host galaxy. These events are considered ``host-less" and are typically seen at distances $30-75\,$kpc in projection from the nearest host galaxy candidate \citep{berger2010short,Tunnicliffe_2014,fong2013locations}. 
Studying the orbital evolution of the progenitor binary pairs relative to their host galaxy, before they merge, may allow us to investigate the frequency of progenitors migrating to distances such that they are classed host-less. Establishing constraints on the expected projected offsets of SGRBs, and the fraction of host-less events, may also aid in the efforts to identify the original host of the progenitor binary where it is uncertain, and by extension an improved understanding of
the distribution of these compact binary mergers over the cosmological history. 
This combination of properties supports the compact binary merger hypothesis, in which the binary systems can have long lifetimes before merger, and can receive large kicks during the supernovae in which the compact remnants were formed \citep{hobbs2005statistical,Nakar2007Review}.

In order to understand the relationship between the offsets of detected SGRBs and their host galaxies, we created a population of synthetic binaries and evolved them within hydrodynamically simulated galaxies. 
The compact binaries of interest discussed in this paper are primarily NSNS and BHNS as these 
are the most likely progenitors of SGRBs 
\citep{narayan1992gamma,narayan2001accretion,katz1996long,lee2007progenitors}. 
Previous studies have also explored the relationship between galaxies and the offsets of GRBs within static potentials \citep[e.g.][]{Bloom1999,Bulik1999,bloom2002,Belczynski2006,Church2011}. In this study, we use updated stellar evolution prescriptions and evolving galaxy potentials to contrast with and predict observational SGRB offsets. The search for host galaxies and EM counterparts to GW events can be challenging as seen with the follow up of GW190814 \citep{ENGRAVE190814_2020,DES2020GW190814,GROWTH2020GW190814}. Studies such as the aforementioned and the one detailed in this paper are beneficial in improving the efficiency of follow-up campaigns by enabling prediction of the most likely host populations and offsets.  
We also comment on the merger rates predicted by our analysis, scaled by LIGO/VIRGO gravitational wave constraints for NSNS systems. 

In this work, Section \ref{sec:pipeline} introduces our code, \texttt{zELDA}, that is used for the majority of the data analysis; in Section \ref{sec:results}, we outline the results; in Section \ref{sss:sgrb-env}, we compare our results with observations of SGRBs; in Section \ref{ss:host-sgrbs}, we discuss the relationship of the host galaxies and SGRB localisations; and finally, in Section \ref{sec:conc}, we summarise our key findings.

\section{\texttt{zELDA}: Redshift Electromagnetic Localisation \& Deduction Algorithm}\label{sec:pipeline}
\begin{figure*}
    \centering
    \includegraphics[width = 1.\textwidth,trim={1cm 5cm 5cm 1cm},clip]{{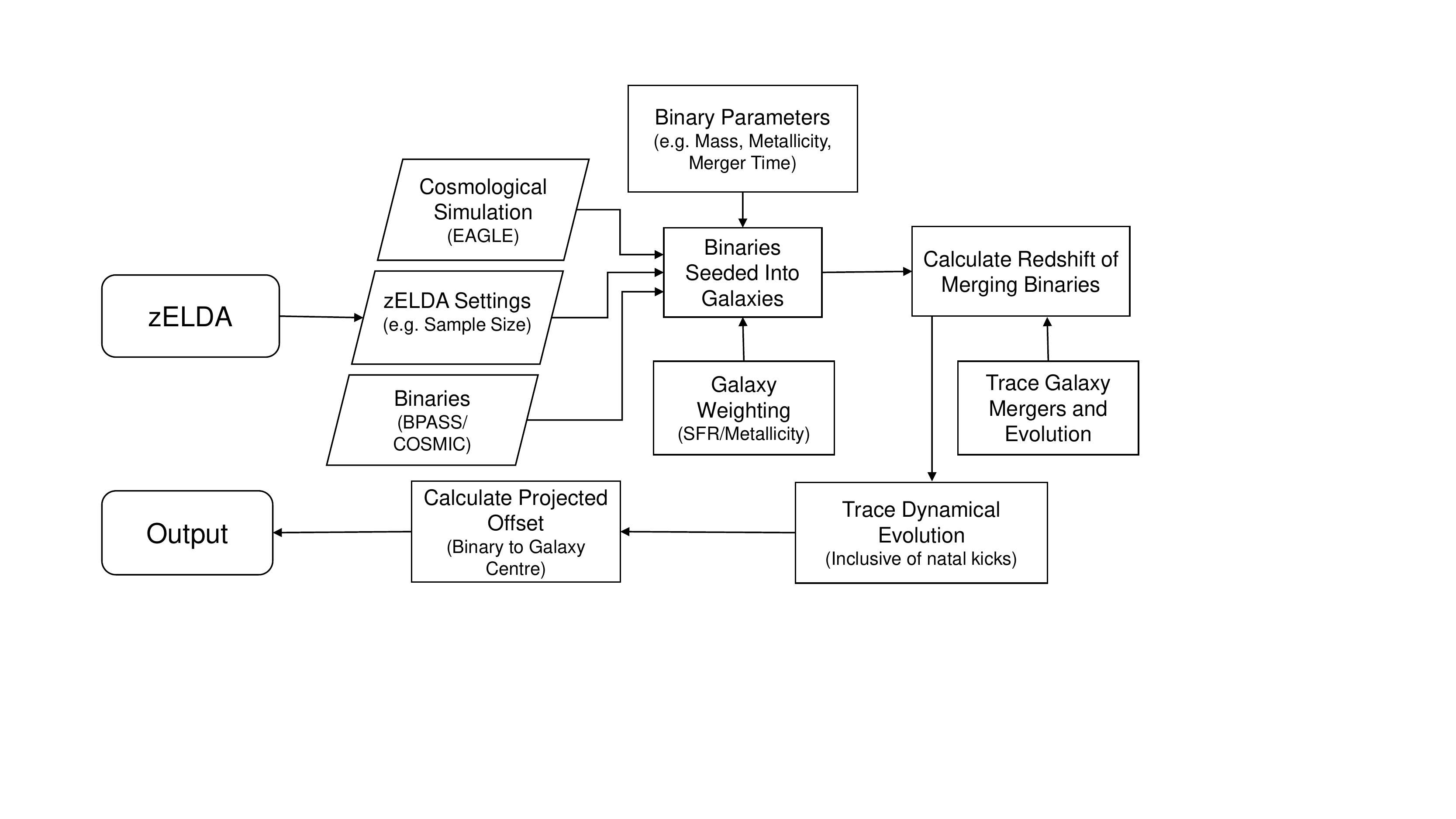}}
    \caption{A schematic breakdown of the Python based \texttt{zELDA} coding suite used within this study. Package dependencies include: 
    NumPy \citep{oliphant2006guide}, 
    Matplotlib \citep{Hunter:2007}, 
    SciPy \citep{2020SciPy-NMeth}, 
    CosmoloPy,  
    Pandas \citep{mckinney-proc-scipy-2010}, 
    AstroPy \citep{astropy:2013,astropy:2018}, 
    and GalPy \citep{Bovy_2015}.
    }
    \label{fig:zELDA-code}
\end{figure*}

The Redshift Electromagnetic Localisation \& Deduction Algorithm (\texttt{zELDA})\footnote{Repository: \url{https://github.com/hirizon/zELDA-COMBIN}} is a collection of scripts designed to process and evolve a population of compact binaries within synthetic galaxies. 
An overview of the \texttt{zELDA} code schematic is shown in Figure \ref{fig:zELDA-code} \footnote{CosmoloPy: \url{http://roban.github.io/CosmoloPy/}}. 
The objective of \texttt{zELDA} is to determine the orbital and stellar evolution of binaries relative to their host galaxy environment. 

Throughout, we have adopted the same flat ($k=0$), $\Lambda$ Friedmann cosmology as used for the \texttt{EAGLE} simulation. Specifically
the density parameters for matter and dark energy 
are $\Omega_{M0} = 0.307$ and $\Omega_{\Lambda 0} = 0.693$
respectively,
and the value for the Hubble constant used is $H_0\, = 67.77\,\text{km\,s$^{-1}\,$Mpc$^{-1}$}$ \citep{ade2014planck,furlong2015evolution}. 

\subsection{Compact Binaries}\label{ss:bin-samp}

\begin{figure*}
\vbox{
\hbox{
\includegraphics[width=0.50\textwidth]{{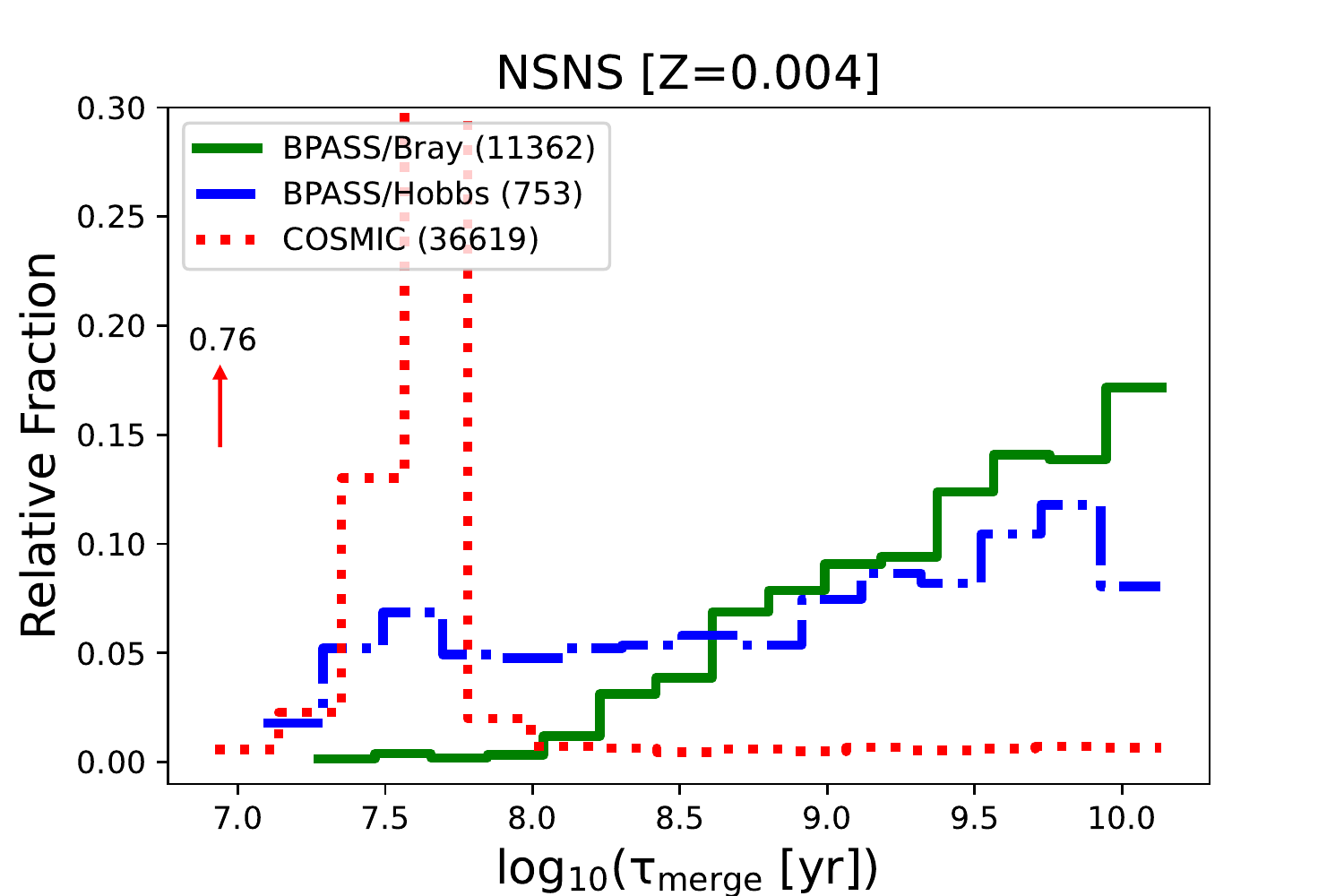}}
\includegraphics[width=0.50\textwidth]{{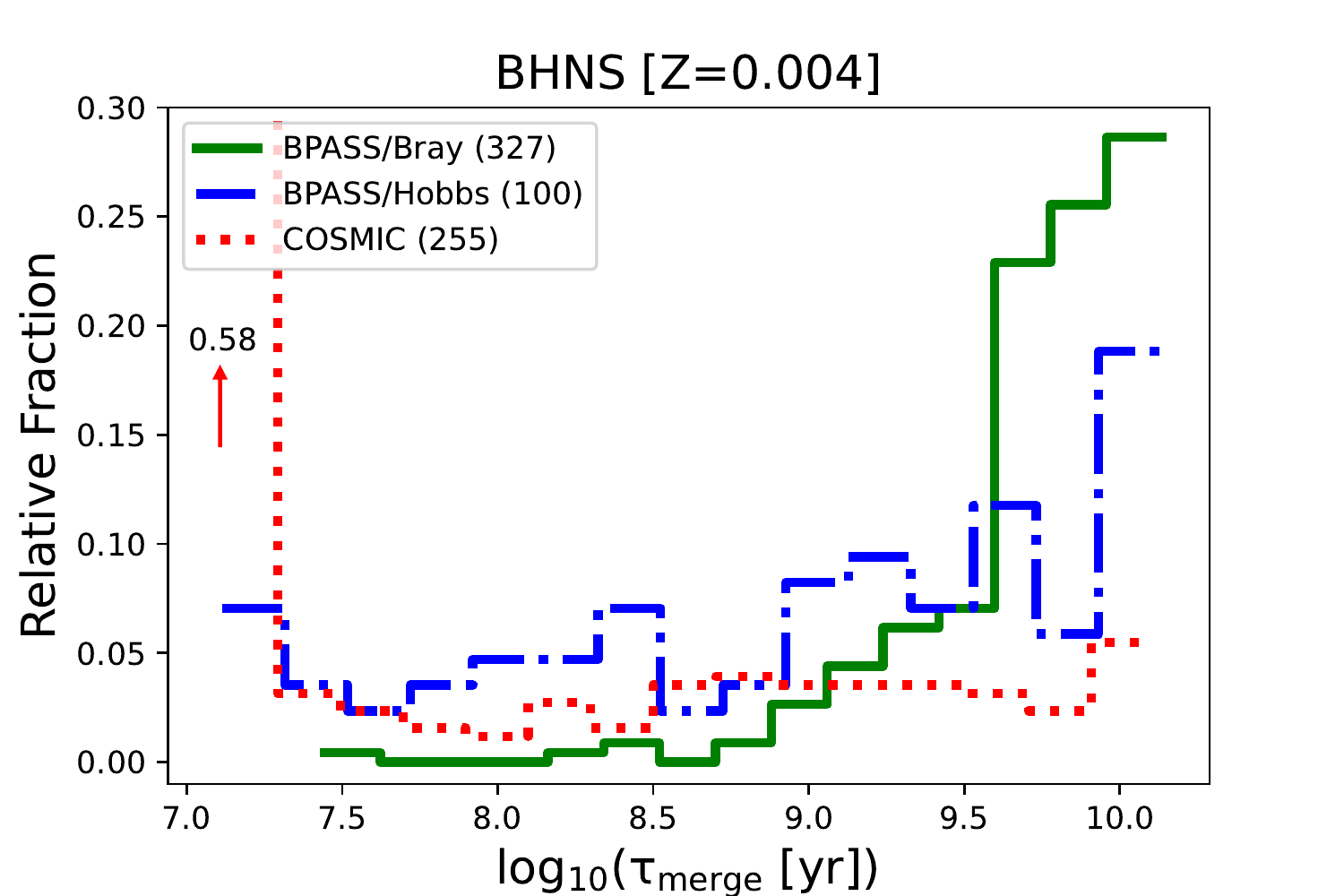}}

}
\hbox{
\includegraphics[width=0.50\textwidth]{{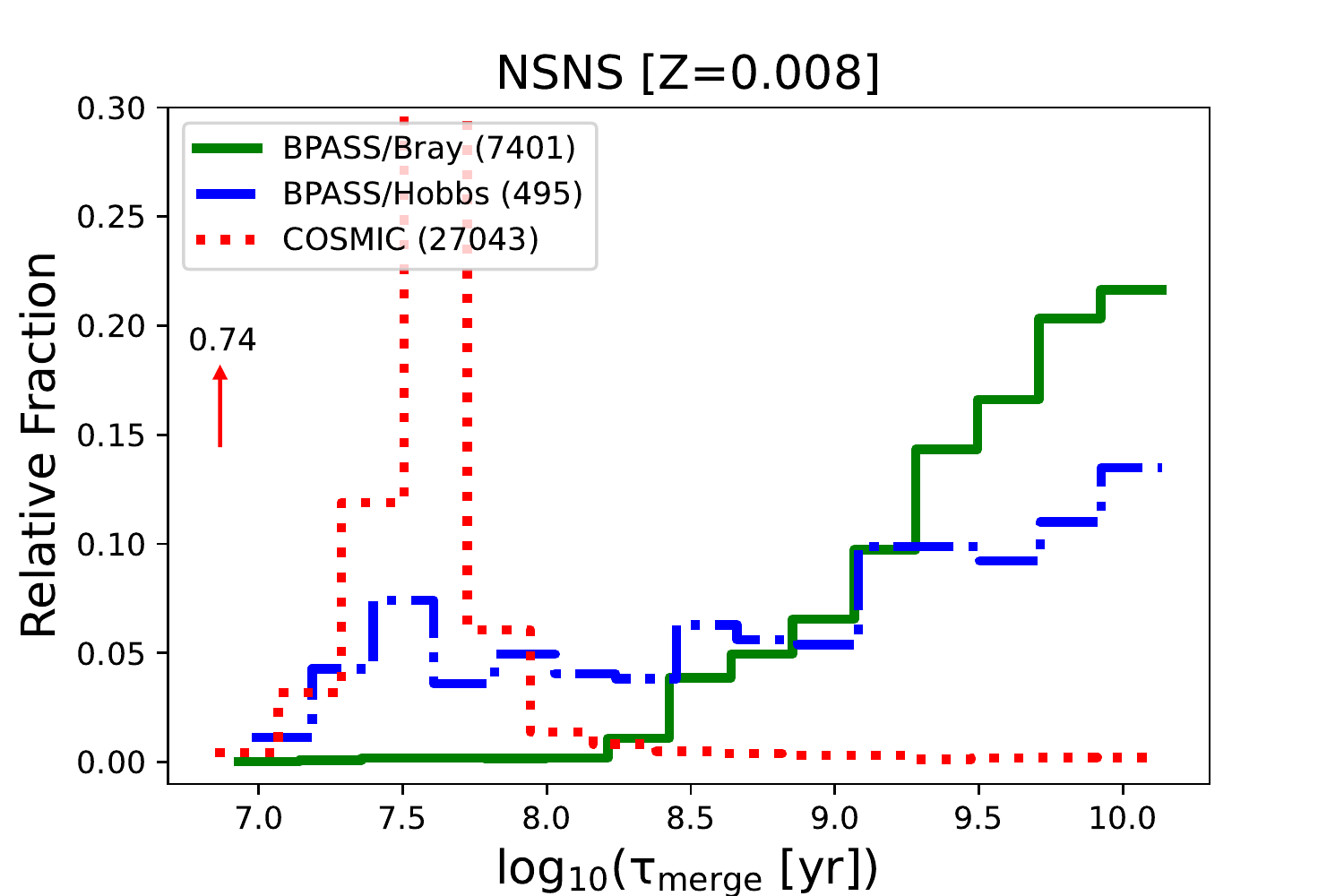}}
\includegraphics[width=0.50\textwidth]{{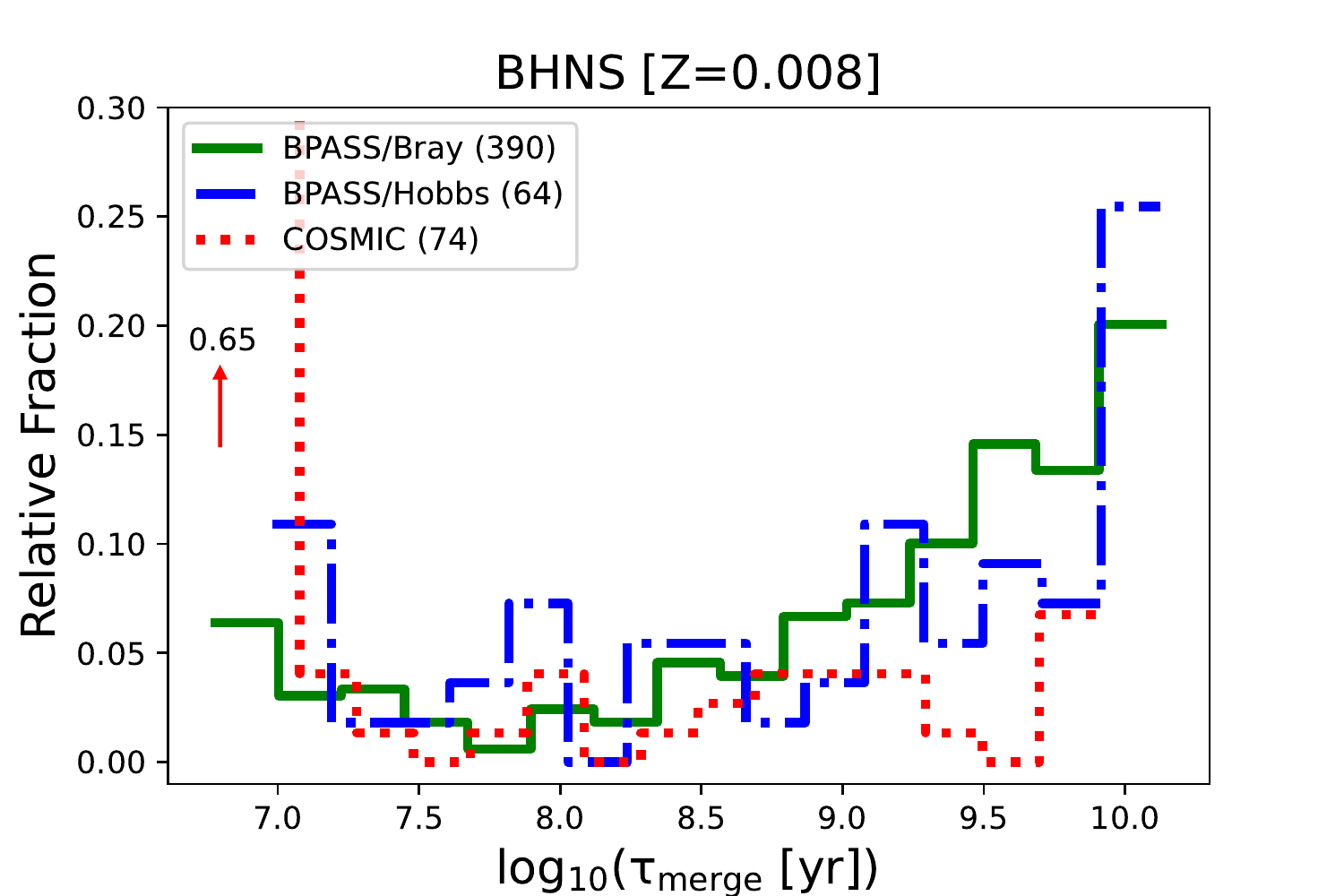}}

}

\hbox{
\includegraphics[width=0.50\textwidth]{{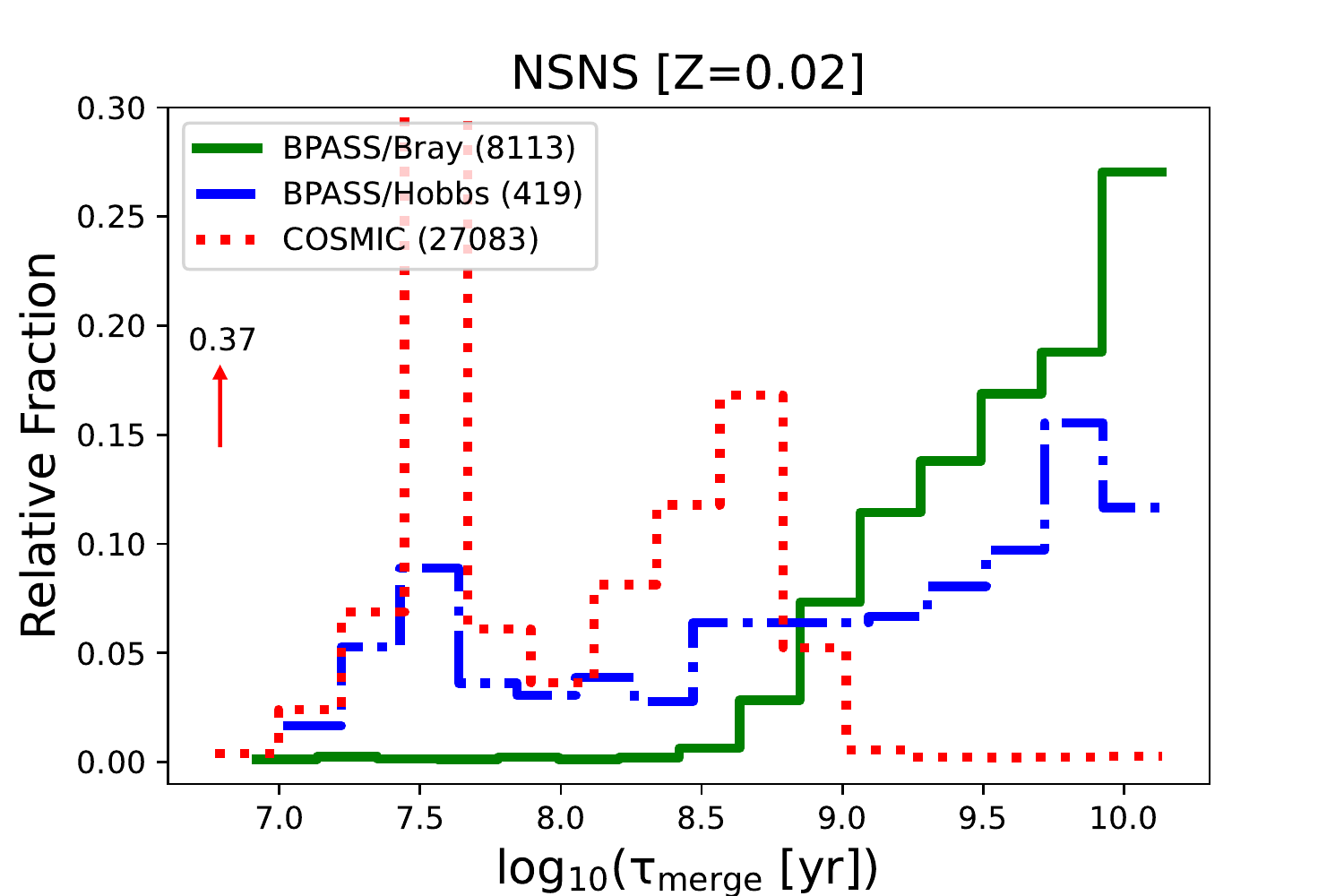}}
\includegraphics[width=0.50\textwidth]{{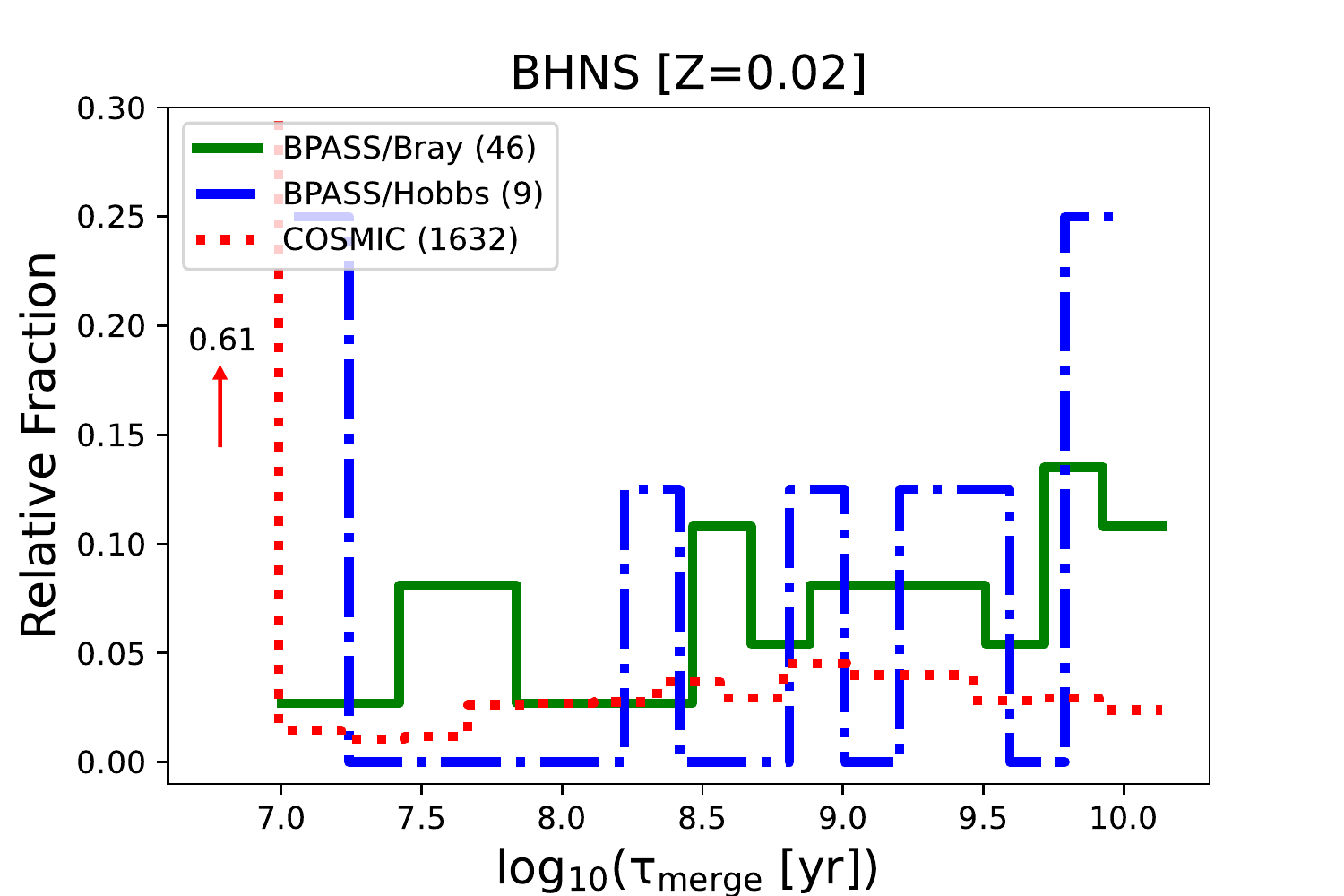}}

}

\caption{The normalised (by population) merger time distributions (i.e. $N_{\rm bin}/N_{\rm tot}$)
for the seeded NSNS and eBHNS type binaries produced using different binary population synthesis configurations/packages. The top, middle, and bottom panels correspond to binaries with metallicities of  Z $= 0.004$, $0.008$, and $0.02$, respectively. 
The numbers in brackets correspond to the total number of EM-bright unique models that are used for the binary seeding. We note an increase in systematic noise resulting from the relative rarity of eBHNS systems within the binary simulations used.
}
\label{fig:bin-rates}

}
\end{figure*}
\begin{figure}
    \centering
    \vbox{
    \includegraphics[width=0.5\textwidth]{{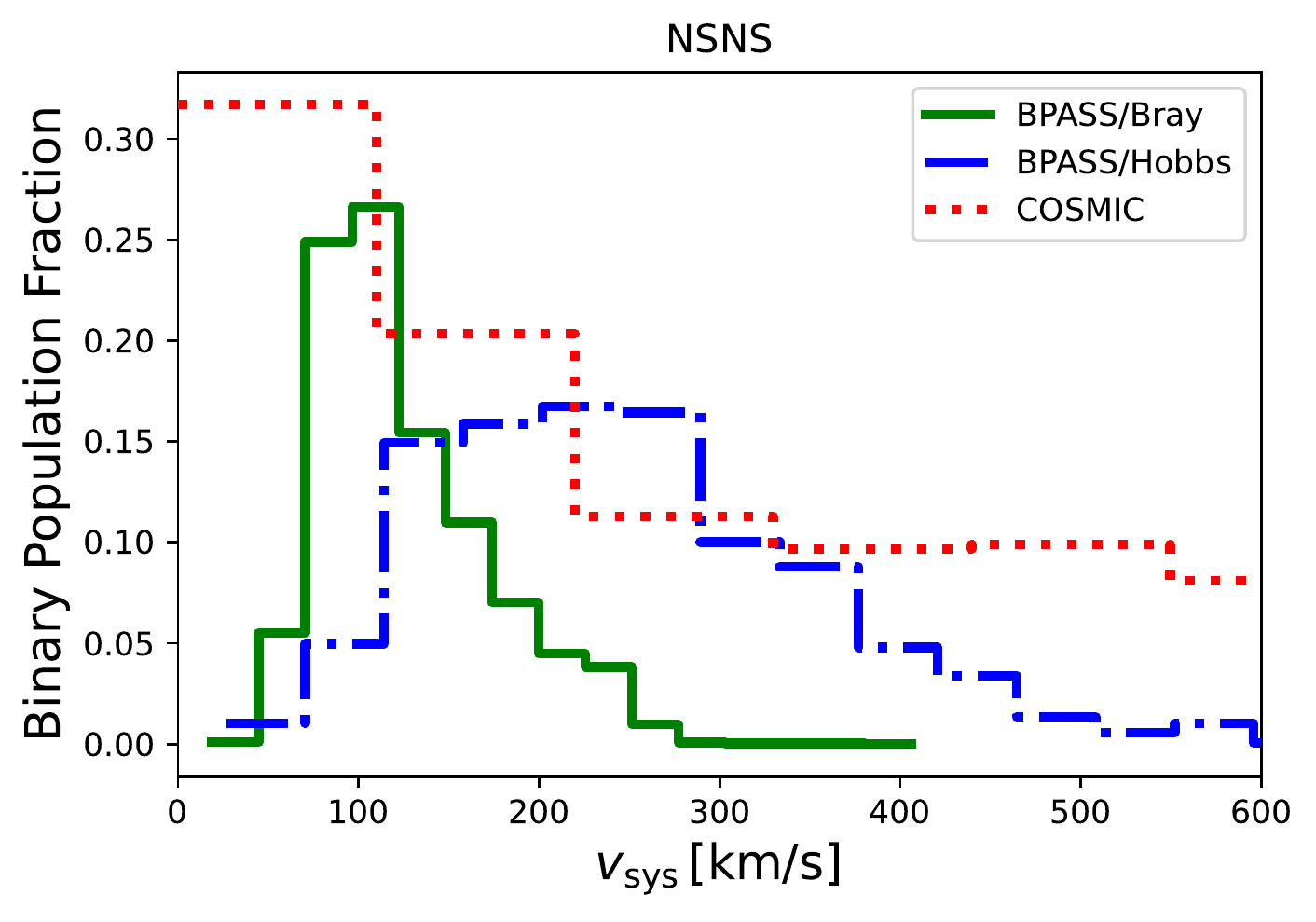}}
    \includegraphics[width=0.5\textwidth]{{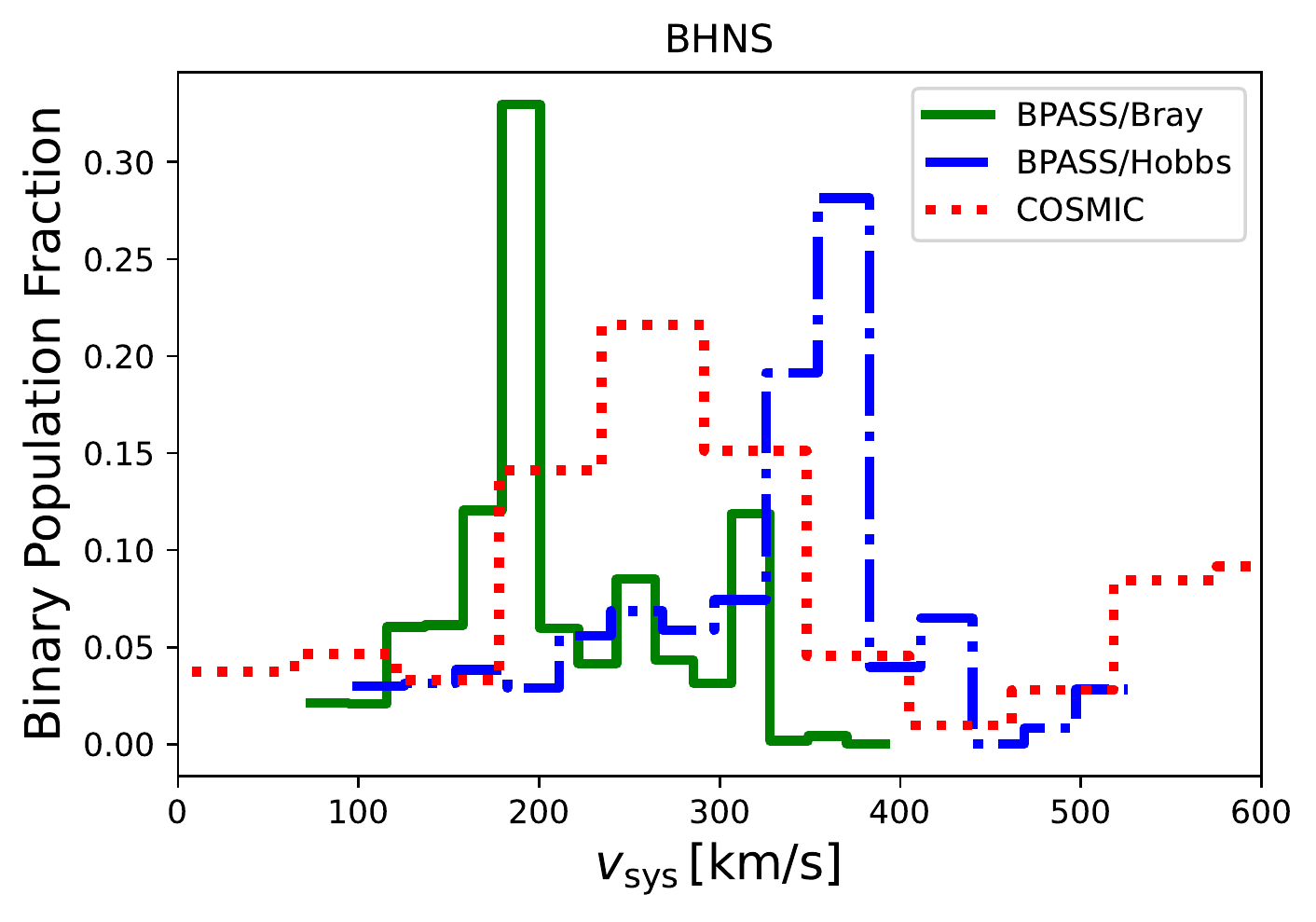}}
    }
    \caption{The final systematic velocity distributions corresponding to the seeded NSNS [upper] and eBHNS [lower] binaries for each simulation used. 
    }
    \label{fig:vel-comp}
\end{figure}

\subsubsection{Formation channels}

The evolution of binary stars with masses $>8\ \textrm{M}_{\odot}$ may result in the formation of compact binary systems that consist of a neutron star or a black hole
paired with a neutron star, i.e. neutron star-neutron star (NSNS), or black hole-neutron star (BHNS). 

For this study, we have simulated a population of synthetic compact binaries that are formed from the end-point evolution of regular, isolated, stellar binary systems.
Compact binaries may also form 
through dynamical channels 
\citep[e.g.][]{Rodriguez_2015, Rodriguez_2016,  Rodriguez_2018, Fragione_2018, Samsing_2018, Choksi_2019, Fernandez_2019,Andrews2019}, for example
in globular or nuclear clusters, or 
during galactic mergers \citep{Palmese_2017}, although the merger rate of such binaries has been estimated to be small when compared to isolated binaries \citep{Belczynski_2018, Ye_2019}.

Recently \citet{Santoliquido2020} have argued that compact
binaries formed in young stellar clusters will
have a more significant contribution to global merger rates, increasing the rate
of NSNS mergers by $\sim50$\% compared to
isolated binaries alone. 
However, this conclusion also depends  on their assumption of typically low  natal-kicks for neutron stars, which itself seems inconsistent with the frequent large impact parameters seen  for SGRBs.
Therefore, we do not explicitly include dynamically formed systems within our calculations.

\subsubsection{Simulated binary samples}\label{sss:bin-sam}
Separate simulations were performed
using two independent stellar evolution and population synthesis codes. For the first case, the binaries were produced by the Binary Population and Spectral Synthesis  \citep[BPASS v2.1]{Eldridge2017} code\footnote{Whilst \texttt{BPASS\,v2.2} is available \citep{StanwayEldridgeBPASS2017}, we use an older version for consistency with the REAPER code.} and processed through the Remnant Ejecta and Progenitor Explosion Relationship (\texttt{REAPER}) population synthesis code \citep{bray2016neutron}.
We use two configurations of compact remnant binaries from \texttt{BPASS} which correspond to the prescriptions highlighted in \citet{Bray_2018}. Henceforth, these are referred to as BPASS/Bray\footnote{The demonstrative binary sample used for the figures and numerical values in the main text.} and BPASS/Hobbs, respectively. In short, both configurations use the detailed BPASS stellar evolution models, however, the net systematic kick velocities 
for each compact object in BPASS/Bray are calculated from a kick prescription based on the mass ratio of the supernova ejecta mass and the final mass of the compact object remnant while for the BPASS/Hobbs dataset, the kick was randomly selected from the Maxwell-Boltzmann distribution outlined in \citet{hobbs2005statistical}. 
In each case, the kicks are applied in random directions relative to the orbital plane of the binaries. The relative number of compact binaries that are non-disrupted is dependent on the kick prescription used.

The third set of simulated binaries were produced using the \texttt{COSMIC v3.2.0}/\texttt{BSE} code \citep{breivik2019cosmic,hurley2002}. Binaries produced with \texttt{BPASS} are formed using detailed stellar evolution models and prescriptions, whilst \texttt{COSMIC} uses analytical scalings and is best suited for rapid-population synthesis of binaries. 
Including binaries from \texttt{COSMIC} in our analysis allows us to more densely populate the parameter space, and reduce simulation noise by increasing the number of models sampled over, in comparison to the \texttt{BPASS} binaries used.
Use of several codes to describe the stellar and binary evolution also allows us to
compare and contrast results stemming from varied approaches. 
The net kick velocities 
for the \texttt{COSMIC} compact objects are obtained from a Maxwellian distribution with a velocity dispersion of $265\, \rm{km s^{-1}}$ which is reduced by the factor of $1-f_{\rm fb}$, where $f_{\rm fb}$ is the fraction of supernova ejecta falling back onto the remnant object \citep{Fryer2012}.

The stellar evolution codes provide outputs for a range of discrete metallicities. We use binaries corresponding to a
sub-solar, metal poor ($Z=0.004$) stellar population, a moderately metal abundant ($Z=0.008$) population, and a roughly solar, metal rich ($Z=0.02$) population. The role of metallicity in the formation of the binaries is discussed in Section\,\ref{sss:bin-seed}.  Noticeably, the \texttt{COSMIC} binaries exhibit lower merger times than the \texttt{BPASS} systems likely owing to the smaller separations between the compact objects within the sample. This can be cumulatively attributed to the differences in the adopted stellar evolutionary prescriptions between both simulations. \texttt{COSMIC} binaries initially have elliptical orbits and incorporate fallback on to the compact objects from ejected supernova material. Conversely for \texttt{BPASS}, the orbits of the binaries are circular and the supernova ejecta fallback is not considered. Other differences in simulation input fractions are also present and are detailed in \citet[BPASS][]{StanwayEldridgeBPASS2017} and \citet[COSMIC][]{breivik2019cosmic}. 
Within the work presented in this paper, BPASS/Bray is used as a primary fiducial model.

Whilst we expect the majority of NSNS binaries to produce an EM counterpart such as an SGRB, 
only BHNS systems with a  comparatively small mass ratio are expected to. 
 The formation of an EM counterpart depends on the ejecta mass released from the neutron star when merging with its companion black-hole \citep{Tanaka2014, Kawaguchi2016, Hinderer_2019,Foucart_2019}. 
For more massive black holes the NS can pass the event horizon before it is tidally disrupted, therefore, stifling the formation of an EM counterpart. 
The mass ratio of the black-hole to neutron star, $Q=M_{BH}/M_{NS}$ is, therefore, crucial in governing the mass remaining following the coalescence of the binary \citep{Shibata_2011,Ruiz2020}. 
However, a higher black-hole spin increases the value of $Q$ for which significant NS ejecta is expected \citep{Barbieri_2019}.
For this study, we use a $Q<3$ threshold for the mass ratio \citep{Hayashi2020}. The notation of ``eBHNS" within this work refers to the potentially EM bright sub-population of systems that satisfy our condition.   This applies to $\sim 30-37\%$ 
and $\lesssim\,1\%$ of the \texttt{BPASS} and \texttt{COSMIC} BHNS binaries, respectively. Notably, \texttt{COSMIC} imposes a mass-gap between $3-5\,\rm{M_\odot}$ which results in a smaller fraction of BHNS binaries within the $Q$ threshold. 
 The formation of these systems relative to the total BHNS population is rare. The number of unique eBHNS models produced by \texttt{BPASS} and \texttt{COSMIC} are relatively smaller than the simulated NSNS binaries. Regardless, we have included these binaries to account for their contribution towards the formation of SGRBs, although it does lead to a more sparse sampling of the merger  time and kick velocity distributions for eBHNS systems, as is evident in Figures \ref{fig:bin-rates} and \ref{fig:vel-comp}.

\subsection{Cosmological simulations}\label{ss:cosmo-sim}

We place the simulated binaries into environments representative of the real universe. To achieve this, we allocate these systems to host galaxies produced by a cosmological simulation.
We use the publicly available output from the Evolution and Assembly of GaLaxies and their Environments (\texttt{EAGLE} ) project\footnote{http://icc.dur.ac.uk/Eagle/database.php} \citep{Schaye_2014,crain2015eagle,McAlpine2015}, specifically the RefL0100N1504 sample; the box size of $(100\,$Mpc$)^3$ gives us a large, cosmologically representative volume to work with. 
Galaxies with stellar masses in the range, $ \,\textrm{M}_*/\textrm{M}_\odot> 10^8$
, are considered well resolved 
and successfully reproduce the observed stellar galaxy mass function to $\lesssim 0.2\,$dex within this mass range \citep{furlong2015evolution}. We also include poorly resolved galaxies with masses close to the baryonic particle mass used in the simulation, $\sim 10^6\,\textrm{M}_\odot$ \citep{crain2015eagle}, to enable us to gauge the fraction of binaries that are seeded (see Section \ref{sss:bin-seed}) into very small host galaxies, while recognising that individual galaxy properties are not well determined in these cases. 

The simulation is split into 28 non-uniformly spaced 
epochs
spanning from $z\sim15$ to $z\sim0$.
The use of \texttt{EAGLE} cosmological simulations allows binaries to be seeded in galaxies over cosmic time.
These galaxies will undergo mergers and produce typically larger galaxies of varied morphologies \citep{Qu_2016} before the binaries coalesce. Thus, by tracing the evolution of the binary systems within the evolving potential of the host,
we can extract the galaxy properties when the compact binary coalesces.

\subsection{Algorithm functionality}

\subsubsection{Binary seeding in \texttt{EAGLE} galaxies}\label{sss:bin-seed}

For each cosmological epoch available in our selected \texttt{EAGLE} sample, we
seed the simulated binaries in galaxies in proportion to their star formation rates 
and the averaged global metallicity. As the binaries form in discrete snapshots, a birth time offset is applied to create a continuous distribution of binary formation redshifts. 
Previous studies have used similar techniques to predict the morphology of binary merger host galaxies  \citep[e.g.][]{Mapelli_2018,Artale_2019,jiang2020simulating}.

Within \texttt{zELDA}, the binaries are seeded within their host galaxies at a distance that is weighted by the mass profile of the stellar disc.
The evolution of the binary is traced from the birth of the progenitor massive stars to the point at which the final compact binary coalesces. 
The stellar evolutionary lifetime of each star is typically short ($\sim10$\,Myr), and usually ends with a core-collapse supernova, leaving behind a compact remnant.
Each supernova provides the system with a natal-kick velocity \citep{belczynski1999effect,fragos2009understanding,giacobbo2018progenitors}. In some cases, the imparted impulse may disrupt the system entirely, and therefore, a compact binary is never created.  
For surviving compact binaries,
the direction of the kick is applied to the entire system randomly 
with respect to the galactic plane.
For sufficiently large kicks and long lifetimes, systems may migrate to great distances from their hosts before merging, consistent with the prevalence of SGRBs that are either apparently host-less or at large offsets on the sky from their likely hosts 
\citep[e.g.][]{fong2013demographics,Tunnicliffe_2014,berger2014short,zevin2019forward, gompertz2020search}. 

The overall star formation rate density within the \texttt{EAGLE} simulation is noted to be $0.2-0.5\,$dex lower than the observed rate density \citep{furlong2015evolution}. 
For the purpose of this study, we assume the star formation in different galaxy types reflects 
the real universe \citep{Madau_2014}, despite the offset in the overall normalisation.
We stack the ordered EAGLE galactic star formation rates within a snapshot to obtain the cumulative star formation rate (CSFR). For each binary, a random value between zero and the maximum CSFR is then selected, and the corresponding galaxy is taken as the host. Galaxies with higher star formation rates, are therefore, more likely to host binaries. We use three sub-samples for the synthesised binaries. These consist of a low-metallicity ($Z=0.004$), an intermediate-metallicity ($Z=0.008$), and a high-metallicity ($Z=0.02$) population (see Section \ref{sss:bin-sam}). The binaries are seeded into galaxies based on the averaged star forming gas metallicity, $Z_{\rm gal}$. This quantity is allocated to bins corresponding to the three binary metallicity populations used. An associated binary from these populations is then selected and seeded into a binned host galaxy, i.e., low-met binaries are seeded in galaxies with $Z_{\rm gal}< 0.006$, intermediate binaries in galaxies with $0.006\leq Z_{\rm gal} \leq 0.014$, and high metallicity binaries in $Z_{\rm gal}> 0.014$.
The binaries are drawn according to a Kroupa IMF \citep{Kroupa2001} profile for the primary stellar component.

Figure \ref{fig:bin-rates} shows the population density of the seeded binaries as a function of $\tau_{\rm merge}$, the total system lifetime (i.e. stellar evolution plus in-spiral lifetimes),  
for all metallicities. The relative fraction of binaries merging with a given lifespan varies between the type (i.e. NSNS or eBHNS) and the metallicity of the binary. 
The total natal-kick velocity distribution of the seeded NSNS and eBHNS binaries is shown in Figure \ref{fig:vel-comp}. 

\begin{figure*}
\vbox{
\includegraphics[width=\textwidth]{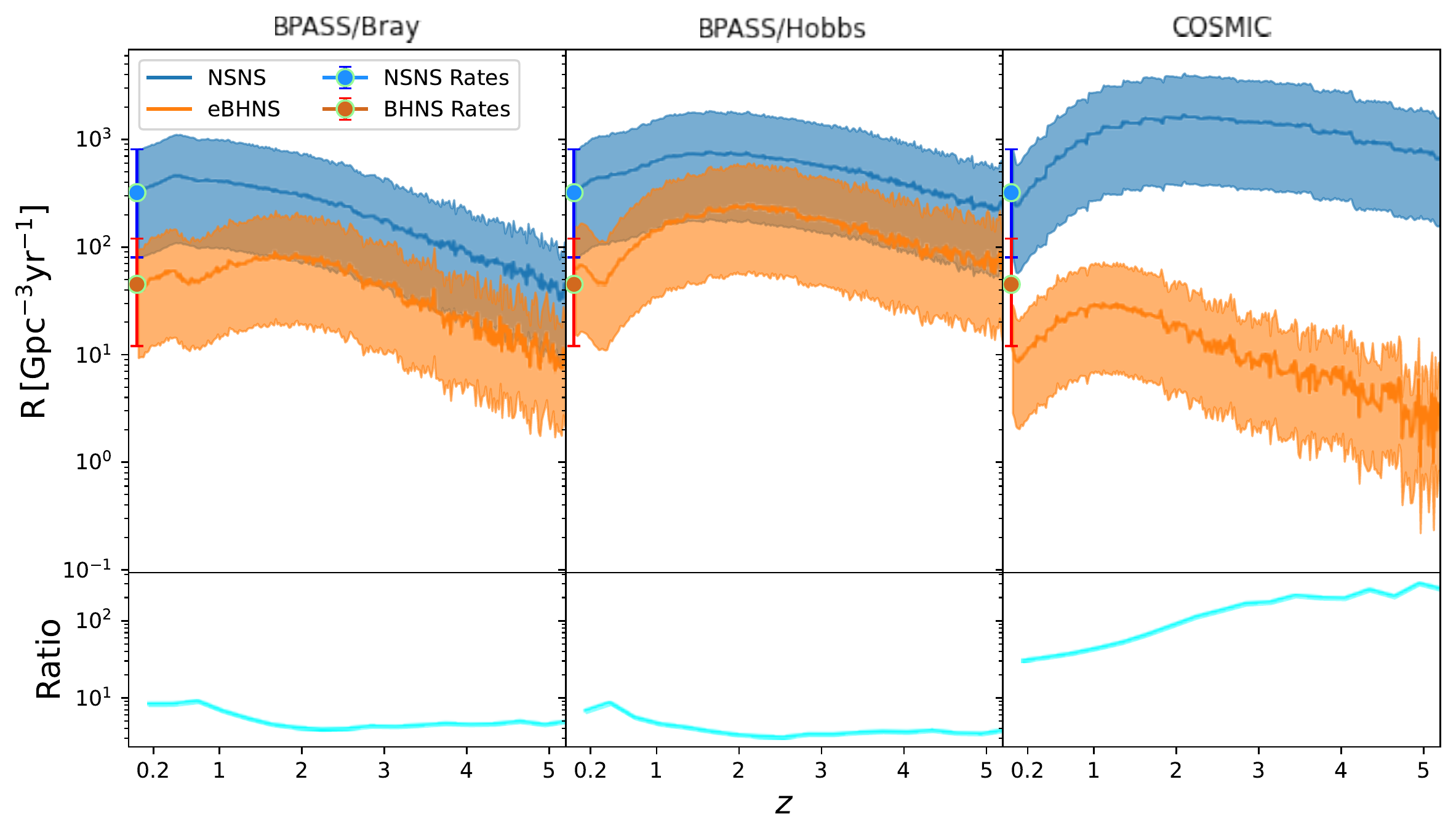}
}
\caption{Comoving volumetric rate of NSNS (blue) and eBHNS (orange) compact binary mergers normalised by the local LIGO/Virgo gravitational wave rate, $320^{+490}_{-240}\, \rm{Gpc^{-3}\,yr^{-1}}$  for NSNS systems \citep{LIGOGWTC2-20201}. 
The shaded regions correspond to the rates contained within the upper and lower bound of the LIGO/Virgo estimate used. The cyan lines in the bottom panels show the relative ratio of NSNS against eBHNS systems. The plotted points correspond to the local volumetric rate estimates for NSNS, and BHNS ($45^{+75}_{-33}\, \rm{Gpc^{-3}\,yr^{-1}}$) \citep{LIGO2021BHNS} mergers.
}
\label{fig:rates-obs-comp}
\end{figure*}

For the different binary evolution codes, the rate of merging compact binaries as a function of redshift is 
shown in Figure\,\ref{fig:rates-obs-comp}. 
These rates have been normalised using the LIGO/Virgo O3 estimate of 
$320^{+490}_{-240}\, \rm{Gpc^{-3}\,yr^{-1}}$
for the merger rate density of NSNS systems \citep{LIGOGWTC2-20201}. 
The  NSNS merger rate density always exceeds that of EM-bright eBHNS, but the ratio varies significantly between our three simulations, 
as demonstrated by the cyan lines in the figure. 

\subsubsection{Orbital evolution of kicked compact binaries within their host potentials}\label{sec:orb-ev}

For binaries that remain intact, we trace their orbital evolution within their host galaxies. The morphology and potential of the galaxy evolves according to its descendant in the following \texttt{EAGLE} snapshots encountered during the lifetime of the orbit. The gravitational potential is modelled as a static well until the next snapshot, where it is updated 
if the host evolves or merges.

In our study, the global potential 
is composed of three contributors, a Navarro-Frenk-White \citep[NFW;][]{Navarro96} profile  
for the dark matter halo, and two Miyamoto-Nagai (MN) \citep{MiyamotoNagai95} potentials 
for the stellar and gas bodies. 
The prescriptions for the galactic potential are scaled accordingly by the component masses for the \texttt{EAGLE} galaxies. 
The potential for the NFW profile is described by Equation\,\ref{eq:NFW-pot}. The \texttt{EAGLE} simulation provides particle information corresponding to the galaxies which can be used to construct the gravitational potentials. However, this would be significantly more computationally expensive with little practical gain.  
As such, we do not expect a significant difference between the two methods when averaged over the total population of galaxies.

\begin{center}
\begin{equation}
    \Phi(r) = -\frac{4\pi G\rho_0R_s^3}{r} {\rm ln}\Bigg(1+\frac{r}{R_s}\Bigg) ,
    \label{eq:NFW-pot}
\end{equation}   
\end{center}
where $G$ is the gravitational constant, $\rho_0$ is the central density, $R_s$ the scale radius, and $r$ is the radius from the central axis of the galaxy.
$R_s$ is fixed by requiring that the half mass radius for the halo matches that 
reported in EAGLE. 
The MN potential on the other hand, is constructed with the following general  prescription, 
 \begin{equation}
    \Phi(r,z) =  \frac{-GM}{[r^2 + (a+(z^2+b^2)^{1/2})^2]^{1/2}} ,
    \label{eq:MN-pot}
\end{equation}   
where 
$z$ is the height relative to the galactic plane, $a$ is the scale length, $b$ is the scale height. Increasing the ratio of $b/a$ dictates whether the geometry of the galaxy is more disc-like ($b<a$) or more spheroidal ($b\geq  a$).

The MN potential (see Equation \ref{eq:MN-pot}) for the stellar distributions are created assuming a non-flat potential. 
We can approximate the scale length, $a$, by applying a similar minimisation for the half mass of the component as seen with the estimates of $R_s$. The scale height, $b$, is approximated using the relation, $b = a [(1-\epsilon)/\epsilon]^2$, where $\epsilon$ is the stellar ellipticity for the EAGLE galaxy. A Milky Way like galaxy has an approximate scale length ratio of $b/a\sim0.1$.  

The morphological structure can be quantified using the $\alpha_m\,=\,(\epsilon^2 + 1 -T)/2$ diagnostic described in \citet{Thob2019EAGLE},  where $T$ is the triaxiality of the galaxy.
We broadly distinguish early-/late-type morphologies based on a fiducial value of $\alpha$. Galaxies with $\alpha_m \leq0.4$, are assumed to be early-type galaxies, with late-types assumed to have $\alpha_m >0.4$.  
We note that this is an essentially structural definition, and at higher redshifts other typical characteristics of early-type galaxies, such as low specific star formation rate, will not apply.
For galaxies where $\alpha_m$ can not be determined due to the absence of the required shape parameters, we use the red/blue sequence criterion discussed in \citet{CorreaEagle2017}. This affects $\lesssim 5\%$ of the early-type galaxy classifications made. 

The effects of galaxy mergers on the binary are approximated 
within \texttt{zELDA}. 
If the host merges with a larger galactic dark matter halo, we apply a pseudo-kick equivalent to the maximum circular velocity of the major halo in a random direction to approximately account for the velocity offset entailed in switching to the new origin. Tidal interactions with the central super-massive black-holes of galaxies can also result in the disruption of binaries \citep{Amaro-Seoane2012,Mainetti2016}. Similar effects may also arise from interactions with nearby gravitating bodies such as satellite galaxies. 
However, in the analysis underlined in this paper, we do not consider the effects of the central black-hole or nearby neighbouring galaxies.  As such, the resulting gravitational intricacies such as these are beyond the scope of this paper.

The orbital and galaxy dynamic prescriptions that are used in \texttt{zELDA} incorporate functionality from the \texttt{GalPy}
\footnote{http://github.com/jobovy/galpy} Python package, \citep{Bovy_2015} using the relevant mass, radius, and shape properties from the \texttt{EAGLE} simulation. 

\section{Host Galaxies of Compact Binary Mergers}\label{sec:results}
Table \ref{tab:demographics} shows the host galaxy demographics broken down for each binary simulation. 
Figures \ref{fig:nsns-ev}, \ref{fig:bhns-ev}, \ref{fig:nsns-ip}--\ref{fig:ip_cumal} and Tables \ref{tab:nsns-ip-table}, \ref{tab:bhns-ip-table} correspond to BPASS/Bray binaries. The equivalent results for BPASS/Hobbs and COSMIC are found in Appendices\footnote{The appendices are available online on the journal website.} \ref{app:Hobbs} and \ref{app:COSMIC},  respectively.
\subsection{Galaxy host characteristics of coalescing binaries}\label{ss:gal-merg}

\begin{figure*}

\centering
\vbox{
\hbox{
	\includegraphics[width=0.50\textwidth]{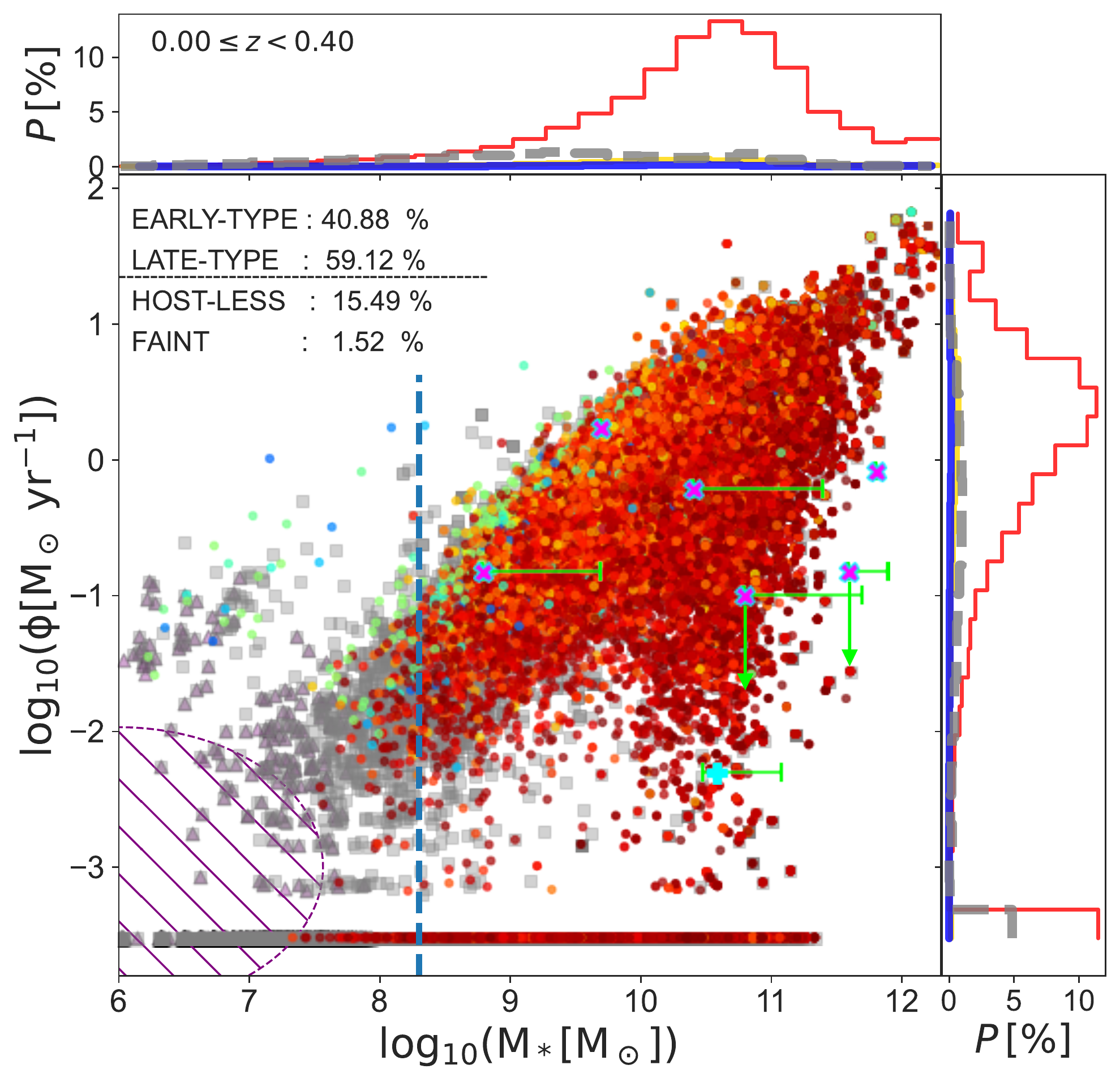}	
	\includegraphics[width=0.50\textwidth]{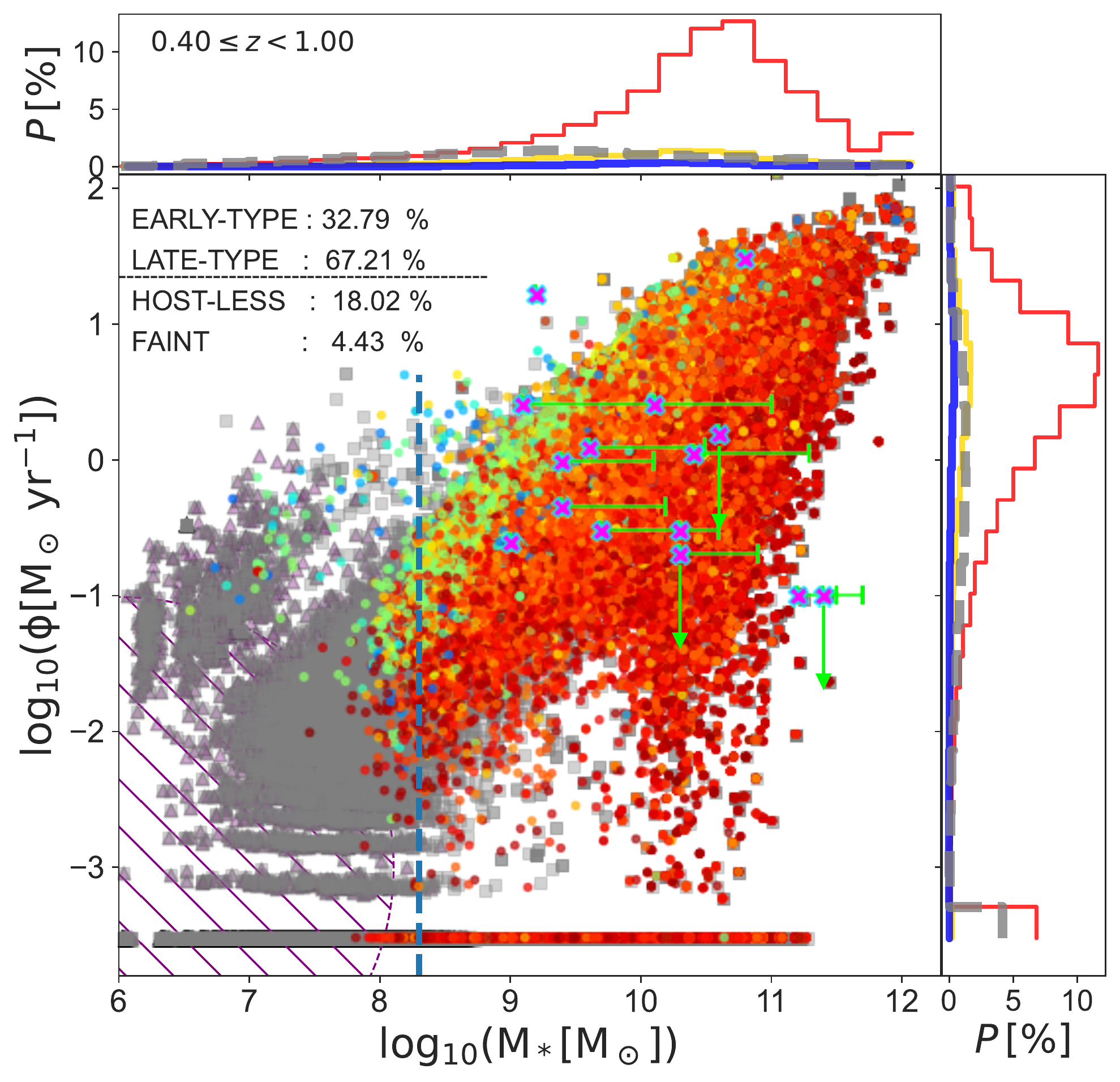}	
		}
\hbox{
	\includegraphics[width=0.50\textwidth]{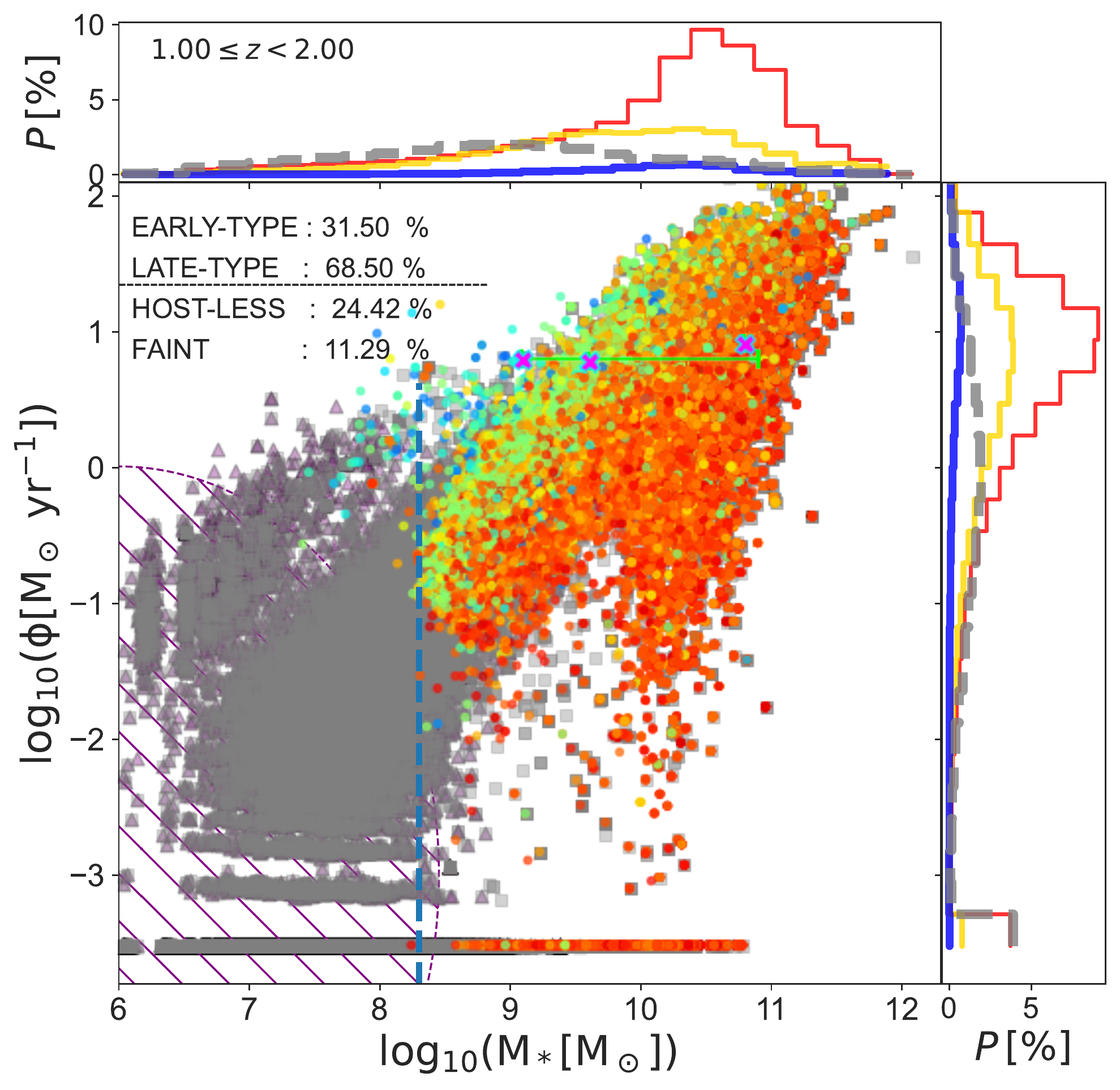}
	\includegraphics[width=0.50\textwidth]{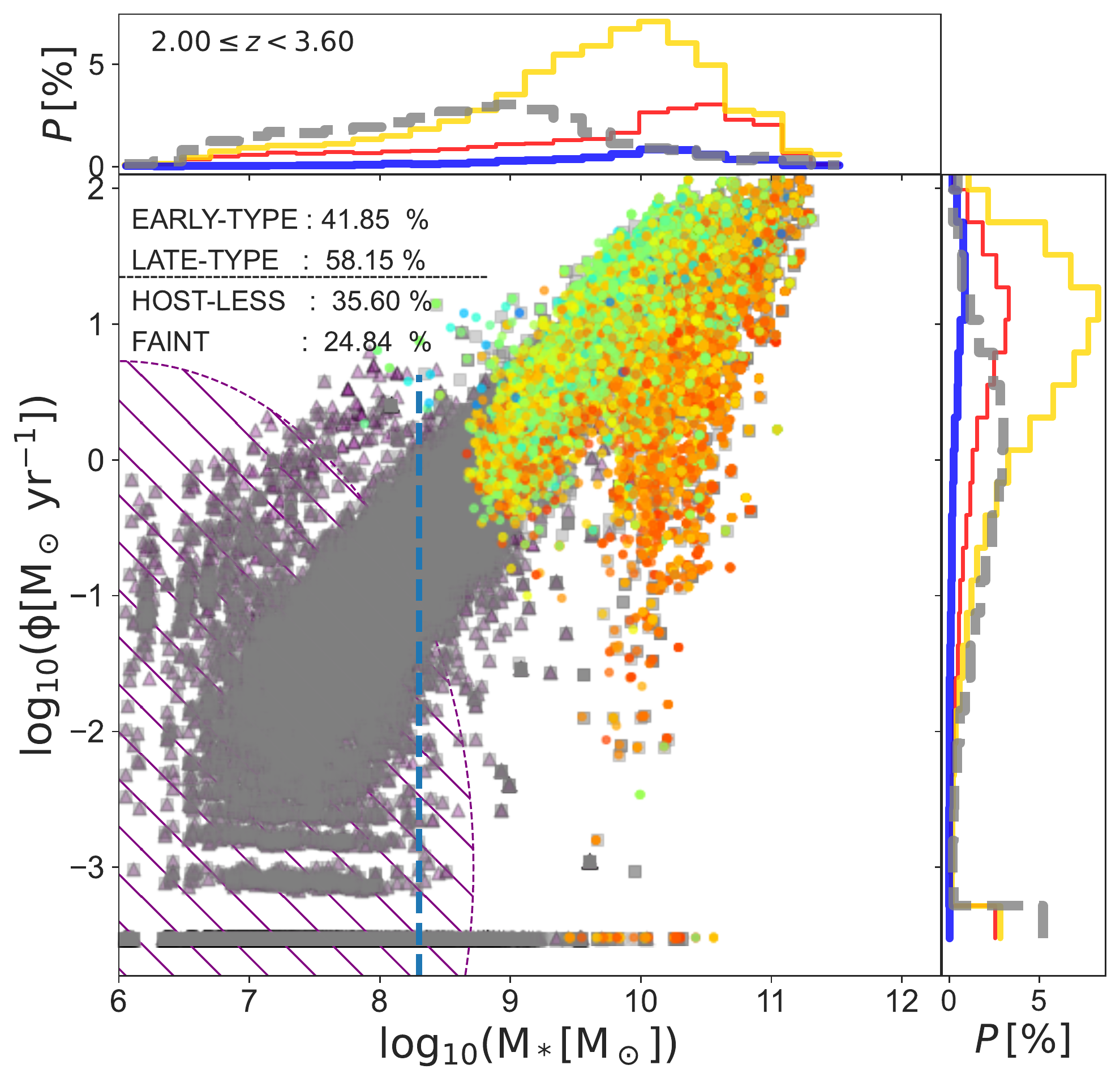}	
		}
\hbox{

\includegraphics[width=0.7\textwidth,trim={0 0 0.2cm 0.4cm},clip]{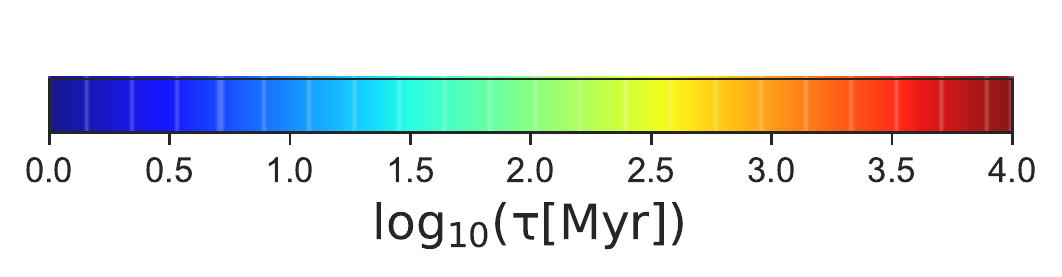}
\hsize=.2\linewidth
\vbox{
\includegraphics[scale=0.6,trim={0 0.3cm 0.4cm 0.2cm},clip]{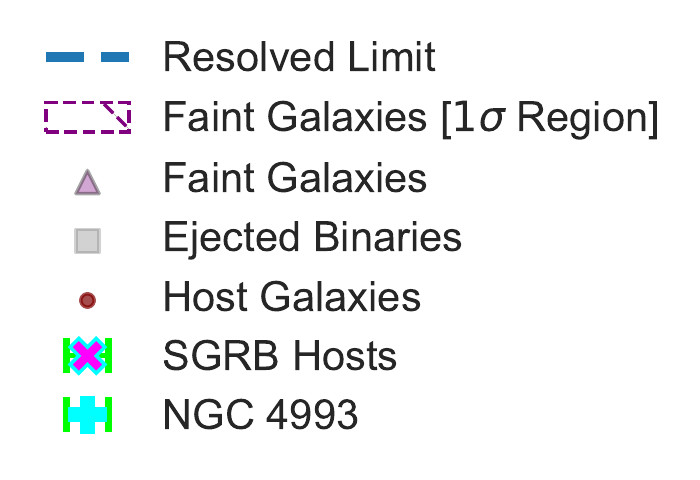}\\
\includegraphics[scale=0.6,trim={0 0.4cm 0.4cm 0.4cm},clip]{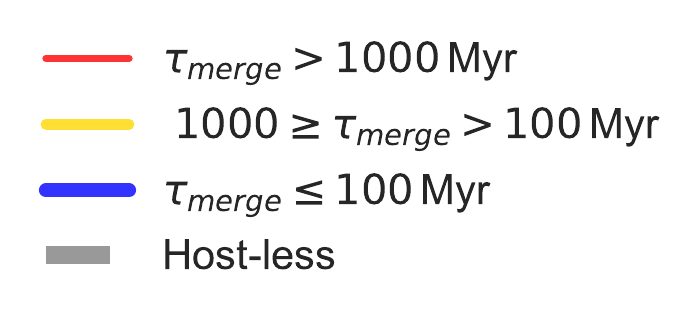}
}

}
		}
 \caption{[BPASS/Bray - NSNS] The star formation rate ($\phi$) against stellar mass ($\rm{M_*}$) evolution of NSNS merger hosts. In each panel, the point colours represent the binary merger time. The vertical dashed line indicates the resolved threshold for \texttt{EAGLE} galaxies (see Section \ref{ss:cosmo-sim}).
The hosts of ejected binaries are represented in grey. 
Galaxies that are considered
observationally faint (apparent magnitude $H>26$) are depicted as purple triangles, the indicated hatched region within $1\sigma$ deviation from the mean $\phi$ and $\rm{M_*}$ of this population. 
  The purple crosses indicate the host properties of localised SGRBs and their $\phi$ and $\rm{M_*}$ uncertainties, \citep{berger2014short, Leibler2010}.
  The cyan point corresponds to the galaxy (NGC 4993) that is the known of host of SGRB 170817A \citep{Levan2017,Myungshin2017} - the EM counterpart of the binary neutron star GW detection, GW170817. In each panel, the percentage of predicted late/early merger hosts are indicated. The estimated total population of host-less binaries (see Section \ref{sss:hostless}) and faint hosts are also stated. 
  }
 \label{fig:nsns-ev}
\end{figure*}

\begin{figure*}

\centering
\vbox{
\hbox{
	\includegraphics[width=0.50\textwidth]{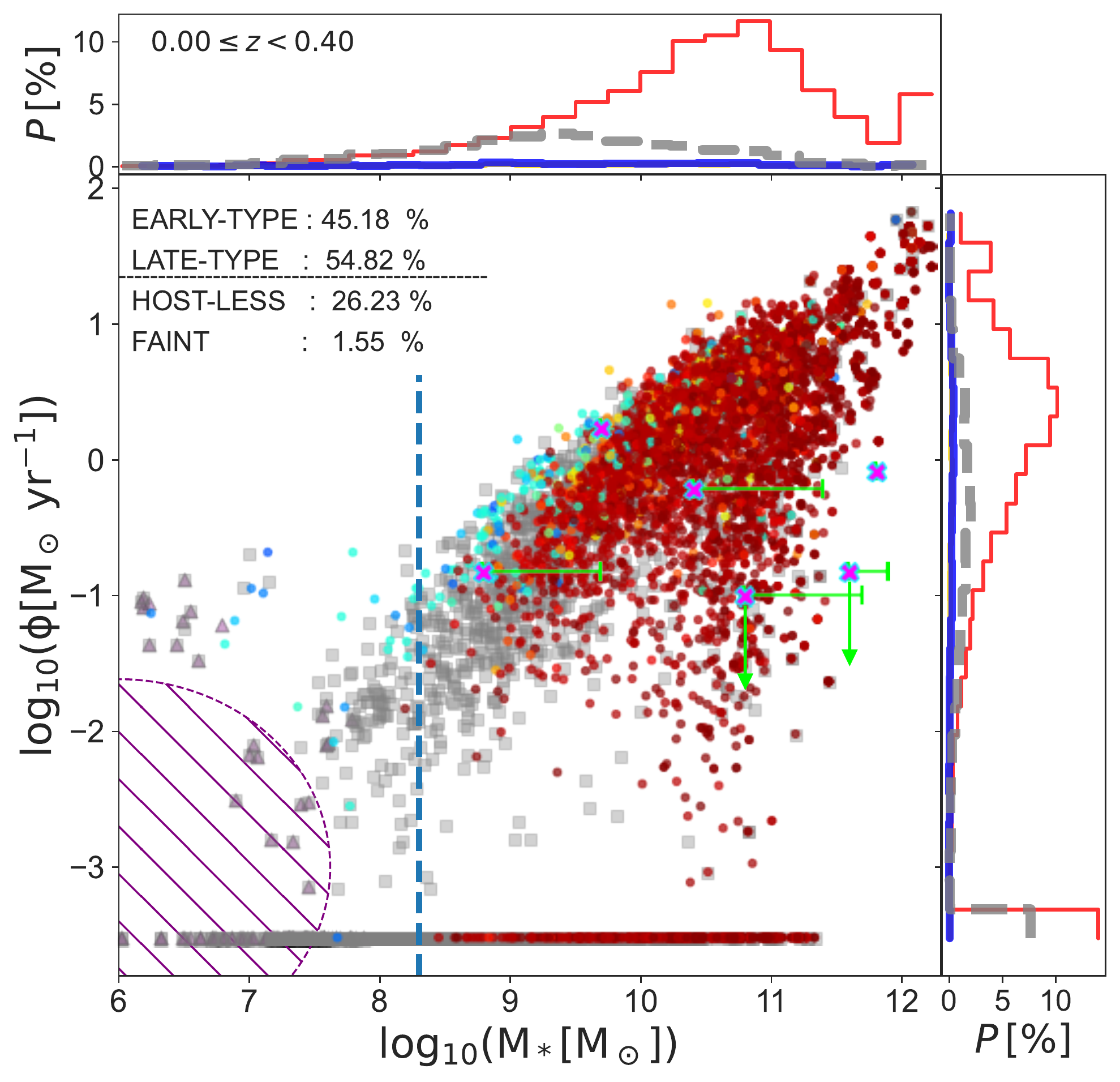}	
	\includegraphics[width=0.50\textwidth]{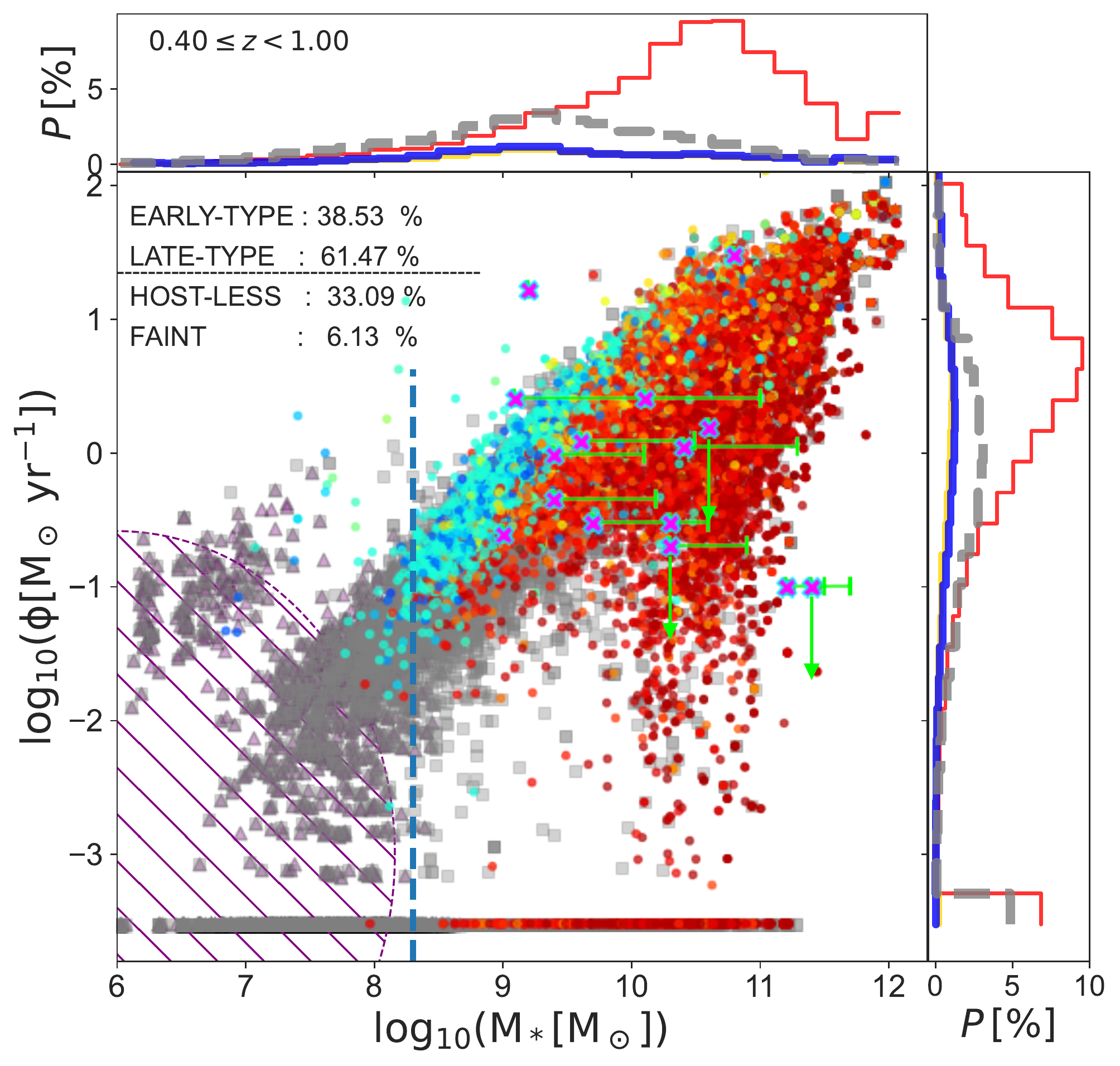}	
		}
\hbox{
	\includegraphics[width=0.50\textwidth]{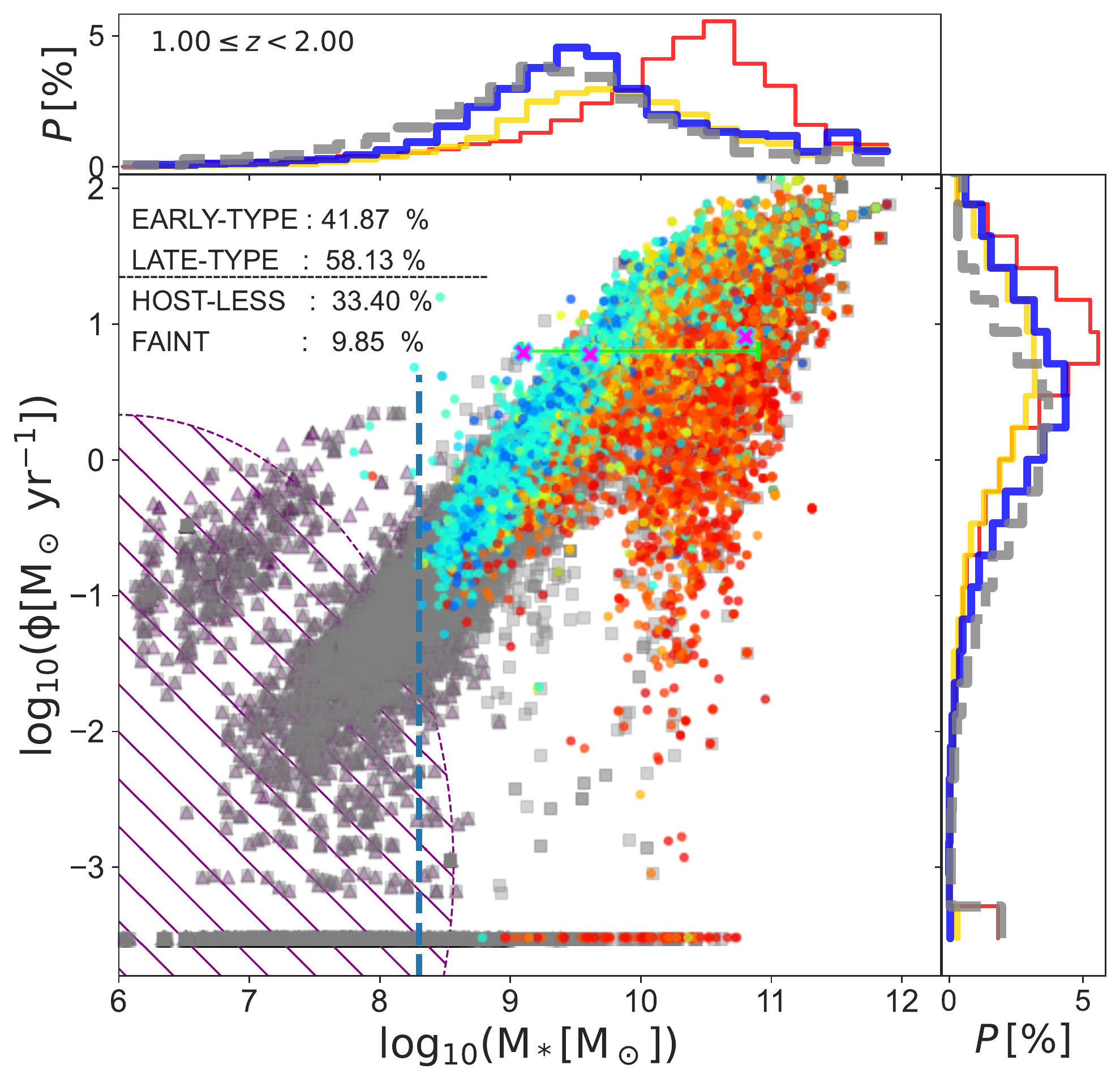}
	\includegraphics[width=0.50\textwidth]{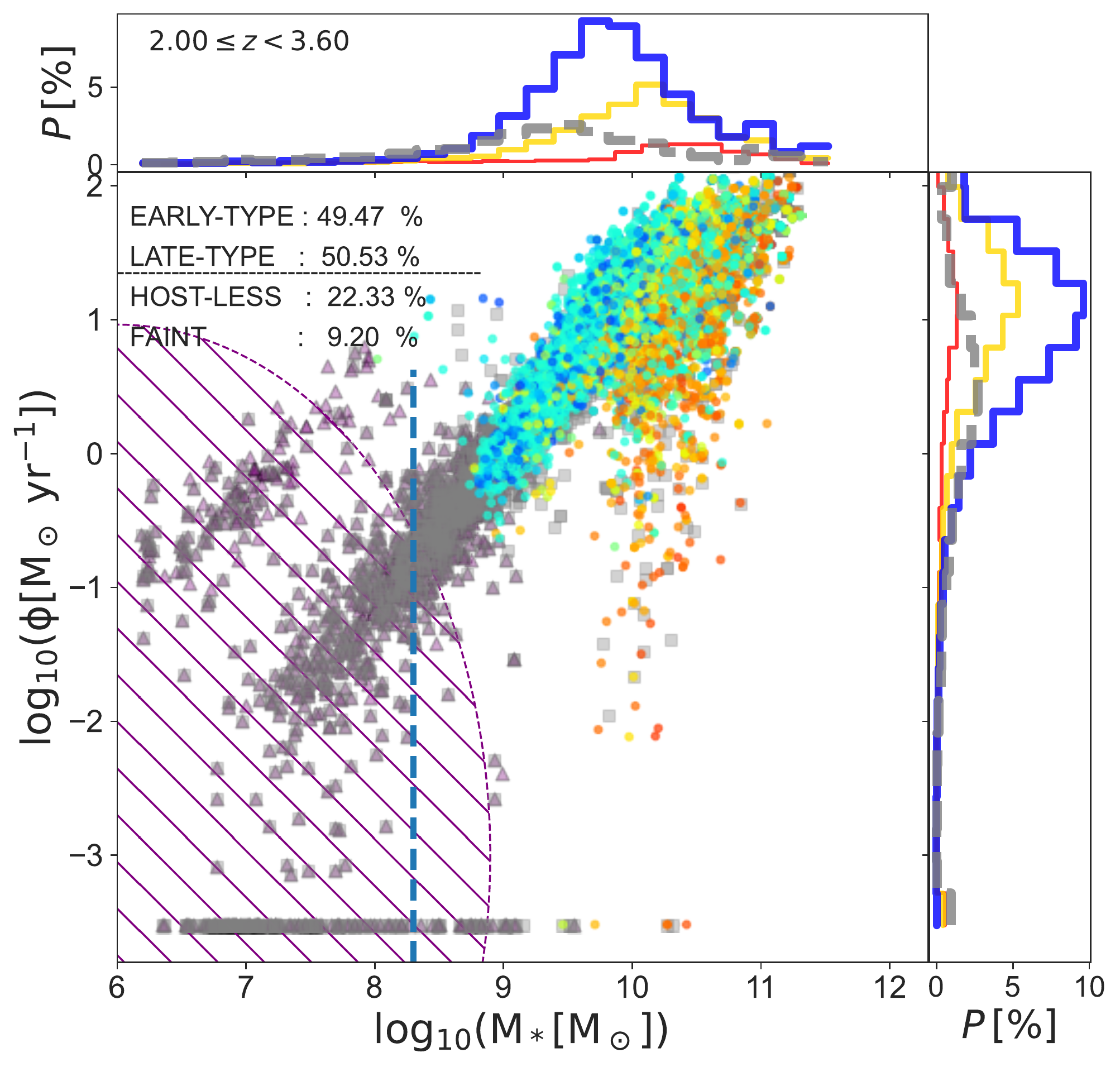}	
		}

\hbox{

\includegraphics[width=0.7\textwidth,trim={0 0 0.2cm 0.4cm},clip]{plots/colorbar.pdf}
\hsize=.2\linewidth
\vbox{
\includegraphics[scale=0.6,trim={0 0.3cm 0.4cm 0.2cm},clip]{plots/m-sfr-legend.pdf}\\
\includegraphics[scale=0.6,trim={0 0.4cm 0.4cm 0.4cm},clip]{plots/main_legend.pdf}
}

}

		}
 \caption{[BPASS/Bray - eBHNS] The star formation rate ($\phi$) against stellar mass ($\rm{M_*}$) evolution of eBHNS merger hosts. In each panel, the point colours represent the binary merger time. The vertical dashed line indicates the resolved threshold for \texttt{EAGLE} galaxies (see Section \ref{ss:cosmo-sim}).
The hosts of ejected binaries are represented in grey. 
Galaxies that are considered
observationally faint (apparent magnitude $H>26$) are depicted as purple triangles, the indicated hatched region within $1\sigma$ deviation from the mean $\phi$ and $\rm{M_*}$ of this population. 
  The purple crosses indicate the host properties of localised SGRBs and their $\phi$ and $\rm{M_*}$ uncertainties, \citep{berger2014short, Leibler2010}.
In each panel, the percentage of predicted late/early merger hosts are indicated. The estimated total population of host-less binaries (see Section \ref{sss:hostless}) and faint hosts are also stated.
 }
  
 \label{fig:bhns-ev}
\end{figure*}

The majority of binaries seeded within the \texttt{EAGLE} galaxies are born within $0.5 \lesssim z  \lesssim 3$ and merge within $0\lesssim z \lesssim 1$. The distribution of the merger time varies between different prescriptions for binary formation. This enables us to gauge the typical age and relative binary population as a function of redshift \citep{Chruslinska_2018}. The understanding of binaries merging within their host galaxy is also pivotal in providing insight into the chemical enrichment of the host galaxy \citep{Shen2015,vandeVoort2020}. Similarly,  ejected systems 
have relevance for 
our understanding of the enrichment of the intergalactic medium. 

The successful association of a compact binary and its host galaxy can further yield constraints on the system's merger timescale, if the star formation history of the host is known \citep{McCarthy2020}.
In Figures \ref{fig:nsns-ev} (NSNS), and  \ref{fig:bhns-ev} (eBHNS), we show the relationship between the host galaxy mass and SFR at various redshift ranges for merging binaries. Colour coding shows the binary lifetimes. 
which range from $10$s to $1000$s of Myr. 
At lower redshift ($z\,<\,1$), 
a large fraction of BPASS/Bray and BPASS/Hobbs merging binaries 
have lifespans of $\tau_{\rm merge}>1000\,$Myr. 
These systems formed at earlier epochs ($z>1$) within actively star forming galaxies, but ultimately merge when the host is in a more quiescent phase. 

Coalescing binaries within $z<0.4$
have host galaxies with a median stellar mass between $M_*\sim1.5-3\times 10^{10}\,\textrm{M}_{\odot}$, and a star formation rate within the range $\phi\sim 1.3-2.1\,\textrm{M}_{\odot}\,\textrm{yr}^{-1}$. 
These galaxies tend to be massive, star-forming, and disk dominated 
\citep[e.g.][]{Calvi2018}.
For comparison, the Milky Way has 
a stellar mass, $ M_*\sim 6 \times10^{10}\,\textrm{M}_{\odot}$, and a 
star formation rate of, $ \phi\sim1.6\,\textrm{M}_{\odot}\,\textrm{yr}^{-1} $  \citep{Licquia_2015}. 
The purple crosses marked in Figures \ref{fig:nsns-ev} and \ref{fig:bhns-ev} indicate the properties of galaxies associated with localised SGRB detections, 
for the  subset of cases for which these host parameters have been estimated.
Generally, the majority of these galaxies are consistent with the expected hosts of merging compact objects predicted by \texttt{zELDA}. For the lowest redshift slice, the over-plotted properties of the host galaxies reside below the peak of the host distributions. 
We note that intrinsically very faint hosts may be missed, or unidentified, in follow-up observations, so their absence in the observed sample may not be surprising. 
The properties of the NSNS merger host associated with GW170817, NGC 4993, falls within a region where very few galaxies are expected to host coalescing NSNS systems. 
An event such as this is rare based on the \texttt{zELDA} predictions.

We distinguish between early and late-type host galaxies at the point of coalescence based 
on the morphological criteria described 
in Section \ref{sec:orb-ev}. Approximately $50-70\%$ of mergers for \texttt{BPASS} compact binaries appear to originate from late-type galaxies 
at all redshifts in each simulation. For the COSMIC sample, the fraction of binary mergers originating from late-type hosts can range up to $\sim 80\%$. 
We note that \citet{Artale2020a} performed a similar simulation based on EAGLE, and concluded that at low redshifts ($z<0.1$)
early-type galaxies dominate the host population. However,  the discrepancy between our studies is due to a different approach in distinguishing galaxy morphologies. Specifically, they define early-type galaxies to have specific star formation rates, ${\rm SFR}/M_* <10^{-10}\,{\rm yr}^{-1}$. This means that many galaxies that we classify as late-type disks at low redshifts are considered early-types within their work based on their star formation rate criterion.

A non-negligible fraction of these binaries are shown to migrate to distances where they are considered to be host-less (elaborated on in Section \ref{sss:hostless}); the host galaxies for these systems are depicted as grey squares in Figures \ref{fig:nsns-ev} and \ref{fig:bhns-ev}. The figures also indicate galaxies that could be missed with an observational follow-up, as they are faint (H$>$26; purple triangles). Small spiral/dwarf galaxies, and more distant intrinsically brighter galaxies,  are likely to fall into this category due to their low apparent luminosity. The overall galaxy demographics including the contribution of host-less systems is shown in Table \ref{tab:demographics}.

\subsection{Projected galactocentric distance at merger}\label{ss:proj-dist}

\subsubsection{The impact parameter}\label{sss:ip-overview}

By following the evolution of binaries from their birth to their coalescence, we are able to observe the overall migration distance during their lifetimes.
In our simulations, these are turned into projected on-sky distances by choosing random orientations for each host galaxy and calculating the corresponding impact parameter.

\begin{figure*}

\includegraphics[width=\textwidth]{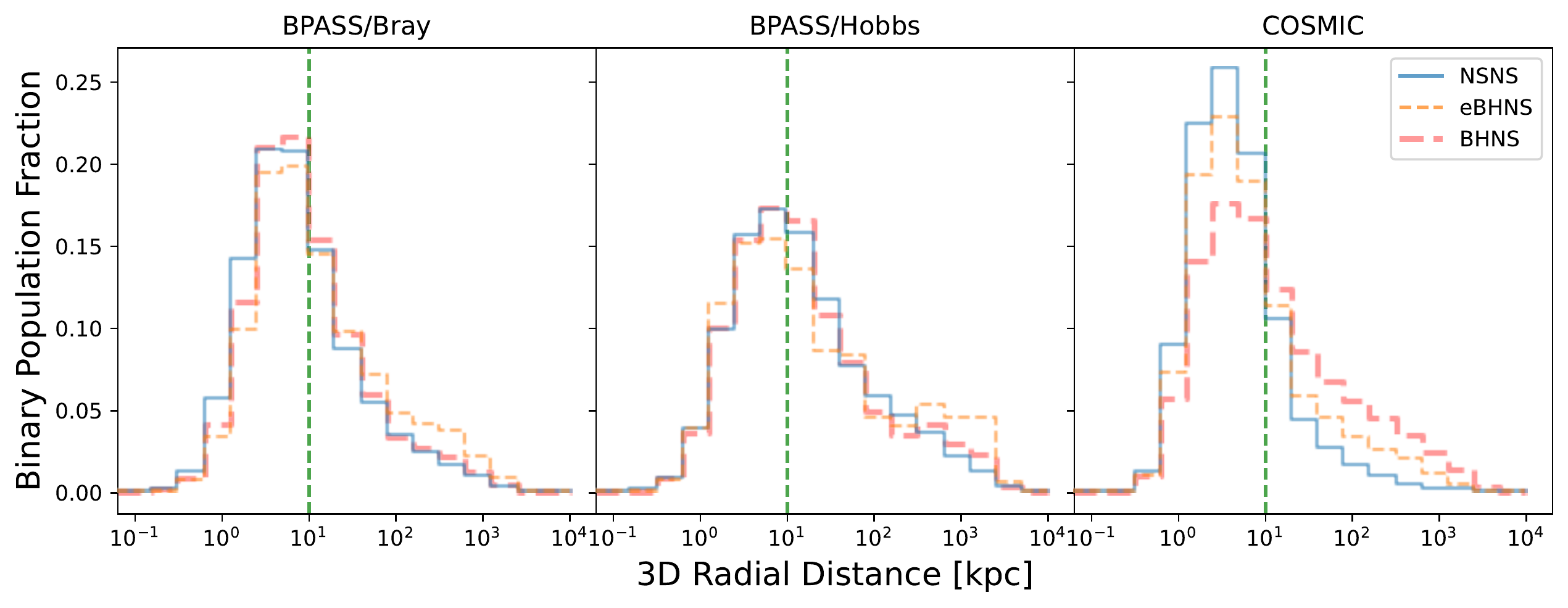}

\caption{
The population normalised ($N/N_{\rm bin}$) migrated distance distributions over all redshifts, for the BPASS/Bray, BPASS/Hobbs, and COSMIC simulated NSNS (blue) and eBHNS (orange) compact binaries. The fainter red lines correspond to the total population of seeded binaries, inclusive of non-bright EM systems (refer to Section \ref{sss:bin-sam}).
Relative to the \texttt{COSMIC} binary models sampled over, fewer unique \texttt{BPASS} models are used within our simulation. The resolution of the binned population is therefore limited*. \newline*Secondary and tertiary peaks are artefacts of noise. The green dashed line corresponds to a distance of $10\,$kpc. 
}
\label{fig:ip-dist}
\end{figure*}

Figure \ref{fig:ip-dist} 
shows the global distribution of the migration distances that the NSNS and eBHNS mergers exhibit for the various population synthesis codes used. 
We find $\sim 60-70\%$ of NSNS and eBHNS mergers occur within  $\lesssim 10 \,\textrm{kpc}$ (indicated by the vertical dashed line) from the centre of their host galaxies, for BPASS/Bray binaries. For BPASS/Hobbs binaries, $\sim50\%$ of binaries are retained within this distance. For COSMIC binaries, $\sim 70-80\%$ of systems merge within this distance. 
In extreme cases, we find $\lesssim1\%$  of these binaries merging at galactocentric distances $> 1\, \rm{Mpc}$ for BPASS/Bray and COSMIC binaries, and $\lesssim 5\%$ for the BPASS/Hobbs population. 

Figures \ref{fig:nsns-ip} and \ref{fig:bhns-ip} show the impact parameter distributions, where the impact parameter corresponds to the projected on-sky separations,  split by the host galaxy characteristics (stellar mass and star formation rate) and redshift.  
Generally, binaries with high velocities have a higher likelihood of overcoming the gravitational potential of their host galaxy and thus travelling further before coalescing. For low mass galaxies such as small disk or dwarf galaxies, the likelihood increases as the gravitational potential is weaker and therefore, easier to escape. As such, these galaxies are expected to retain very few binaries. We see this for both NSNS and eBHNS systems in the top panels of both figures.  

For binaries merging 
in larger galaxies such as large spirals (middle and bottom panels) or ellipticals (bottom panel), the coalescence is likely to occur within $\sim 10\,\rm{kpc}$ and therefore, close to or within the host galaxy.
Tables \ref{tab:nsns-ip-table} and \ref{tab:bhns-ip-table}, show the fractions of binaries merging in each panel per redshift slice, corresponding to the aforementioned figures. 

\subsubsection{Chance alignment of binaries and their host galaxies}\label{sss:hostless}
A binary is considered host-less if the offset from its host is likely to be observationally registered as a chance alignment, with a probability, $P_{\rm chance}>0.05$ \citep[e.g.][]{fong2013demographics} and/or if the host is fainter than $H>26$. The $P_{\rm chance}$ quantity is calculated using the prescription for chance alignments from \citet{bloom2002} and defined as
\begin{equation}
    P_{\rm chance} = 1 - e^{-\eta} , 
\end{equation}
where
\begin{equation}\label{eq:eta}
    \eta = \pi r_i^2 \sigma(\leq m).
\end{equation}
Here, $r_i$ is an angular distance that is determined based on the offset of a binary merger relative to the centre of its host, and the size of the galaxy. 
Specifically, if a binary merger occurs 
at a projected radius less than twice the projected half mass stellar radius ($R_{*}$) of its host galaxy, it is assumed to be ``within" the host and $r_i$ is set to $r_i = 2R_{*} $. 
For mergers occurring further out from their host galaxy centre $r_i = \sqrt{R_0^2+4R_{*}^2}$, where $R_0$ is the angular offset between the centre of the host and the site of the merger.
$\sigma(<m)$ is the  mean surface density of galaxies 
brighter than the $H$-band magnitude ($m$) of the host galaxy. 

Our procedure differs slightly from the original \citet{bloom2002} prescription in that we use the projected half-mass stellar radius  in place of the half-light radius; and we use $H$-band rather than $r$-band magnitudes as early-type galaxies in particular tend to become rapidly fainter in the $r$-band at $z\gtrsim1$ due to the redshifting of the 4000\,\AA\ break through the filter passband. 
$H$-band number counts for 
field galaxies are taken from \citet{Metcalf2006,Frith2006}, and we use a polynomial interpolation between the values to extract the mean surface density for any $H$-band magnitude. 
In the context of our simulations, we use the given apparent magnitude of the host galaxies reported by  \texttt{EAGLE}.

The $H>26$ 
criterion accounts for hosts that appear faint
(particularly those at higher redshift)  and therefore are likely to be missed when following up SGRBs,  
resulting in them being classified as host-less by observers.  
Recognising the existence of these cases is essential when accounting for potential observational biases that can arise with SGRB samples.

\begin{table*}
    \centering
    \caption{A population breakdown
    of EM bright compact binary mergers according to the probable observed host classification, for the binary simulations used. 
    The bottom segment of the table compares the total simulated ``BAT-detectable" SGRB progenitors  against the demographics breakdown of observed SGRBs from \citet{fong2013demographics}. The square brackets indicate the demographics of systems expected to exhibit bright afterglows.
    A further breakdown of the host-less population is highlighted beneath the divisions for the ``NSNS", ``eBHNS", and ``BAT-detectable SGRB Progenitors" segments. Faint galaxies are defined as sources with a magnitude, $H>26$, while binaries are considered remote if $P_{\rm chance}>0.05$. 
    \newline *
    The demographics fractions corresponding to a sub-population of SGRB progenitors merging in afterglow-bright environments (see Section \ref{sss:afterglows}).
    }
    \label{tab:demographics}
\scalebox{1.0}[1.0]{
\begin{tabular}{|c|c|c|c|c|}
\hline 
Binary Source & BPASS/Bray ($\%$)  & BPASS/Hobbs ($\%$) & COSMIC ($\%$) & \citet{fong2013demographics} ($\%$)\tabularnewline
\hline 
\multicolumn{5}{|c|}{NSNS}\tabularnewline
\hline 
Early Type Galaxies  & 26.5 & 20.8 & 26.5 & -\tabularnewline
Late Type  Galaxies  & 50.7 & 40.5 & 51.5 & -\tabularnewline
Total Obs. Host-less  & 22.9 & 38.7 & 22.1 & -\tabularnewline
\hline 
Faint Galaxies  & 0.9 & 1.4 & 5.8 & -\tabularnewline
Remote Binaries  & 14.1 & 25.7 & 9.1 & -\tabularnewline
Faint Host + Remote  & 7.8 & 11.7 & 7.2 & -\tabularnewline
\hline 
\multicolumn{5}{|c|}{eBHNS}\tabularnewline
\hline 
Early Type Galaxies  & 29.6& 25.0 & 19.3& -\tabularnewline
Late Type  Galaxies  & 38.0 & 28.8 & 63.9 & -\tabularnewline
Total Obs. Host-less & 32.4 & 46.2 & 16.8 & -\tabularnewline
\hline 
Faint Galaxies  & 1.5 & 2.9 & 1.8& -\tabularnewline
Remote Binaries  & 24.5 & 31.7 & 13.1 & -\tabularnewline
Faint Host + Remote  & 6.4 & 11.6 & 1.8& -\tabularnewline
\hline 
\multicolumn{5}{|c|}{BAT-detectable SGRB Progenitors [*]}\tabularnewline
\hline 
Early Type Galaxies  &30.2 [31.4] &23.8 [30.0] &23.6 [25.1] & $\gtrapprox$17\tabularnewline
Late Type  Galaxies  &50.6 [63.1] &41.5 [59.7] &61.0 [67.6] & $\approx47$\tabularnewline
Total Obs. Host-less  &19.2 [5.6] &34.8 [10.3] &15.5 [7.3] & $\gtrapprox$17\tabularnewline
\hline 
Faint Galaxies  &0.3 [0.3] & 0.6 [0.6] &2.7 [1.6] & -\tabularnewline
Remote Binaries  &15.1 [4.5] &28.8 [8.6] &10.1 [4.3] & -\tabularnewline
Faint Host + Remote  &3.7 [0.8] &5.4 [1.2] &2.7 [1.5] & -\tabularnewline
\hline 
\end{tabular}
}
\end{table*}

\begin{figure*}
\includegraphics[width =  1.\textwidth]{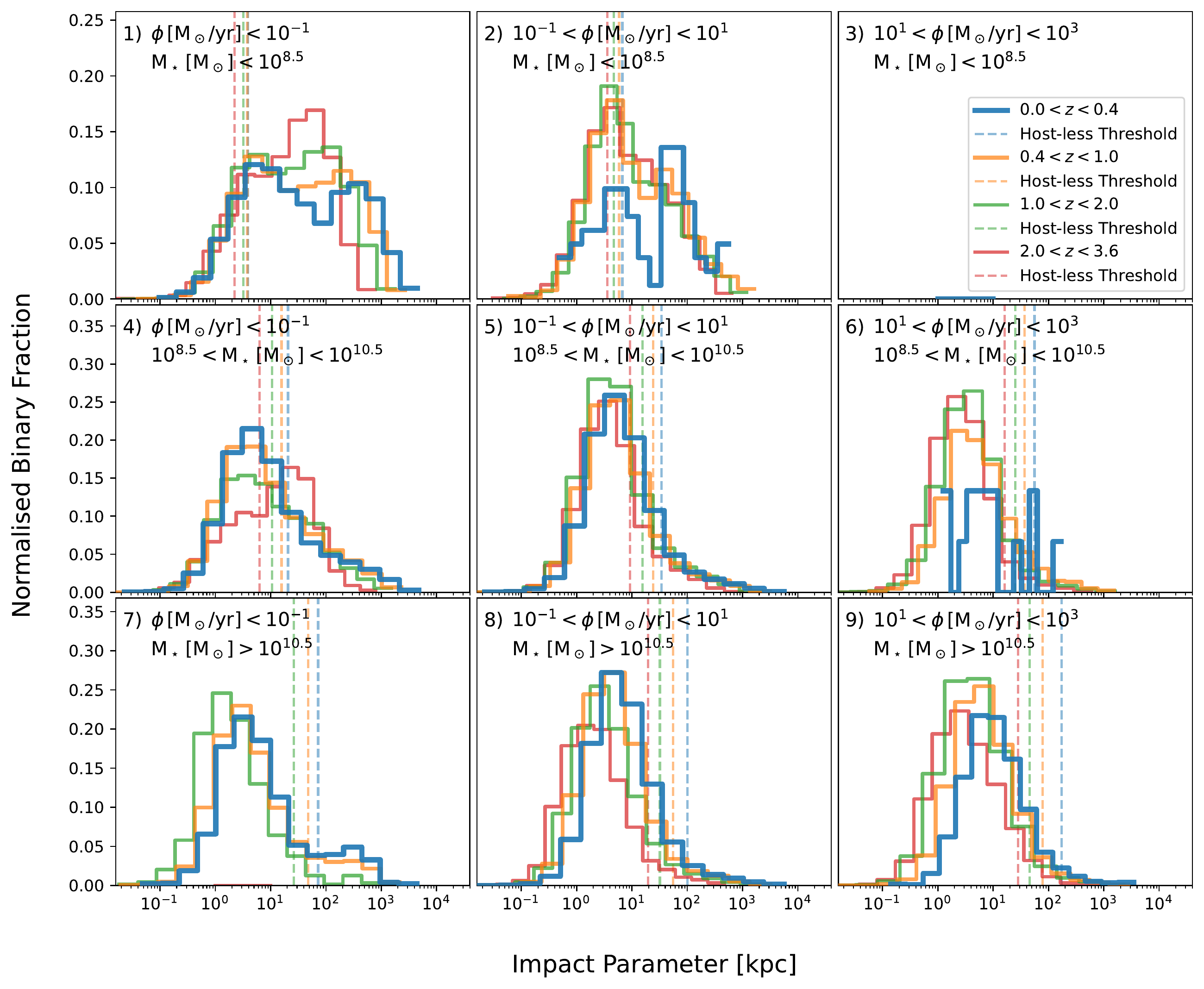}

\caption{[BPASS/Bray - NSNS] For varied redshift slices, each panel corresponds to a breakdown of the 
impact parameter distributions for different galaxy masses (row), and star formation rates (column).
Each histogram is composed of the number of binaries with a given impact parameter normalised over the total summation of merging systems within the respective panel and redshift. 
The dashed vertical line corresponds to the galaxy averaged impact parameter corresponding to the $P_{\rm chance}$ limit used to identify systems that may be potentially classed as host-less. 
A numerical breakdown of each panel is shown in Table\,\ref{tab:nsns-ip-table}. }
\label{fig:nsns-ip}
\end{figure*}

\begin{figure*}

\includegraphics[width = 1.\textwidth]{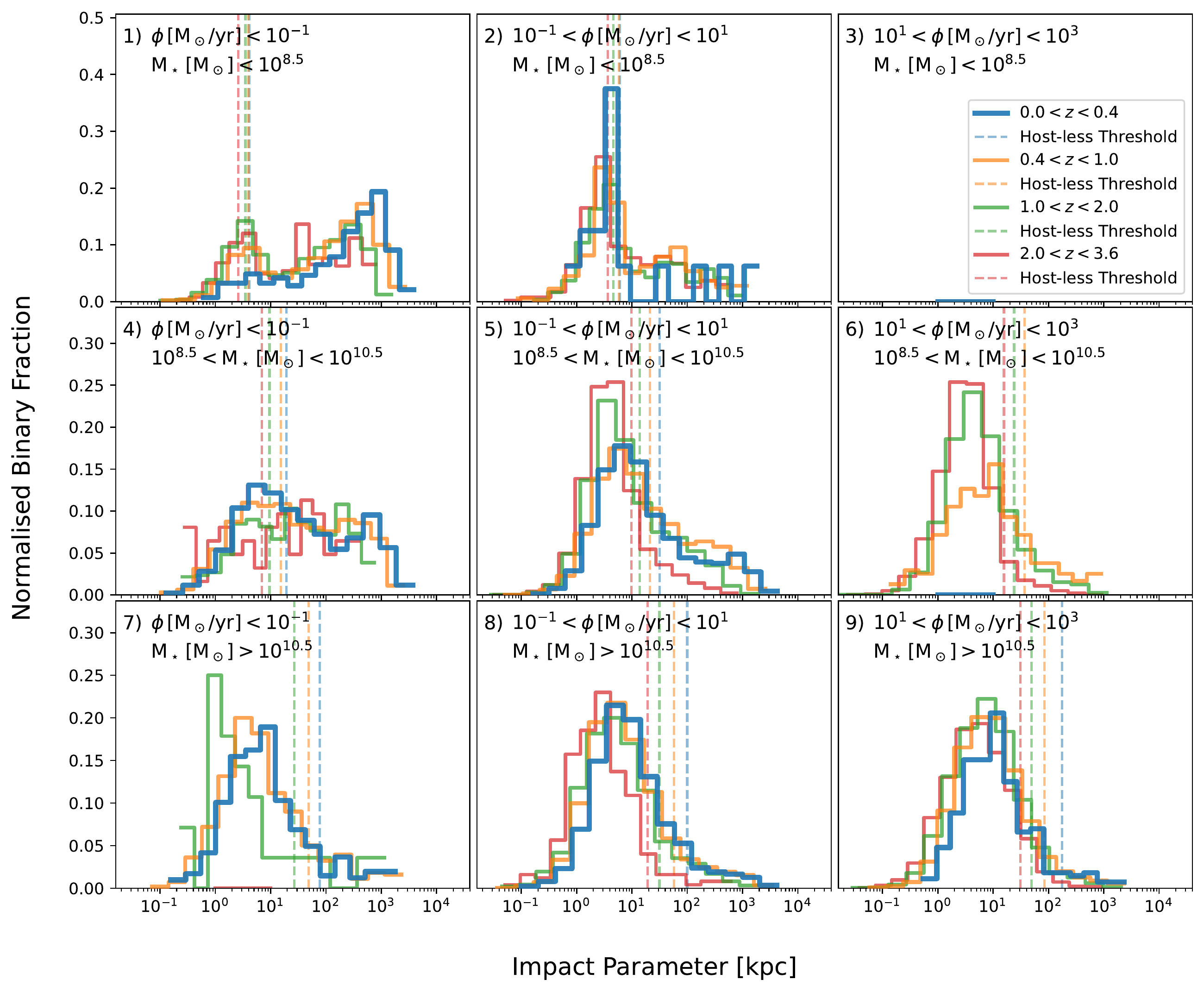}

\caption{[BPASS/Bray - eBHNS] 
For varied redshift slices, each panel corresponds to a breakdown of the
impact parameter distributions for different galaxy masses (row), and star formation rates (column).
Each histogram is composed of the number of binaries with a given impact parameter normalised over the total summation of merging systems within the respective panel and redshift. 
The dashed vertical line corresponds to the galaxy averaged impact parameter corresponding to the $P_{\rm chance}$ limit used to identify systems that may be potentially classed as host-less.  
A numerical breakdown of each panel is shown in Table\,\ref{tab:bhns-ip-table}.
}
\label{fig:bhns-ip}
\end{figure*}

 The total fraction of NSNS and eBHNS that will appear to be host-less is shown in Table \ref{tab:demographics} for each binary simulation used. 
 A breakdown of the fraction of binaries that fulfil our host-less criteria is shown for each panel of Figures \ref{fig:nsns-ip} and \ref{fig:bhns-ip} within the second sections of Tables \ref{tab:nsns-ip-table} and \ref{tab:bhns-ip-table}. 

\begin{table*}

\caption{[BPASS/Bray - NSNS] A breakdown of the relative binary mergers occurring within each panel of Figure \ref{fig:nsns-ip}. The top division refer to the relative fractions of binaries coalescing for a given redshift slice. The last column indicates the population of the binaries contained within each slice as a percentage of the cumulative population. The second section of the table shows the estimated host-less population per panel [overall fraction corresponding to the redshift slice] based on the criteria defined in Section \ref{ss:proj-dist}. 
}
\label{tab:nsns-ip-table}
\centering
\scalebox{0.87}[0.87]{
\begin{tabular}{|c|c|c|c|c|c|c|c|c|c|c|}
\hline 
\multirow{2}{*}{z} & 1  & 2  & 3  & 4  & 5  & 6  & 7  & 8  & 9  & \multirow{2}{*}{Total Population (\%)}\tabularnewline
\cline{2-10} \cline{3-10} \cline{4-10} \cline{5-10} \cline{6-10} \cline{7-10} \cline{8-10} \cline{9-10} \cline{10-10} 
 & \multicolumn{9}{c|}{Binary Fraction (\%)} & \tabularnewline
\hline 
0.0\textless z\textless 0.4  & 4.6 & 0.1 & 0.0 & 9.9 & 33.9 & 0.0 & 5.5 & 38.0 & 8.0 & 10.6\tabularnewline
0.4\textless z\textless 1.0  & 5.4 & 0.5 & 0.0 & 5.5 & 39.8 & 0.9 & 2.2 & 28.8 & 17.0 & 42.4\tabularnewline
1.0\textless z\textless 2.0  & 7.8 & 2.2 & 0.0 & 2.1 & 43.8 & 7.6 & 0.2 & 11.8 & 24.5 & 36.5\tabularnewline
2.0\textless z\textless 3.6  & 11.4 & 6.0 & 0.0 & 0.9 & 41.4 & 22.4 & 0.0 & 2.4 & 15.6 & 10.5\tabularnewline
\hline 
 & \multicolumn{9}{c|}{Host-less Fraction/Panel {[}Relative to z-slice{]} (\%) } & Total Host-less Fraction (\%)\tabularnewline
\hline 
0.0\textless z\textless 0.4  & 86.8 {[}4.0{]} & 67.9 {[}0.1{]} & 0.0 {[}0.0{]} & 34.2 {[}3.4{]} & 14.9 {[}5.0{]} & 13.3 {[}0.0{]} & 14.8 {[}0.8{]} & 5.3 {[}2.0{]} & 2.8 {[}0.2{]} & 15.5\tabularnewline
0.4\textless z\textless 1.0  & 90.5 {[}4.9{]} & 65.9 {[}0.3{]} & 0.0 {[}0.0{]} & 40.8 {[}2.2{]} & 18.9 {[}7.5{]} & 9.4 {[}0.1{]} & 12.1 {[}0.3{]} & 6.3 {[}1.8{]} & 5.2 {[}0.9{]} & 18.0\tabularnewline
1.0\textless z\textless 2.0  & 93.1 {[}7.2{]} & 76.8 {[}1.7{]} & 0.0 {[}0.0{]} & 51.3 {[}1.1{]} & 25.9 {[}11.4{]} & 7.8 {[}0.6{]} & 6.0 {[}0.0{]} & 7.3 {[}0.9{]} & 6.5 {[}1.6{]} & 24.4\tabularnewline
2.0\textless z\textless 3.6  & 96.5 {[}11.0{]} & 86.2 {[}5.1{]} & 25.0 {[}0.0{]} & 67.9 {[}0.6{]} & 37.3 {[}15.4{]} & 8.5 {[}1.9{]} & 0.0 {[}0.0{]} & 6.8 {[}0.2{]} & 8.7 {[}1.4{]} & 35.6\tabularnewline
\hline 
\end{tabular}
}
\end{table*}

\begin{table*}

\caption{[BPASS/Bray - eBHNS] 
A breakdown of the relative binary mergers occurring within each panel of Figure \ref{fig:bhns-ip}. The top division refer to the relative fractions of binaries coalescing for a given redshift slice. The last column indicates the population of the binaries contained within each slice as a percentage of the cumulative population. The second section of the table shows the estimated host-less population per panel [overall fraction corresponding to the redshift slice] based on the criteria defined in Section \ref{ss:proj-dist}.
}
\label{tab:bhns-ip-table}
\centering
\scalebox{0.87}[0.87]{
\begin{tabular}{|c|c|c|c|c|c|c|c|c|c|c|}
\hline 
\multirow{2}{*}{z} & 1  & 2  & 3  & 4  & 5  & 6  & 7  & 8  & 9  & \multirow{2}{*}{Total Population (\%)}\tabularnewline
\cline{2-10} \cline{3-10} \cline{4-10} \cline{5-10} \cline{6-10} \cline{7-10} \cline{8-10} \cline{9-10} \cline{10-10} 
 & \multicolumn{9}{c|}{Binary Fraction (\%)} & \tabularnewline
\hline 
0.0\textless z\textless 0.4  & 5.5 & 0.2 & 0.0 & 12.3 & 31.4 & 0.0 & 5.3 & 35.2 & 10.1 & 8.5\tabularnewline
0.4\textless z\textless 1.0  & 7.3 & 1.2 & 0.0 & 5.8 & 42.4 & 0.8 & 1.9 & 24.0 & 16.6 & 31.5\tabularnewline
1.0\textless z\textless 2.0  & 5.4 & 2.9 & 0.0 & 1.5 & 54.2 & 8.9 & 0.1 & 6.9 & 20.1 & 44.1\tabularnewline
2.0\textless z\textless 3.6  & 2.5 & 2.7 & 0.0 & 0.4 & 43.1 & 32.6 & 0.0 & 1.7 & 17.0 & 15.9\tabularnewline
\hline 
 & \multicolumn{9}{c|}{Host-less Fraction/Panel {[}Relative to z-slice{]} (\%) } & Total Host-less Fraction (\%)\tabularnewline
\hline 
0.0\textless z\textless 0.4  & 93.9 {[}5.2{]} & 56.2 {[}0.1{]} & 0.0 {[}0.0{]} & 55.9 {[}6.9{]} & 31.8 {[}10.0{]} & 0.0 {[}0.0{]} & 10.8 {[}0.6{]} & 8.9 {[}3.1{]} & 3.6 {[}0.4{]} & 26.2\tabularnewline
0.4\textless z\textless 1.0  & 90.0 {[}6.6{]} & 58.9 {[}0.7{]} & 0.0 {[}0.0{]} & 59.5 {[}3.5{]} & 41.4 {[}17.5{]} & 18.6 {[}0.1{]} & 15.7 {[}0.3{]} & 12.1 {[}2.9{]} & 8.7 {[}1.4{]} & 33.1\tabularnewline
1.0\textless z\textless 2.0  & 87.7 {[}4.8{]} & 69.2 {[}2.0{]} & 0.0 {[}0.0{]} & 69.6 {[}1.0{]} & 38.7 {[}21.0{]} & 14.5 {[}1.3{]} & 17.9 {[}0.0{]} & 13.7 {[}0.9{]} & 11.7 {[}2.4{]} & 33.4\tabularnewline
2.0\textless z\textless 3.6  & 91.5 {[}2.3{]} & 68.2 {[}1.9{]} & 0.0 {[}0.0{]} & 64.5 {[}0.3{]} & 29.6 {[}12.8{]} & 8.3 {[}2.7{]} & 0.0 {[}0.0{]} & 8.1 {[}0.1{]} & 13.5 {[}2.3{]} & 22.3\tabularnewline
\hline 
\end{tabular}
}
\end{table*}

\section{Simulating Observational Selection Effects}\label{sss:sgrb-env}

\subsection{Luminosity function and selection}\label{ss:lum-fun}

We produce a simulated population of observable SGRBs assuming the broken power-law intrinsic luminosity function for SGRB peak luminosities from \cite{paul2018},
\begin{equation}
    \Phi(L) \propto \begin{cases} 
    \left(\frac{L}{10^{52.18}{\rm erg/s}}\right)^{-0.5} & L\leq 10^{52.18}{\rm erg/s},    \\
    \left(\frac{L}{10^{52.18}{\rm erg/s}}\right)^{-1.86} & L>10^{52.18}{\rm erg/s}.
    \end{cases} 
    \label{eq:LF}
\end{equation} 
A broken power-law is consistent with the findings in \cite{wanderman2015,ghirlanda2016}; alternatively, see \cite{yonetoku2014, sun2015} for examples of single power-law luminosity functions. As the distribution of intrinsically low-luminosity SGRBs is poorly constrained, we fix the cut-off at $10^{49}\,\rm{erg\ s^{-1}}$ \citep{Mogushi2019}.
As such, the luminosities, $L$, are drawn from a distribution within the range $10^{49}\leq L\leq 10^{55}\,\rm{erg\ s^{-1}}$. 
The ascribed SGRB luminosity
has no dependence on the binary characteristics, since we lack any theoretical or observational mapping that would motivate this. 
Whilst some SGRBs, such as GRB 170817A, exhibit luminosities below our minimum, these events are expected to account for a small fraction of the cumulative observed population \citep{Tan2020}.
We use a 
sensitivity of $0.2$\,ph\,s$^{-1}$\,cm$^{-2}$ to estimate the fraction of GRBs that would be detectable by \textit{Swift}/BAT. We apply a Band function \citep{Band1993} with an index $\alpha=0.5$ and $\beta=2.25$ for the distribution below and above $E_P$ respectively, where $E_p$ is the spectral peak energy given by the relation for SGRBs found by \citep{Tsutsui2013}.

\subsection{Afterglow dependence on environment}\label{sss:afterglows}

GRB afterglows are thought to arise when the 
relativistic jet impacts and shocks the external
ambient medium, producing long-lived 
emission from radio to X-ray \citep{Zhang2006,Zhang2007}.
Identification of candidate host galaxies, and hence redshifts and impact parameters, relies on the $\sim$arcsec localisations that come from X-ray and optical/nIR detections, rather than the $\sim$arcmin positions that are produced by the {\em Swift}/BAT.
For some SGRBs, particularly those with ``extended soft emission'' \citep{Meszaros1997,Sari1998}, the X-ray localisation may be found independently of afterglow detection, however, more commonly observations of the afterglow is necessary.  Thus there is the possibility that SGRBs in low density environments may be selectively lost from the observational sample we consider here.
In this section we outline an approach to introduce this into the simulations, which may indicate a plausible maximum size of this selection bias.

The average gas density, $n$, surrounding a galaxy will decrease with increasing distance from the galaxy core, therefore, the afterglows from bursts at large separations from their host will typically have fainter afterglow emission \citep{Perna_2002}.
The maximum luminosity distance, $D_{\rm max}$, for the peak afterglow emissions to be above a given threshold 
depends on the ambient density as $D_{\rm max} \propto n^{(p+1)/8}$ assuming slow-cooling 
and an observed frequency above the characteristic synchrotron frequency at the peak, where $p$ is the relativistic electron distribution index in the shock -- typically $p\sim2.4$ \citep{Fong_2015}. 
The afterglows of SGRBs can be used to put constraints on the ambient density, where the earliest afterglow detection can be used to constrain the minimum density.
Using this minimum ambient density and an assumed galaxy gas density profile, \citet{oconnor2020constraints} 
infer a maximum impact parameter distribution for the population of observed SGRBs.

For binaries capable of producing an SGRB e.g. NSNS, and eBHNS systems,
(see Section \ref{sss:bin-sam}), we estimate the fraction 
that merge in a very low-density environment, 
defined by \citet{Fong_2015} as $n<10^{-4}\,\rm{cm^{-3}}$. 
Based on the location of the merger with respect to its host galaxy, we estimate the ambient density by considering the gas mass contained within an \texttt{EAGLE} aperture. These apertures are composed of spheres with sizes of $1$, $3$, $5$, $10$, $20$, $30$, $40$, $50$, $70$, and $100$\,kpc. We split these spheres into shells based on the radii of neighbouring apertures and calculate the spherically averaged gas density. 
We added an additional hot gas component, similar to the approach described in \citet{oconnor2020constraints}. This secondary component is described by the Maller \& Bullock gas model \citep{Maller2004,Fang2013}, 
\begin{equation}
    \rho^{MB}_{g}(x) = \rho_v\bigg[ 1+\frac{3.7}{x}\rm{ln}(1+x)-\frac{3.7}{c_v}\rm{ln}(1+c_v)\bigg]^{3/2},
\end{equation}
where $\rho_v$ is the density at the virial radius, $r_{v}$ \citep{Carlberg1997}, corresponding to an overdensity parameter of $\Delta_v=200$; $x$ is the ratio of $r/R_s$
; and $c_v$ is the concentration parameter, \citep[e.g.][]{Bullock2001,Comerford2007}. 
Using this prescription, we identify the fraction 
of the SGRB progenitor mergers that occur in environments with $n<10^{-4}$\,cm$^{-3}$ as afterglow-faint. Mergers occurring in denser environments are considered to be afterglow-bright.

The total population of the simulated SGRB progenitors merging in environments with low ambient densities is shown in the first half of Table \ref{tab:sgrb-aft}. 
This is further split according to whether the merger fulfils our criteria for being host-less or not. 
The proportion of mergers occurring in low density media is therefore somewhat increased over the whole population. 
Notably, the proportion of host-less BAT-detectable SGRBs formed by COSMIC and BPASS/Bray binaries merging within afterglow-faint environments are broadly consistent, within errors, with the estimated fraction of SGRBs occurring in a $n<10^{-4}\,{\rm cm}^{-3}$ ambient density medium by 
\citet{oconnor2020constraints},
whereas the $\approx35$\% fraction of SGRBs in low density environments predicted by the BPASS/Hobbs simulation is rather in excess of the observations.

\begin{figure}
\vbox{
\includegraphics[width=\columnwidth]{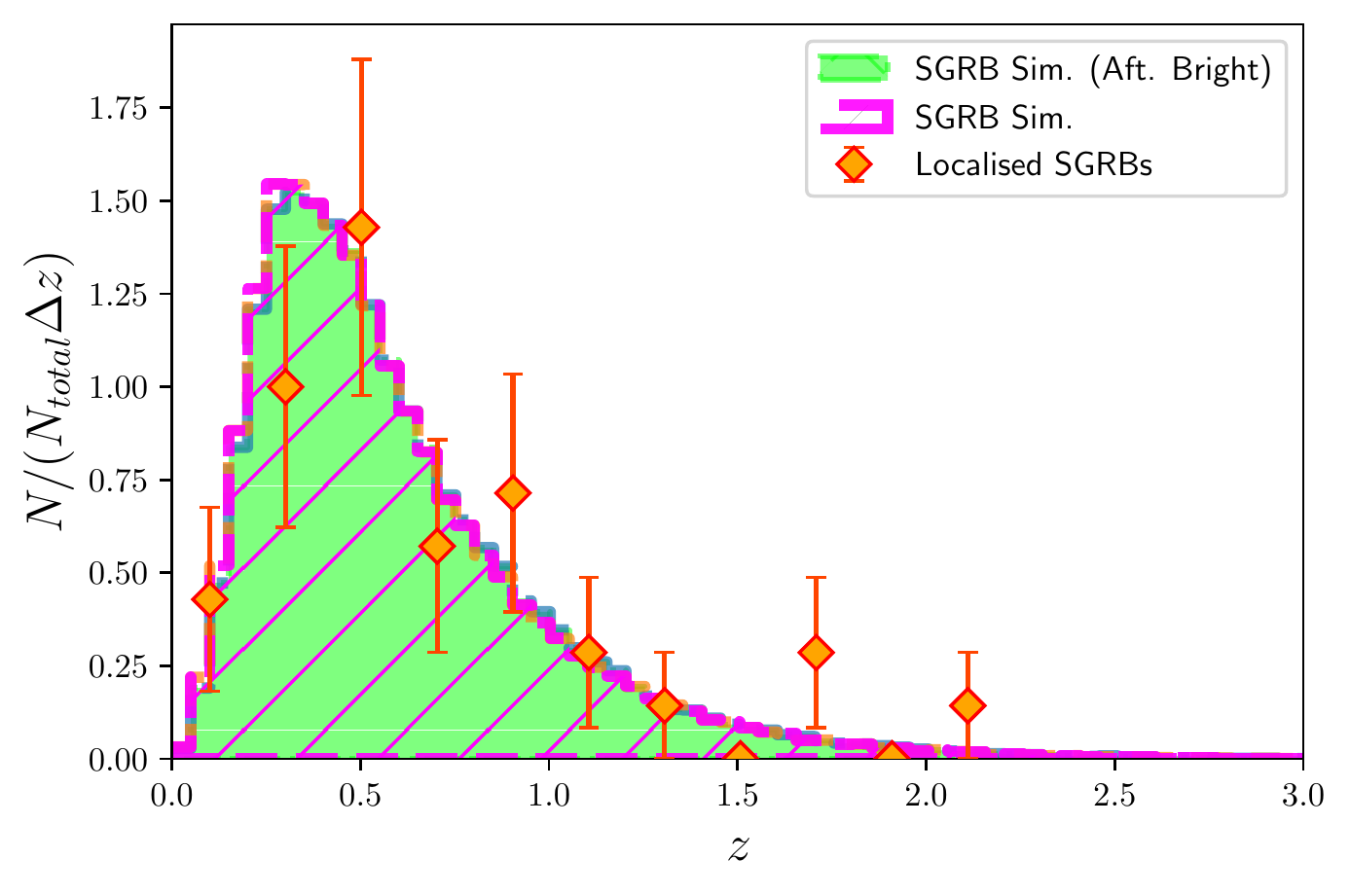}
}

\caption{[BPASS/Bray] The distribution of the NSNS and eBHNS binary mergers from our study that are likely to produce SGRBs detectable by \textit{Swift}/BAT, excluding host-less systems, refer to Section \ref{sss:hostless}.  The magenta histogram outlines the total population of BAT-detectable SGRBs with identifiable hosts. The lime histograms corresponds to the subset of events merging in afterglow-bright environments, ($n>10^{-4}\,\rm{cm^{-3}}$). 
The orange points correspond to the population of \textit{Swift} observed SGRBs within redshift bin sizes of $z = 0.2$, with a Poisson uncertainty. 
}

\label{fig:sgrb-obvs-dist}
\end{figure}

In Figure \ref{fig:sgrb-obvs-dist} we compare the redshift distribution of observed SGRBs against our predictions for the BAT-detectable mergers with host associations. 
The agreement is generally reasonable, although the finding of several SGRBs at $z\gtrsim1.7$ is harder to reconcile with the small numbers predicted at those redshifts. This may suggest that the bright-end slope of our adopted luminosity function (Eqn.~\ref{eq:LF}) is rather too steep.

\begin{table*}
\caption{Summary of the fractions of SGRB progenitors merging in environments with ambient densities $<10^{-4}\,\rm{cm^{-3}}$, for each binary simulation used. Systems are considered host-less if the host galaxy has a magnitude, $H>26$, and/or if the offset of the binary is such that $P_{\rm chance}>0.05$. 
}
\label{tab:sgrb-aft}
    \centering
\scalebox{0.9}[0.9]{
\begin{tabular}{|c||c|c|c||c|c|c|}
\hline 
\multirow{2}{*}{Binary Simulation } & \multicolumn{3}{c||}{All SGRB Progenitors ($\%$)} & \multicolumn{3}{c|}{BAT-detectable ($\%$)}\tabularnewline
\cline{2-7} 
 & \multirow{1}{*}{With Host} & Host-less & Total & With Host & Host-less & Total\tabularnewline
\hline 
BPASS/Bray  & 4.9 & 14.9 & 19.9  & 9.3 & 14.8 & 24.2\tabularnewline
\hline 
BPASS/Hobbs  & 2.9 & 26.1 & 29.0 & 7.0 & 28.1 & 35.0\tabularnewline
\hline 
COSMIC  & 1.7 & 8.9 & 10.6  & 3.6 & 9.1 & 12.7\tabularnewline
\hline 
\end{tabular}
}
\end{table*}

\section{Identifying the host galaxies of SGRBs}\label{ss:host-sgrbs}
\subsection{Comparing study predictions against observations}

\begin{figure}
\vbox{
\includegraphics[width=\columnwidth]{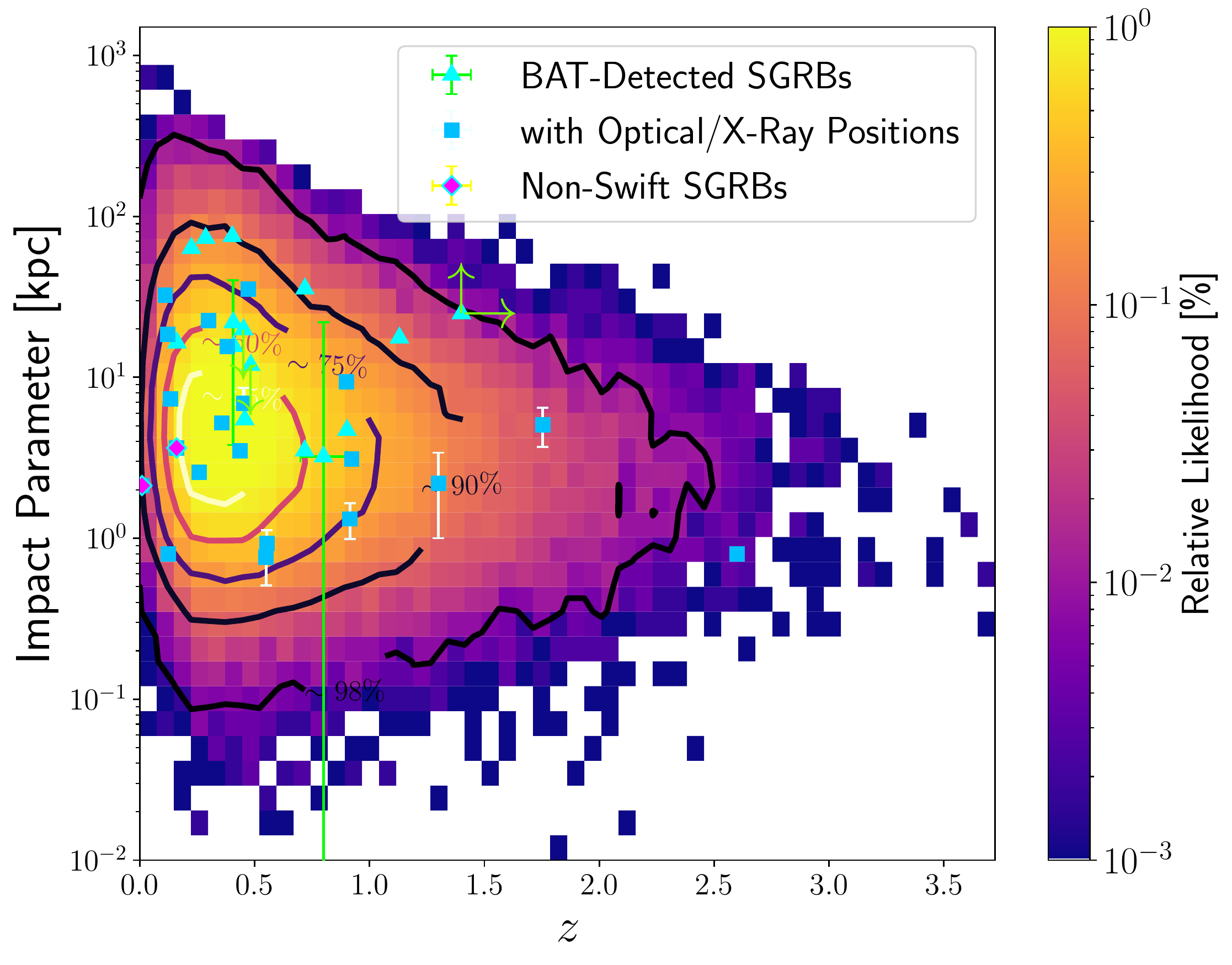}
}

\caption{[BPASS/Bray] A density map of the impact parameter and redshift for binaries that are expected to produce a SGRB within \textit{Swift}/BAT's detection sensitivity range and are not considered host-less (see Section \ref{sss:hostless}). 
In both cases, offsets and redshifts corresponding to SGRB-galaxy from literature 
are plotted with their relevant uncertainties, refer to Table \ref{tab:sgrb-tab}. The contour levels shown, proceeding outwards, indicate the $\sim\,25,\,50,\,75,\,90$ and $98\%$ enclosed synthetic merging binary population.  The total binary population is comprised of both NSNS and eBHNS binaries. 
}
\label{fig:obs-ip-z-dist}
\end{figure}

Binaries capable of producing SGRBs within the sensitivity threshold of \textit{Swift}/BAT are shown in Figure\,\ref{fig:obs-ip-z-dist}.
We have overlaid offset and redshift information for SGRBs noted in literature. There is a relatively high concentration of \texttt{zELDA} processed binaries at redshift $\sim0.5$ with relatively low impact parameters, $\sim 10\,$kpc. The observed localised SGRBs tend to cluster at low redshift ($z\lesssim1$) with spatial offsets $<100\,$kpc from their likely hosts. This is found to be consistent with the peak concentration predicted by the binary simulations used.

The identification of the parent galaxy associated with an SGRB is often non-trivial, especially for events with no 
bright galaxies in or near their localisation uncertainty region, or in cases where no arcsec level localisation was obtained in the first place. 
Campaigns aiming to successfully isolate and identify the hosts of these events have been ongoing throughout the {\em Swift} era 
\citep[e.g.][]{levan07,berger2009host,fong2013demographics,berger2014short}. 
A host that 
has a low-apparent luminosity may easily be missed when searching within or around the localisation uncertainty of a given SGRB detection. 
Apart from the rare cases where the SGRB is located on or close to an obvious bright (typically low redshift) galaxy, the first requirement is to conduct a deep-imaging of the field of sky surrounding the region of the burst. 
This is often provided by follow-up imaging of the afterglow following the SGRB. Ultimately, associations made between the burst and the host galaxy is probabilistic given that the galaxy could be within the localised field by chance, see Section \ref{sss:hostless}.

\citet{fong2013demographics} analysed the host demographics for 36 well associated SGRBs (i.e. events with convincing host galaxies backed by afterglow observations), finding that $\sim47$\% have late-type hosts, $\sim17$\% early-type, $\sim17$\% host-less, and $\sim19$\% are ``inconclusive" owing to their uncertain association with a nearby galaxy. 
Comparing our predictions (refer to the final segment of Table\,\ref{tab:demographics}) with these values
we find reasonably consistent fractions for SGRBs arising from late-type galaxies (e.g. BPASS/Bray: $\sim51\%$).
For the early-type hosts we predict a higher fraction
(e.g. BPASS/Bray: $\sim30\%$), although 
given the limited statistics, and the possibility that
some of the ``inconclusive" cases could in fact be early-type hosts, we do not think this is a serious discrepancy. Similarly, 
the host-less  population fraction 
of  BAT-detectable SGRBs 
only slightly varies from the \citet{fong2013demographics} numbers for the BPASS/Bray and COSMIC simulations, which could easily be accounted for if some ``inconclusive" cases are ultimately considered host-less, combined with the observational selection bias against localising afterglow-faint SGRBs since they are also more likely to be host-less (cf. Table~\ref{tab:sgrb-aft}).  
For BPASS/Hobbs, the $\sim35$\% host-less fraction is only consistent with the observed population if the majority of the ``inconclusive'' population from \citet{fong2013demographics} is considered host-less.

\subsection{Implications for host searches for GW detected events}
\begin{figure}
    \centering
    \vbox{
    \includegraphics[width=\columnwidth]{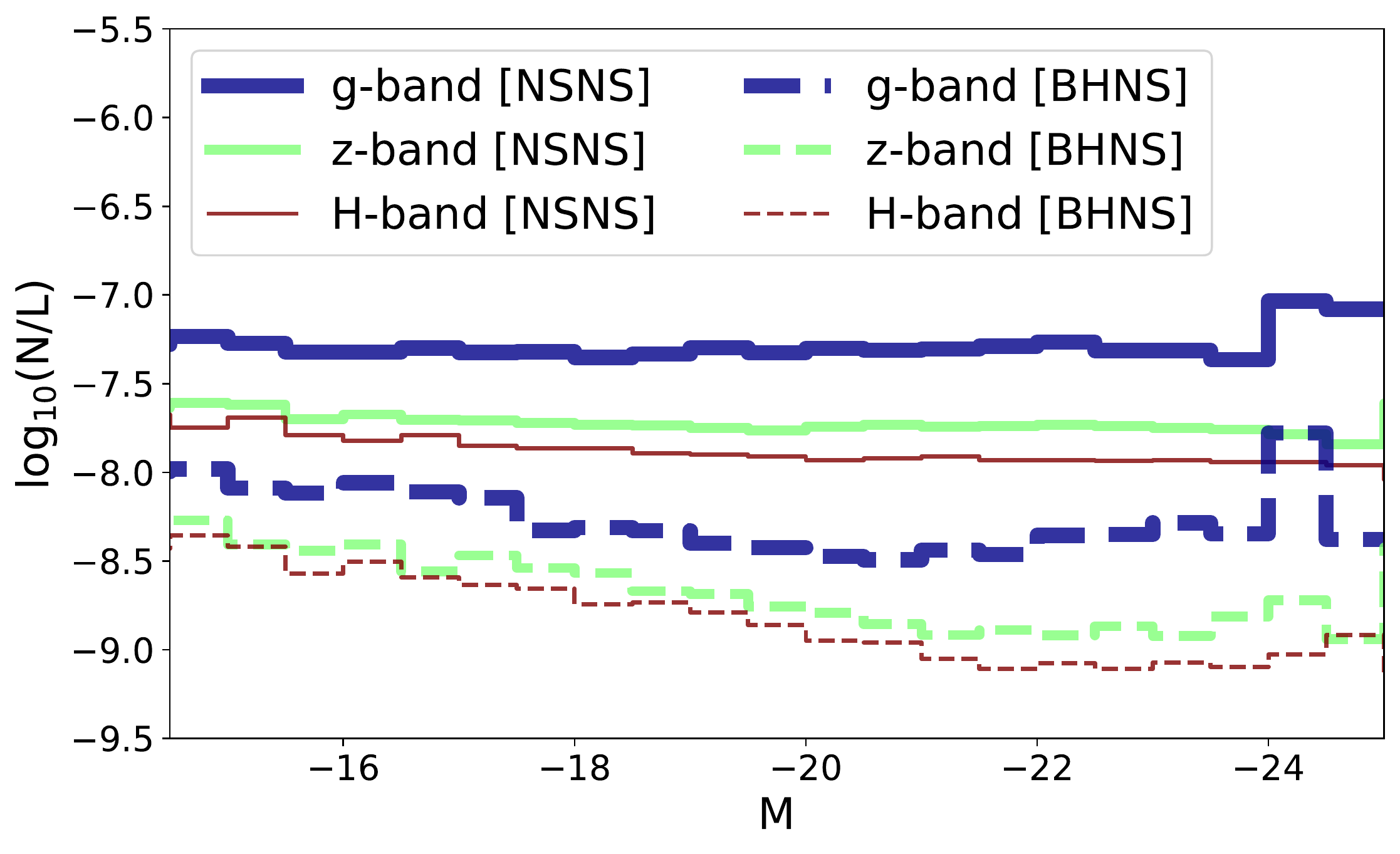}    
    }

    \caption{The ratio of NSNS (solid) and eBHNS (dashed) mergers ($N$) per unit luminosity ($L$) in bins of absolute magnitude, for all galaxies in the simulation volume (restricted to redshift $z<0.05$, and the BPASS/Bray model).  Shown are results for the $g$/$z$/$H$-band: of these, the $g$-band is generally flatter, indicating that merger rate is, on average, more nearly directly proportional to luminosity in this band. Note that the luminosity scale here is arbitrary, so there is no particular significance to the relative values of $N/L$.}
    \label{fig:power-comp}
\end{figure}

The rich science returns from observations of neutron-star compact binary mergers in both gravitational waves and electromagnetic radiation has been exemplified by the case of GW170817 \citep[e.g.][]{abbott2017H0,abbott2017MMA}.
However, even with the current three advanced GW detectors, positional error regions are typically 10s or 100s of sq-deg in size, 
while the distance range for such events is $\sim$\,{\rm few}\,$\times10^2\,$Mpc  \citep{GWProspectsv6}. 
This is challenging both in terms of sky area and depth required to locate EM counterparts, apart from the rare cases of on-axis SGRBs.
To aid in mapping these error regions, particularly with narrow-field optical, near-IR or X-ray instruments, various strategies have been suggested to prioritise observations based on the known positions and properties of potential host galaxies \citep[e.g.][]{gehrels2016,evans2016,Artale2020b}. 

A question arises as to how to weight the galaxies in such schemes, most simply by making use of their absolute luminosities in some pass-band (since that information is more widely available than, for example, more detailed characterisation of their star formation histories).
With this in mind, in Figure~\ref{fig:power-comp} we plot the ratio of the number of mergers (restricted to $z<0.05$) occurring per unit luminosity (averaged over all galaxies in the simulation box), as a function of host absolute luminosity, in various pass-bands.
A situation in which binary merger rate was exactly proportional to luminosity would result in completely flat curves in this diagram.  As can be seen, this is most nearly the case for the $g$-band (both NSNS and eBHNS), which may reflect 
the balance between an enhanced rate of mergers from relatively young (and therefore bluer) stellar populations and the overall dominance of intermediate and older (redder) populations.
In contrast, the (near-IR) $H$-band magnitude does not provide such a good tracer of merger likelihood, presumably due to large early-type (elliptical/lenticular) galaxies being bright in that band, but producing comparatively few mergers.

\subsection{Implications for SGRB host identification}

\begin{figure*}
    \centering
    \hbox{
    \includegraphics[width=\textwidth]{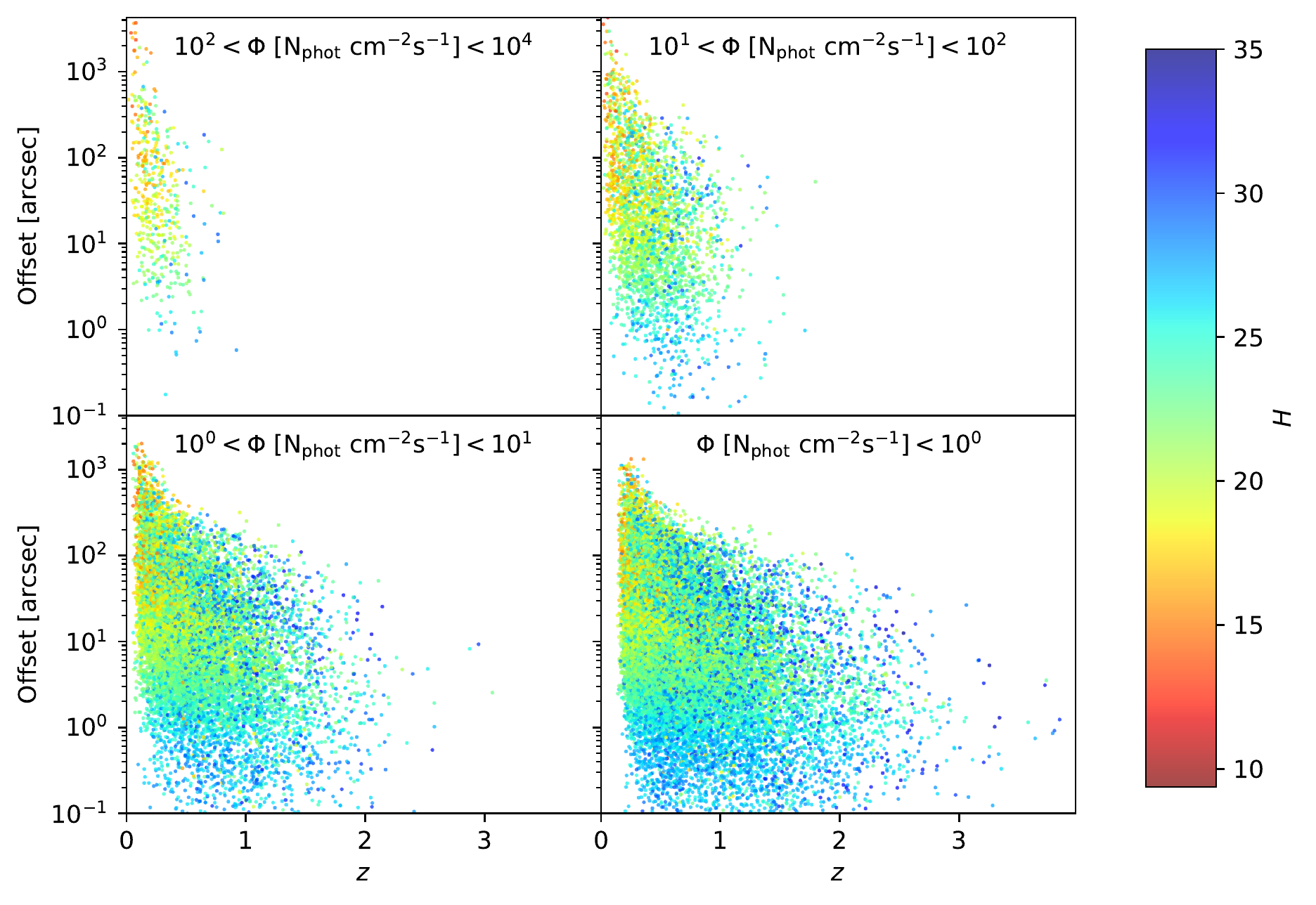}    
    }

    \caption{[BPASS/Bray] The distribution of host-less (see Section \ref{sss:hostless}) binary  merger offsets against redshift, $z$, corresponding to different ranges of photon flux, $\Phi$. The colour of the points indicate the brightness of the host galaxy within the $H$-band, with redder points indicating brighter host galaxies than the bluer points.}
    \label{fig:phot-z-grid}
\end{figure*}

\begin{figure}
    \centering
    \hbox{
    \includegraphics[width=\columnwidth]{{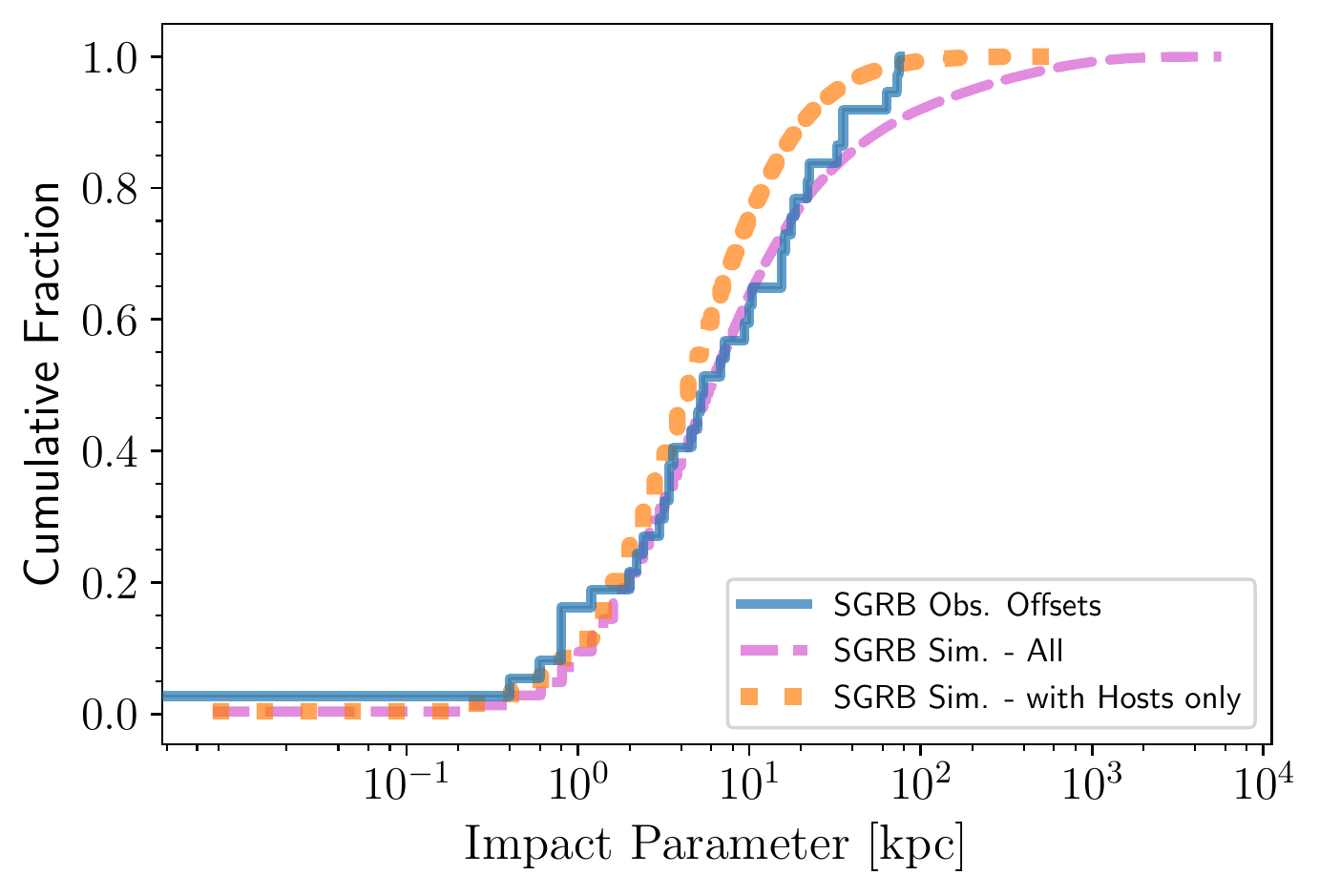}}
    }
    \caption{[BPASS/Bray] The impact parameter distribution for the population of SGRB compatible NSNS and eBHNS binaries expected to be observable by \textit{Swift}/BAT 
    for i) all systems (lilac dashed), ii) systems with hosts only (orange dotted). The blue line corresponds to a sample of SGRBs with known redshifts and offsets (blue) inclusive of non-\textit{Swift} detected events. Refer to Table \ref{tab:sgrb-tab} for sources used.
    }
    \label{fig:ip_cumal}
\end{figure}

\begin{table}

\caption{Redshift and impact parameter values for the SGRBs used within this study. The E$_{\rm iso}$ is calculated from \textit{Swift}/BAT's $15-350\,\rm{keV}$ energy band.
\newline *Calculated using \textit{Fermi}/GBM's $10-10 000\,\rm{keV}$ energy band. 
}
\label{tab:sgrb-tab}
\centering
\resizebox{\columnwidth}{!}{\begin{tabular}
{|c|c|c|c|c|}

\hline 
GRB  & z  & IP {[}kpc{]}  & E$_{\rm{iso}}$ {[}$\times10^{48}$erg{]} & Refs.\tabularnewline
\hline 
050509B  & 0.225  & 63.40  & 7.42 & {[}1–5{]}\tabularnewline
050709  & 0.161  & 3.64  & 29.6* & {[}1–4, 6{]} \tabularnewline
050724  & 0.258  & 2.57$_{-0.8}^{+0.8}$  & 83.9 & {[}1–4, 7{]} \tabularnewline
050813  & 0.719  & 35.50  & 219 & {[}1, 3, 8{]}\tabularnewline
051210  & $>$1.400  & $>$24.90  & 2480 & {[}1, 3{]}\tabularnewline
051221A  & 0.550  & 0.76$_{-0.25}^{+0.25}$  & 2680 & {[}1–4, 9–10{]} \tabularnewline
060502B  & 0.287  & 73.30  & 49.2 & {[}1–4, 11{]} \tabularnewline
060801  & 1.130  & 17.60  & 2290 & {[}1–4, 12–13{]} \tabularnewline
061006  & 0.436  & 3.50  & 1070 & {[}1–3, 14{]}\tabularnewline
061210  & 0.410  & 15.60  & 552 & {[}1–4, 15–17{]} \tabularnewline
070429B  & 0.902  & 4.70  & 460 & {[}1–4, 15, 18–20{]} \tabularnewline
070707  & $<$3.6  & -  & 178000* & {[}1, 18, 21{]} \tabularnewline
070714B  & 0.923  & 3.10  & 3340 & {[}1–4, 15, 18, 22{]} \tabularnewline
070724A  & 0.457  & 5.46 $_{-0.14}^{+0.14}$  & 39.7 & {[}1–4, 15, 18, 23–24{]}\tabularnewline
070729  & 0.800 $_{-0.1}^{+0.1}$  & 3.21 $_{-18.80}^{+18.80}$  & 897 & {[}3{]}\tabularnewline
070809  & 0.473  & 35.40  & 140 & {[}1–2, 15, 18{]}\tabularnewline
071227  & 0.381  & 15.50$_{-0.24}^{+0.24}$  & 204 & {[}1–3, 14–15, 18{]} \tabularnewline
080503  & $<$4.000  & -  & 16200 & {[}1, 18{]}\tabularnewline
080905A  & 0.122  & 18.50  & 16.1 & {[}1–4, 18, 25{]} \tabularnewline
090305A  & $<$4.100  & -  & 36000 & {[}18{]}\tabularnewline
090510  & 0.903  & 9.40  & 4980 & {[}1–3, 15, 18, 26{]}\tabularnewline
100117A  & 0.915  & 1.32$_{-0.33}^{+0.33}$  & 1100 & {[}1–3, 15, 18{]}\tabularnewline
100206A  & 0.407  & 21.90$_{-18.10}^{+18.10}$  & 263 & {[}1–2, 27–28{]} \tabularnewline
100625A  & 0.452  & $<$19.80  & 525 & {[}1–2, 27{]} \tabularnewline
100816A  & 0.804  & 10.07  & 7730 & {[}29–30{]} \tabularnewline
101219A  & 0.718  & 3.50  & 3500 & {[}1–2, 27, 31–32{]}\tabularnewline
111117A  & 2.211  & 10.6$_{-1.70}^{+1.70}$  & 19800 & {[}2, 33–34{]} \tabularnewline
120804A  & 1.300  & 2.20$_{-1.20}^{+1.20}$  & 18200 & {[}2, 27, 35{]} \tabularnewline
130603B  & 0.356  & 5.21$_{-0.17}^{+0.17}$  & 803 & {[}2, 18, 27, 36–37{]} \tabularnewline
140903A  & 0.351  & 0.5$_{-0.2}^{+0.2}$ & 90.7 & {[}2, 38–40{]} \tabularnewline
150101B  & 0.134  & 7.35$_{-0.07}^{+0.07}$  & 5.84 & {[}2, 41–43{]} \tabularnewline
150424A  & 0.300  & 22.50  & 1080 & {[}2, 44{]} \tabularnewline
160410A  & 1.717  & $\sim0$  & 1590 & {[}45–46{]}\tabularnewline
160624A  & 0.483  & \textless{}11.91  & 192 & {[}2, 47–48{]} \tabularnewline
160821B  & 0.162  & 16.40  & 13.9 & {[}2, 49–51{]} \tabularnewline
170428A  & 0.454  & 6.90$_{-1.70}^{+1.70}$  & 663 & {[}2, 52{]} \tabularnewline
170817A  & 0.001  & 2.13  & 0.031* & {[}2, 53{]} \tabularnewline
181123B  & 1.754  & 5.08$_{-1.38}^{+1.38}$  & 10400 & {[}54{]} \tabularnewline
200522A  & 0.554  & 0.93$_{-0.19}^{+0.19}$  & 287 & {[}48, 55{]} \tabularnewline
200826A  & 0.748  & -  & 13000* & {[}56{]} \tabularnewline
201221D  & 1.046  & -  & 6660* & {[}57{]} \tabularnewline
\hline 
\end{tabular}
}
\\
 References: [1] \citet{Tunnicliffe_2014} [2] \citet{gompertz2020search} [3] \citet{Kann2011} [4] \citet{Church2012SGRBs} [5] \citet{Bloom2006}   [6] \citet{Fox2005}   [7] \citet{Prochaska2005GCN} [8] \citet{Foley2005GCN} [9] \citet{Berger2005GCN051221} [10] \citet{Soderberg2006} [11] \citet{bloom2007putative} [12] \citet{Cucchiara2006GCN} [13] \citet{Racusin2006GCN} [14] \citet{Davanzo2009}  [15] \citet{Leibler2010} [16] \citet{Ziaeepour2006GCN} [17] \citet{Berger2006GCN} [18] \citet{fong2013locations}  [19] \citet{Perley2007GCN} [20] \citet{Holland2007GCN}
[21] \citet{Piramonte2008} [22] \citet{Graham09} [23] \citet{Berger2009} [24] \citet{Kocevski2010} [25] \citet{rowlinson2010discovery} [26] \citet{Rau2009GCN}   [27] \citet{berger2014short} [28] \citet{Cenko2010GCN} [29] \citet{Tanvir2010} [30] \citet{ImGCN2010} [31] \citet{Kuin2011GCN} [32] \citet{Chornock2011GCN}  [33] \citet{selsing18} [34] \citet{Margutti12}  [35] \citet{Berger2013GRB120804} [36] \citet{Thone2013GCN} [37] \citet{Levan2013GCN} [38] \citet{Levan2014GCN} [39] \citet{Cucchiara2014GCN} [40] \citet{Troja2016} [41] \citet{Levan2015} [42] \citet{Fong2015GCN} [43] \citet{Fong2016} [44] \citet{CastroTirado2015} 
[45] \citet{Selsing2016GCN} [46] \citet{Selsing19} [47] \citet{Cucchiara2016} [48] \citet{Oconnor21} [49] \citet{lamb2019short} [50] \citet{Troja2019GRB160821B} [51] \citet{MAGIC2021} [52] \citet{Izzo2017} [53] \citet{GW170817LIGO} [54] \citet{Paterson2020} [55] \citet{Fong2020}  [56] \citet{Rothberg2020} [57] \citet{deUgartePostigo2020}.
\end{table}

\begin{table}
    \centering
    \caption{Two sample Kolmogorov-Smirnov test rejection levels for the simulated BAT-detectable SGRB progenitors impact parameters as compared against observed SGRBs. The second half of data includes the KS-test rejection levels for the subset of systems that merge in afterglow-bright environments, refer to Section \ref{sss:afterglows}. }
    \label{tab:ks-test}
    \begin{tabular}{|c|c|c|c|}
    \hline 
    \multirow{2}{*}{BAT-Detectable SGRBs} & \multicolumn{3}{c|}{Kolmogorov-Smirnov Test Rejection ($\%)$}\tabularnewline
    \cline{2-4} 
     & BPASS/Bray & BPASS/Hobbs & COSMIC\tabularnewline
    \hline 
    All & 18 & 95 & 81\tabularnewline
    With Host only & 92 & 64 & 98\tabularnewline
    \hline 
     & \multicolumn{3}{c|}{Afterglow Bright Events}\tabularnewline
    \hline 
    All & 90 & 42 & 98\tabularnewline
    With Host only & 95 & 79 & 99\tabularnewline
    \hline 
    \end{tabular}
\end{table}

Successfully ejected binaries with long lifespans are likely to merge further away from their hosts due to prolonged migration. The greater their galactocentric radius, the lower the likelihood of a successful association with their parent galaxy, which, as we have seen in previous sections, will bias SGRB redshift samples, for example. It is therefore interesting to ask, if a SGRB detected by {\em Swift}/BAT is found to be host-less, what can we say about its likely redshift and offset from its parent galaxy, based solely on the peak gamma-ray luminosity?

Figure \ref{fig:phot-z-grid} shows the distribution of the resulting offsets with redshift for host-less merging systems. Each panel corresponds to a range of peak photon fluxes, $\Phi$, that would be detected by BAT. The colour of each point corresponds to the $H$-band magnitude of the host galaxy. The majority of events with high photon fluxes $\Phi\gtrsim100\,\rm{cm^{-2}s^{-1}}$  occur between $0.1 \lesssim z \lesssim 0.5$. 
At all fluxes there are some mergers that are low redshift and have large offsets ($>100^{\prime\prime}$) on the sky from their hosts, making association very challenging. 
For SGRBs with low photon fluxes ($<1$\,ph\,s$^{-1}$\,cm$^{-2}$; refer to the final panel of Figure \ref{fig:phot-z-grid}) close to the BAT detection threshold, the fraction of these events with faint host galaxies ($H>26$) is found 
to be $\sim 23\%$, $\sim 19\%$, and $\sim 38\%$ for BPASS/Bray, BPASS/Hobbs, and COSMIC simulations respectively.

\subsection{Impact Parameters of SGRBs}

In Figure\,\ref{fig:ip_cumal} we compare the impact parameters for an observed sample of $32$ SGRBs  with well-constrained offsets and redshifts (see Table \ref{tab:sgrb-tab}) against our simulated BPASS/Bray SGRB progenitor systems that are detectable by \textit{Swift}/BAT (for the same comparison with BPASS/Hobbs, and COSMIC binaries see Figures \ref{fig:hobbs-ip_cumal}, \ref{fig:cosmic-ip_cumal}). 
The median of the observed SGRB impact parameters is $\sim5.5\,$kpc. In comparison, for all simulated BAT-detectable SGRBs, we find median impact parameters of $6.1$, $11.6$, and $4.2\,$kpc for BPASS/Bray, BPASS/Hobbs, and COSMIC respectively. When the host-less population of progenitors are excluded, these values are revised to $4.7$, $6.2$, and $3.8\,$kpc.

In Table\,\ref{tab:ks-test}, we present the rejection levels for the two-sample Kolmogorov–Smirnov (KS) tests comparing impact parameters for the observed SGRB sample and each binary simulation. We conduct this test for the whole BAT-detectable SGRB samples as well as the sub-samples of afterglow-bright events (see Section \ref{sss:afterglows}). 
The most relevant comparison is with the ``with host" samples: COSMIC simulations perform poorly in these tests, due to the typically short life-times and consequently low impact parameters that it produces.
Otherwise BPASS/Hobbs matches well, and BPASS/Bray is marginally consistent with the observed sample. The results do not vary significantly if restricted to ``afterglow-bright" systems.
Finally, we note that 
all simulations are reasonably consistent with the $<10\,$kpc median SGRB galactocentric offset estimated by \citet{oconnor2020constraints}.

\section{Summary}\label{sec:conc}

Using our new \texttt{zELDA} code suite along with binary population synthesis codes, specifically \texttt{BPASS} and \texttt{COSMIC}, in addition to the cosmological simulation, \texttt{EAGLE}, we have 
traced the orbital evolution and host properties of a population of simulated isolated compact binaries from birth to coalescence. We use three binary simulated populations with varied prescriptions which are referred to as, BPASS/Bray, BPASS/Hobbs, and COSMIC. The key difference between the samples is the employed natal-kick prescriptions and merger time distributions. Most seeded BPASS/Bray binaries merge on relatively long timescales of 1--$10\, \rm{Gyr}$ with the majority of natal-kick velocities ranging between $\sim100-300 \, \rm{km\,s^{-1}}$; BPASS/Hobbs systems exhibit a broader range of lifespans with higher kick velocities, with the modal velocity at $v\sim350\,\rm{km\,s^{-1}}$; COSMIC systems merge with comparatively short lifespans, $<100\,\rm{Myr}$ and natal-kick velocities that are typically $<100\,\rm{km\,s^{-1}}$. 
Despite these differences, we find some conclusions are common to all the models, specifically:

\begin{itemize}
    \item For NSNS and eBHNS SGRB progenitors at merger, we find $\sim50-70\%$ of the hosts to be late-type galaxies, at all redshifts. For COSMIC binaries, the upper limit on this demographics estimate can extend to $\sim 80\%$.
    \item A consequence of this is that, for low redshift events $z<0.05$ (i.e. within the current range of GW detection), we find the majority of mergers to arise from relatively blue, star forming hosts.
    Thus,  we  find the $g$  band to be best suited for predicting compact binary merger rates in low redshift galaxies, and hence useful for weighting/prioritising galaxies for gravitational wave electromagnetic follow-up observations. 
    \item The modal average of the radial offset from the centre of the host galaxy at the time of merger is $\lesssim 10\,\rm{kpc}$. For each binary population used, these values range between $\sim 5-10 \,\rm{kpc}$ for BPASS/Hobbs systems and eBHNS mergers from the BPASS/Bray sample. For BPASS/Bray NSNS and COSMIC systems, this range varies between $\sim 2.5-5\,\rm{kpc}$. 
    Long-lived systems with $\tau_{\rm merge}\gtrapprox1\,\rm{Gyr}$ can potentially travel well beyond $100\,\rm{kpc}$ from their host galaxy provided that they have natal-kick velocities that are sufficient for them to escape the gravitational potential of their host galaxy.
 
\end{itemize}

To gauge the consistency of the \texttt{zELDA} processed binaries and their hosts we compare against the population of well-localised short-duration gamma-ray bursts (SGRBs) (i.e. events with convincing host galaxies and offsets). We achieve this by simulating \textit{Swift}/BAT observations of NSNS and  eBHNS mergers 
from our study, using an assumed luminosity function consistent with literature estimates. 

Observationally, some SGRBs have been found to be ``host-less", in the sense that there is no clear association with a galaxy, presumably due to either large offsets at merger and/or because the host is very faint. 
GRBs without a confident host association will typically have no redshift and impact parameter estimates, biases of these cases must be 
considered in our comparison. 
To mimic this, we define simulated binaries 
to be "host-less" if either the parent galaxy has a greater than 5\% probability of being simply a chance alignment based on the impact parameter (i.e. projected radial offset from galaxy centre) and apparent magnitude of the galaxy, 
and/or has a host galaxy with a $H$-band magnitude of $H>26$ (i.e. requiring deep follow-up observations, e.g. with {\em HST} to detect). Our findings are:
\begin{itemize}
    \item For both the BPASS/Bray and COSMIC simulations, we find the hosts
    of the simulated SGRBs to be $\sim$50--60\% late-type, and $\sim$15--20\% host-less, in reasonable agreement with the demographic distribution of observed hosts \citep{fong2013demographics}.  In contrast, the BPASS/Hobbs sample predicts  a
    much higher proportion of host-less events ($\sim35$\%), due largely to the typically higher kick velocities, which is likely in excess of the host-less fraction observed.
    
    \item The impact parameter distribution of the simulated BAT-detectable SGRBs for BPASS/Bray and BPASS/Hobbs are reasonably consistent with the observed SGRB offsets. 
    In this comparison, however, the COSMIC simulation under-predicts the offsets that are observed, which is a consequence of the lower kick velocities and shorter lifetimes. 

\end{itemize}

It is interesting to ask the question, if a SGRB is found to be host-less,
can our simulations give a prior expectation for its most likely redshift,
host and projected offset?

\begin{itemize}

    \item 
    
    For the brightest host-less simulated SGRBs (those with peak fluxes $>10^2$ photons\,cm$^{-2}$\,s$^{-1}$), the parent galaxies are found to be always at redshifts $z<1$ and  offsets are typically tens of arcsec, but can range up to in excess of $\sim10^3$ arcsec (as expected, the BPASS/Hobbs simulation typically results in rather larger offsets).
    
    \item At the other extreme, for bursts close to the \textit{Swift}/BAT detection threshold, we find that the majority of events (for all simulations) to be $z<1$, but with a significant tail to higher redshifts.
Notably, for these faint bursts, a sizable fraction of events    
    ($\sim 23\%$, $\sim 19\%$, and $\sim 38\%$  for BPASS/Bray, BPASS/Hobbs, and COSMIC populations, respectively) are found to be host-less by virtue of
    their  hosts being faint ($H>26$) rather than having large impact parameters. 

    \end{itemize}
    
    As a final step, we also considered an additional selection effect, namely 
    that bursts arising in a low  ambient density medium (defined here as $n<10^{-4}\,\rm{cm^{-3}}$) are expected to have
    on average fainter  afterglows, and so would be selected against in 
    observed samples that rely on detection of X-ray or optical counterparts
    for precise localisation.

    \begin{itemize}
    
    \item We estimate the fraction of BAT-detectable SGRBs arising from such low density environments. For the BPASS/Bray, BPASS/Hobbs, and COSMIC samples these fractions are found to be $\sim 24\%$, $\sim35\%$, and $\sim13\%$, respectively.

\end{itemize}

Overall, we conclude that of the three population synthesis models considered, BPASS/Bray gives results most closely aligning with the available observational constraints i.e. host demographics, and impact parameter distributions.
This may be expected as it includes the most sophisticated binary evolution modelling, although this conclusion reflects most strongly the differences in kick velocity than merger time distributions.

Our study does not include binaries that are formed through dynamical processes, for example in globular or open clusters.
In comparison to isolated systems, the merger rate of dynamically formed binaries has generally been estimated to be significantly lower. As such, isolated EM-bright compact binaries are expected to be the dominant channel for producing the observed population of SGRBs. 
If these systems are found to be more common, for instance due to their formation in young stellar clusters, 
then it will be important to account for their contribution. 
With further host associations of SGRBs and/or with gravitational waves produced by NSNS and BHNS compact binary mergers, we will be able to further refine and improve constraints on the demographic and offset estimates derived and discussed in this work. Similarly, with a greater sample of observed SGRBs, predictions for the physical input parameters such as the total net-kick velocity and merger time distributions can be tested from the comparison with associated impact parameters.

\section*{Acknowledgements}

S.F.M. is supported by a PhD studentship funded by the College of Science
and Engineering at the University of Leicester; G.P.L. is supported by STFC grant ST/S000453/1; N.R.T. acknowledge support from ERC Grant 725246 TEDE; C.J.N is supported by the Science and Technology Facilities Council grant number ST/M005917/1 and has received funding from the European Union’s Horizon 2020 research and innovation program under the Marie Sk\l{}odowska-Curie grant agreement No 823823 (Dustbusters RISE project); R.A.J.E-F. is supported by an STFC PhD studentship. Furthermore, this research used the ALICE High Performance Computing Facility at the University of Leicester. We acknowledge the Virgo Consortium for making their simulation data available. 
The \texttt{EAGLE} simulations were performed using the DiRAC-2 facility at Durham, managed by the ICC, and the PRACE facility Curie based in France at TGCC, CEA, Bruy\'eres-le-Ch\^atel.

We would like to extend our gratitude to the numerous colleagues who have provided valuable comments and meaningful discussion over the course of the work encapsulated within this paper. We would like to extend our gratitude to Rob Crain and Adrian Thob for their useful discussions regarding \texttt{EAGLE}. We would also like to thank Kim Page for discussions regarding \rm{Swift}/XRT observations.  

\section*{Data Availability}

The cosmological simulation data underlying this article are available at \href{http://icc.dur.ac.uk/Eagle/database.php}{EAGLE}, for the galaxies used. The BPASS/Bray and BPASS/Hobbs datasets are derived using the BPASS models from \href{https://bpass.auckland.ac.nz/8.html}{BPASS 2.1}. The COSMIC binaries used are produced using the COSMIC python package available at \href{https://cosmic-popsynth.github.io/}{COSMIC Homepage}. 

\bibliographystyle{mnras}
\bibliography{biblio}

\begin{thebibliography}{}
\makeatletter
\relax
\def\mn@urlcharsother{\let\do\@makeother \do\$\do\&\do\#\do\^\do\_\do\%\do\~}
\def\mn@doi{\begingroup\mn@urlcharsother \@ifnextchar [ {\mn@doi@}
  {\mn@doi@[]}}
\def\mn@doi@[#1]#2{\def\@tempa{#1}\ifx\@tempa\@empty \href
  {http://dx.doi.org/#2} {doi:#2}\else \href {http://dx.doi.org/#2} {#1}\fi
  \endgroup}
\def\mn@eprint#1#2{\mn@eprint@#1:#2::\@nil}
\def\mn@eprint@arXiv#1{\href {http://arxiv.org/abs/#1} {{\tt arXiv:#1}}}
\def\mn@eprint@dblp#1{\href {http://dblp.uni-trier.de/rec/bibtex/#1.xml}
  {dblp:#1}}
\def\mn@eprint@#1:#2:#3:#4\@nil{\def\@tempa {#1}\def\@tempb {#2}\def\@tempc
  {#3}\ifx \@tempc \@empty \let \@tempc \@tempb \let \@tempb \@tempa \fi \ifx
  \@tempb \@empty \def\@tempb {arXiv}\fi \@ifundefined
  {mn@eprint@\@tempb}{\@tempb:\@tempc}{\expandafter \expandafter \csname
  mn@eprint@\@tempb\endcsname \expandafter{\@tempc}}}

\bibitem[\protect\citeauthoryear{{Abbott} et~al.,}{{Abbott}
  et~al.}{2016}]{LIGO2016}
{Abbott} B.~P.,  et~al., 2016, \mn@doi [\prl] {10.1103/PhysRevLett.116.061102},
  \href {https://ui.adsabs.harvard.edu/abs/2016PhRvL.116f1102A} {116, 061102}

\bibitem[\protect\citeauthoryear{{Abbott} et~al.,}{{Abbott}
  et~al.}{2017a}]{abbott2017H0}
{Abbott} B.~P.,  et~al., 2017a, \mn@doi [\nat] {10.1038/nature24471}, \href
  {https://ui.adsabs.harvard.edu/abs/2017Natur.551...85A} {551, 85}

\bibitem[\protect\citeauthoryear{{Abbott} et~al.,}{{Abbott}
  et~al.}{2017b}]{abbott2017MMA}
{Abbott} B.~P.,  et~al., 2017b, \mn@doi [\apjl] {10.3847/2041-8213/aa91c9},
  \href {https://ui.adsabs.harvard.edu/abs/2017ApJ...848L..12A} {848, L12}

\bibitem[\protect\citeauthoryear{{Abbott} et~al.,}{{Abbott}
  et~al.}{2017c}]{GW170817LIGO}
{Abbott} B.~P.,  et~al., 2017c, \mn@doi [\apjl] {10.3847/2041-8213/aa920c},
  \href {https://ui.adsabs.harvard.edu/abs/2017ApJ...848L..13A} {848, L13}

\bibitem[\protect\citeauthoryear{{Abbott} et~al.,}{{Abbott}
  et~al.}{2018}]{GWProspectsv6}
{Abbott} B.~P.,  et~al., 2018, \mn@doi [Living Reviews in Relativity]
  {10.1007/s41114-018-0012-9}, \href
  {https://ui.adsabs.harvard.edu/abs/2018LRR....21....3A} {21, 3}

\bibitem[\protect\citeauthoryear{{Abbott} et~al.,}{{Abbott}
  et~al.}{2021a}]{LIGOGWTC2-20201}
{Abbott} R.,  et~al., 2021a, \mn@doi [Physical Review X]
  {10.1103/PhysRevX.11.021053}, \href
  {https://ui.adsabs.harvard.edu/abs/2021PhRvX..11b1053A} {11, 021053}

\bibitem[\protect\citeauthoryear{{Abbott} et~al.,}{{Abbott}
  et~al.}{2021b}]{LIGO2021BHNS}
{Abbott} R.,  et~al., 2021b, \mn@doi [\apjl] {10.3847/2041-8213/ac082e}, \href
  {https://ui.adsabs.harvard.edu/abs/2021ApJ...915L...5A} {915, L5}

\bibitem[\protect\citeauthoryear{{Acciari} et~al.,}{{Acciari}
  et~al.}{2021}]{MAGIC2021}
{Acciari} V.~A.,  et~al., 2021, \mn@doi [\apj] {10.3847/1538-4357/abd249},
  \href {https://ui.adsabs.harvard.edu/abs/2021ApJ...908...90A} {908, 90}

\bibitem[\protect\citeauthoryear{{Ackley} et~al.,}{{Ackley}
  et~al.}{2020}]{ENGRAVE190814_2020}
{Ackley} K.,  et~al., 2020, arXiv e-prints, \href
  {https://ui.adsabs.harvard.edu/abs/2020arXiv200201950A} {p. arXiv:2002.01950}

\bibitem[\protect\citeauthoryear{{Alexander} et~al.,}{{Alexander}
  et~al.}{2017}]{Alexander2017}
{Alexander} K.~D.,  et~al., 2017, \mn@doi [\apjl] {10.3847/2041-8213/aa905d},
  \href {https://ui.adsabs.harvard.edu/abs/2017ApJ...848L..21A} {848, L21}

\bibitem[\protect\citeauthoryear{{Amaro-Seoane}, {Miller}  \&
  {Kennedy}}{{Amaro-Seoane} et~al.}{2012}]{Amaro-Seoane2012}
{Amaro-Seoane} P.,  {Miller} M.~C.,   {Kennedy} G.~F.,  2012, \mn@doi [\mnras]
  {10.1111/j.1365-2966.2012.21162.x}, \href
  {https://ui.adsabs.harvard.edu/abs/2012MNRAS.425.2401A} {425, 2401}

\bibitem[\protect\citeauthoryear{{Andreoni} et~al.,}{{Andreoni}
  et~al.}{2020}]{GROWTH2020GW190814}
{Andreoni} I.,  et~al., 2020, \mn@doi [\apj] {10.3847/1538-4357/ab6a1b}, \href
  {https://ui.adsabs.harvard.edu/abs/2020ApJ...890..131A} {890, 131}

\bibitem[\protect\citeauthoryear{{Andrews} \& {Mandel}}{{Andrews} \&
  {Mandel}}{2019}]{Andrews2019}
{Andrews} J.~J.,  {Mandel} I.,  2019, \mn@doi [\apjl]
  {10.3847/2041-8213/ab2ed1}, \href
  {https://ui.adsabs.harvard.edu/abs/2019ApJ...880L...8A} {880, L8}

\bibitem[\protect\citeauthoryear{{Arcavi} et~al.,}{{Arcavi}
  et~al.}{2017}]{Arcavi2017}
{Arcavi} I.,  et~al., 2017, \mn@doi [\apjl] {10.3847/2041-8213/aa910f}, \href
  {https://ui.adsabs.harvard.edu/abs/2017ApJ...848L..33A} {848, L33}

\bibitem[\protect\citeauthoryear{{Artale}, {Mapelli}, {Giacobbo}, {Sabha},
  {Spera}, {Santoliquido}  \& {Bressan}}{{Artale} et~al.}{2019}]{Artale_2019}
{Artale} M.~C.,  {Mapelli} M.,  {Giacobbo} N.,  {Sabha} N.~B.,  {Spera} M.,
  {Santoliquido} F.,   {Bressan} A.,  2019, \mn@doi [\mnras]
  {10.1093/mnras/stz1382}, \href
  {https://ui.adsabs.harvard.edu/abs/2019MNRAS.487.1675A} {487, 1675}

\bibitem[\protect\citeauthoryear{{Artale}, {Mapelli}, {Bouffanais}, {Giacobbo},
  {Pasquato}  \& {Spera}}{{Artale} et~al.}{2020a}]{Artale2020a}
{Artale} M.~C.,  {Mapelli} M.,  {Bouffanais} Y.,  {Giacobbo} N.,  {Pasquato}
  M.,   {Spera} M.,  2020a, \mn@doi [\mnras] {10.1093/mnras/stz3190}, \href
  {https://ui.adsabs.harvard.edu/abs/2020MNRAS.491.3419A} {491, 3419}

\bibitem[\protect\citeauthoryear{{Artale}, {Bouffanais}, {Mapelli}, {Giacobbo},
  {Sabha}, {Santoliquido}, {Pasquato}  \& {Spera}}{{Artale}
  et~al.}{2020b}]{Artale2020b}
{Artale} M.~C.,  {Bouffanais} Y.,  {Mapelli} M.,  {Giacobbo} N.,  {Sabha}
  N.~B.,  {Santoliquido} F.,  {Pasquato} M.,   {Spera} M.,  2020b, \mn@doi
  [\mnras] {10.1093/mnras/staa1252}, \href
  {https://ui.adsabs.harvard.edu/abs/2020MNRAS.495.1841A} {495, 1841}

\bibitem[\protect\citeauthoryear{{Astropy Collaboration} et~al.,}{{Astropy
  Collaboration} et~al.}{2013}]{astropy:2013}
{Astropy Collaboration} et~al., 2013, \mn@doi [\aap]
  {10.1051/0004-6361/201322068}, \href
  {http://adsabs.harvard.edu/abs/2013A%26A...558A..33A} {558, A33}

\bibitem[\protect\citeauthoryear{{Band} et~al.,}{{Band}
  et~al.}{1993}]{Band1993}
{Band} D.,  et~al., 1993, \mn@doi [\apj] {10.1086/172995}, \href
  {https://ui.adsabs.harvard.edu/abs/1993ApJ...413..281B} {413, 281}

\bibitem[\protect\citeauthoryear{{Barbieri}, {Salafia}, {Perego}, {Colpi}  \&
  {Ghirlanda}}{{Barbieri} et~al.}{2019}]{Barbieri_2019}
{Barbieri} C.,  {Salafia} O.~S.,  {Perego} A.,  {Colpi} M.,   {Ghirlanda} G.,
  2019, \mn@doi [\aap] {10.1051/0004-6361/201935443}, \href
  {https://ui.adsabs.harvard.edu/abs/2019A&A...625A.152B} {625, A152}

\bibitem[\protect\citeauthoryear{{Belczy{\'n}ski} \& {Bulik}}{{Belczy{\'n}ski}
  \& {Bulik}}{1999}]{belczynski1999effect}
{Belczy{\'n}ski} K.,  {Bulik} T.,  1999, \aap, \href
  {https://ui.adsabs.harvard.edu/abs/1999A&A...346...91B} {346, 91}

\bibitem[\protect\citeauthoryear{{Belczynski}, {Perna}, {Bulik}, {Kalogera},
  {Ivanova}  \& {Lamb}}{{Belczynski} et~al.}{2006}]{Belczynski2006}
{Belczynski} K.,  {Perna} R.,  {Bulik} T.,  {Kalogera} V.,  {Ivanova} N.,
  {Lamb} D.~Q.,  2006, \mn@doi [\apj] {10.1086/505169}, \href
  {https://ui.adsabs.harvard.edu/abs/2006ApJ...648.1110B} {648, 1110}

\bibitem[\protect\citeauthoryear{Belczynski et~al.,}{Belczynski
  et~al.}{2018}]{Belczynski_2018}
Belczynski K.,  et~al., 2018, \mn@doi [Astronomy & Astrophysics]
  {10.1051/0004-6361/201732428}, 615, A91

\bibitem[\protect\citeauthoryear{{Berger}}{{Berger}}{2006}]{Berger2006GCN}
{Berger} E.,  2006, GRB Coordinates Network, \href
  {https://ui.adsabs.harvard.edu/abs/2006GCN..5965....1B} {5965, 1}

\bibitem[\protect\citeauthoryear{{Berger}}{{Berger}}{2009}]{berger2009host}
{Berger} E.,  2009, \mn@doi [\apj] {10.1088/0004-637X/690/1/231}, \href
  {https://ui.adsabs.harvard.edu/abs/2009ApJ...690..231B} {690, 231}

\bibitem[\protect\citeauthoryear{{Berger}}{{Berger}}{2010}]{berger2010short}
{Berger} E.,  2010, \mn@doi [\apj] {10.1088/0004-637X/722/2/1946}, \href
  {https://ui.adsabs.harvard.edu/abs/2010ApJ...722.1946B} {722, 1946}

\bibitem[\protect\citeauthoryear{{Berger}}{{Berger}}{2014}]{berger2014short}
{Berger} E.,  2014, \mn@doi [\araa] {10.1146/annurev-astro-081913-035926},
  \href {https://ui.adsabs.harvard.edu/abs/2014ARA&A..52...43B} {52, 43}

\bibitem[\protect\citeauthoryear{{Berger} \& {Soderberg}}{{Berger} \&
  {Soderberg}}{2005}]{Berger2005GCN051221}
{Berger} E.,  {Soderberg} A.~M.,  2005, GRB Coordinates Network, \href
  {https://ui.adsabs.harvard.edu/abs/2005GCN..4384....1B} {4384, 1}

\bibitem[\protect\citeauthoryear{{Berger}, {Cenko}, {Fox}  \&
  {Cucchiara}}{{Berger} et~al.}{2009}]{Berger2009}
{Berger} E.,  {Cenko} S.~B.,  {Fox} D.~B.,   {Cucchiara} A.,  2009, \mn@doi
  [\apj] {10.1088/0004-637X/704/1/877}, \href
  {https://ui.adsabs.harvard.edu/abs/2009ApJ...704..877B} {704, 877}

\bibitem[\protect\citeauthoryear{{Berger} et~al.,}{{Berger}
  et~al.}{2013}]{Berger2013GRB120804}
{Berger} E.,  et~al., 2013, \mn@doi [\apj] {10.1088/0004-637X/765/2/121}, \href
  {https://ui.adsabs.harvard.edu/abs/2013ApJ...765..121B} {765, 121}

\bibitem[\protect\citeauthoryear{{Bloom}, {Sigurdsson}  \& {Pols}}{{Bloom}
  et~al.}{1999}]{Bloom1999}
{Bloom} J.~S.,  {Sigurdsson} S.,   {Pols} O.~R.,  1999, \mn@doi [\mnras]
  {10.1046/j.1365-8711.1999.02437.x}, \href
  {https://ui.adsabs.harvard.edu/abs/1999MNRAS.305..763B} {305, 763}

\bibitem[\protect\citeauthoryear{{Bloom}, {Kulkarni}  \& {Djorgovski}}{{Bloom}
  et~al.}{2002}]{bloom2002}
{Bloom} J.~S.,  {Kulkarni} S.~R.,   {Djorgovski} S.~G.,  2002, \mn@doi [\aj]
  {10.1086/338893}, \href
  {https://ui.adsabs.harvard.edu/abs/2002AJ....123.1111B} {123, 1111}

\bibitem[\protect\citeauthoryear{{Bloom} et~al.,}{{Bloom}
  et~al.}{2006}]{Bloom2006}
{Bloom} J.~S.,  et~al., 2006, \mn@doi [\apj] {10.1086/498107}, \href
  {https://ui.adsabs.harvard.edu/abs/2006ApJ...638..354B} {638, 354}

\bibitem[\protect\citeauthoryear{{Bloom} et~al.,}{{Bloom}
  et~al.}{2007}]{bloom2007putative}
{Bloom} J.~S.,  et~al., 2007, \mn@doi [\apj] {10.1086/509114}, \href
  {https://ui.adsabs.harvard.edu/abs/2007ApJ...654..878B} {654, 878}

\bibitem[\protect\citeauthoryear{{Bovy}}{{Bovy}}{2015}]{Bovy_2015}
{Bovy} J.,  2015, \mn@doi [\apjs] {10.1088/0067-0049/216/2/29}, \href
  {https://ui.adsabs.harvard.edu/abs/2015ApJS..216...29B} {216, 29}

\bibitem[\protect\citeauthoryear{{Branchesi}}{{Branchesi}}{2016}]{Branchesi2016}
{Branchesi} M.,  2016, in Journal of Physics Conference Series. p. 022004,
  \mn@doi{10.1088/1742-6596/718/2/022004}

\bibitem[\protect\citeauthoryear{{Bray} \& {Eldridge}}{{Bray} \&
  {Eldridge}}{2016}]{bray2016neutron}
{Bray} J.~C.,  {Eldridge} J.~J.,  2016, \mn@doi [\mnras]
  {10.1093/mnras/stw1275}, \href
  {https://ui.adsabs.harvard.edu/abs/2016MNRAS.461.3747B} {461, 3747}

\bibitem[\protect\citeauthoryear{Bray \& Eldridge}{Bray \&
  Eldridge}{2018}]{Bray_2018}
Bray J.~C.,  Eldridge J.~J.,  2018, \mn@doi [Monthly Notices of the Royal
  Astronomical Society] {10.1093/mnras/sty2230}, 480, 5657

\bibitem[\protect\citeauthoryear{{Breivik} et~al.,}{{Breivik}
  et~al.}{2020}]{breivik2019cosmic}
{Breivik} K.,  et~al., 2020, \mn@doi [\apj] {10.3847/1538-4357/ab9d85}, \href
  {https://ui.adsabs.harvard.edu/abs/2020ApJ...898...71B} {898, 71}

\bibitem[\protect\citeauthoryear{{Bromberg}, {Nakar}, {Piran}  \&
  {Sari}}{{Bromberg} et~al.}{2013}]{Bromberg2013}
{Bromberg} O.,  {Nakar} E.,  {Piran} T.,   {Sari} R.,  2013, \mn@doi [\apj]
  {10.1088/0004-637X/764/2/179}, \href
  {https://ui.adsabs.harvard.edu/abs/2013ApJ...764..179B} {764, 179}

\bibitem[\protect\citeauthoryear{{Bulik}, {Belczy{\'n}ski}  \&
  {Zbijewski}}{{Bulik} et~al.}{1999}]{Bulik1999}
{Bulik} T.,  {Belczy{\'n}ski} K.,   {Zbijewski} W.,  1999, \mn@doi [\mnras]
  {10.1046/j.1365-8711.1999.02878.x}, \href
  {https://ui.adsabs.harvard.edu/abs/1999MNRAS.309..629B} {309, 629}

\bibitem[\protect\citeauthoryear{{Bullock}, {Kolatt}, {Sigad}, {Somerville},
  {Kravtsov}, {Klypin}, {Primack}  \& {Dekel}}{{Bullock}
  et~al.}{2001}]{Bullock2001}
{Bullock} J.~S.,  {Kolatt} T.~S.,  {Sigad} Y.,  {Somerville} R.~S.,  {Kravtsov}
  A.~V.,  {Klypin} A.~A.,  {Primack} J.~R.,   {Dekel} A.,  2001, \mn@doi
  [\mnras] {10.1046/j.1365-8711.2001.04068.x}, \href
  {https://ui.adsabs.harvard.edu/abs/2001MNRAS.321..559B} {321, 559}

\bibitem[\protect\citeauthoryear{{Calvi}, {Vulcani}, {Poggianti}, {Moretti},
  {Fritz}  \& {Fasano}}{{Calvi} et~al.}{2018}]{Calvi2018}
{Calvi} R.,  {Vulcani} B.,  {Poggianti} B.~M.,  {Moretti} A.,  {Fritz} J.,
  {Fasano} G.,  2018, \mn@doi [\mnras] {10.1093/mnras/sty2476}, \href
  {https://ui.adsabs.harvard.edu/abs/2018MNRAS.481.3456C} {481, 3456}

\bibitem[\protect\citeauthoryear{{Carlberg} et~al.,}{{Carlberg}
  et~al.}{1997}]{Carlberg1997}
{Carlberg} R.~G.,  et~al., 1997, \mn@doi [\apjl] {10.1086/310801}, \href
  {https://ui.adsabs.harvard.edu/abs/1997ApJ...485L..13C} {485, L13}

\bibitem[\protect\citeauthoryear{{Castro-Tirado}, {Sanchez-Ramirez}, {Lombardi}
   \& {Rivero}}{{Castro-Tirado} et~al.}{2015}]{CastroTirado2015}
{Castro-Tirado} A.~J.,  {Sanchez-Ramirez} R.,  {Lombardi} G.,   {Rivero} M.~A.,
   2015, GRB Coordinates Network, \href
  {https://ui.adsabs.harvard.edu/abs/2015GCN.17758....1C} {17758, 1}

\bibitem[\protect\citeauthoryear{{Cenko}, {Bloom}, {Perley}, {Cobb}, {Morgan},
  {Miller}, {Modjaz}  \& {James}}{{Cenko} et~al.}{2010}]{Cenko2010GCN}
{Cenko} S.~B.,  {Bloom} J.~S.,  {Perley} D.~A.,  {Cobb} B.~E.,  {Morgan} A.~N.,
   {Miller} A.~A.,  {Modjaz} M.,   {James} B.,  2010, GRB Coordinates Network,
  \href {https://ui.adsabs.harvard.edu/abs/2010GCN.10389....1C} {10389, 1}

\bibitem[\protect\citeauthoryear{{Choksi}, {Volonteri}, {Colpi}, {Gnedin}  \&
  {Li}}{{Choksi} et~al.}{2019}]{Choksi_2019}
{Choksi} N.,  {Volonteri} M.,  {Colpi} M.,  {Gnedin} O.~Y.,   {Li} H.,  2019,
  \mn@doi [\apj] {10.3847/1538-4357/aaffde}, \href
  {https://ui.adsabs.harvard.edu/abs/2019ApJ...873..100C} {873, 100}

\bibitem[\protect\citeauthoryear{{Chornock} \& {Berger}}{{Chornock} \&
  {Berger}}{2011}]{Chornock2011GCN}
{Chornock} R.,  {Berger} E.,  2011, GRB Coordinates Network, \href
  {https://ui.adsabs.harvard.edu/abs/2011GCN.11518....1C} {11518, 1}

\bibitem[\protect\citeauthoryear{{Chornock} et~al.,}{{Chornock}
  et~al.}{2017}]{Chornock2017}
{Chornock} R.,  et~al., 2017, \mn@doi [\apjl] {10.3847/2041-8213/aa905c}, \href
  {https://ui.adsabs.harvard.edu/abs/2017ApJ...848L..19C} {848, L19}

\bibitem[\protect\citeauthoryear{{Chruslinska}, {Nelemans}  \&
  {Belczynski}}{{Chruslinska} et~al.}{2019}]{Chruslinska_2018}
{Chruslinska} M.,  {Nelemans} G.,   {Belczynski} K.,  2019, \mn@doi [\mnras]
  {10.1093/mnras/sty3087}, \href
  {https://ui.adsabs.harvard.edu/abs/2019MNRAS.482.5012C} {482, 5012}

\bibitem[\protect\citeauthoryear{{Church}, {Levan}, {Davies}  \&
  {Tanvir}}{{Church} et~al.}{2011}]{Church2011}
{Church} R.~P.,  {Levan} A.~J.,  {Davies} M.~B.,   {Tanvir} N.,  2011, \mn@doi
  [\mnras] {10.1111/j.1365-2966.2011.18277.x}, \href
  {https://ui.adsabs.harvard.edu/abs/2011MNRAS.413.2004C} {413, 2004}

\bibitem[\protect\citeauthoryear{{Church}, {Levan}, {Davies}  \&
  {Tanvir}}{{Church} et~al.}{2012}]{Church2012SGRBs}
{Church} R.~P.,  {Levan} A.~J.,  {Davies} M.~B.,   {Tanvir} N.,  2012, Memorie
  della Societa Astronomica Italiana Supplementi, \href
  {https://ui.adsabs.harvard.edu/abs/2012MSAIS..21..104C} {21, 104}

\bibitem[\protect\citeauthoryear{{Comerford} \& {Natarajan}}{{Comerford} \&
  {Natarajan}}{2007}]{Comerford2007}
{Comerford} J.~M.,  {Natarajan} P.,  2007, \mn@doi [\mnras]
  {10.1111/j.1365-2966.2007.11934.x}, \href
  {https://ui.adsabs.harvard.edu/abs/2007MNRAS.379..190C} {379, 190}

\bibitem[\protect\citeauthoryear{{Correa}, {Schaye}, {Clauwens}, {Bower},
  {Crain}, {Schaller}, {Theuns}  \& {Thob}}{{Correa}
  et~al.}{2017}]{CorreaEagle2017}
{Correa} C.~A.,  {Schaye} J.,  {Clauwens} B.,  {Bower} R.~G.,  {Crain} R.~A.,
  {Schaller} M.,  {Theuns} T.,   {Thob} A. C.~R.,  2017, \mn@doi [\mnras]
  {10.1093/mnrasl/slx133}, \href
  {https://ui.adsabs.harvard.edu/abs/2017MNRAS.472L..45C} {472, L45}

\bibitem[\protect\citeauthoryear{{Corsi} et~al.,}{{Corsi}
  et~al.}{2018}]{Corsi2018}
{Corsi} A.,  et~al., 2018, \mn@doi [\apjl] {10.3847/2041-8213/aacdfd}, \href
  {https://ui.adsabs.harvard.edu/abs/2018ApJ...861L..10C} {861, L10}

\bibitem[\protect\citeauthoryear{{Coulter} et~al.,}{{Coulter}
  et~al.}{2017}]{Coulter2017}
{Coulter} D.~A.,  et~al., 2017, \mn@doi [Science] {10.1126/science.aap9811},
  \href {https://ui.adsabs.harvard.edu/abs/2017Sci...358.1556C} {358, 1556}

\bibitem[\protect\citeauthoryear{{Cowperthwaite} et~al.,}{{Cowperthwaite}
  et~al.}{2017}]{Cowperthwaite2017}
{Cowperthwaite} P.~S.,  et~al., 2017, \mn@doi [\apjl]
  {10.3847/2041-8213/aa8fc7}, \href
  {https://ui.adsabs.harvard.edu/abs/2017ApJ...848L..17C} {848, L17}

\bibitem[\protect\citeauthoryear{{Crain} et~al.,}{{Crain}
  et~al.}{2015}]{crain2015eagle}
{Crain} R.~A.,  et~al., 2015, \mn@doi [\mnras] {10.1093/mnras/stv725}, \href
  {https://ui.adsabs.harvard.edu/abs/2015MNRAS.450.1937C} {450, 1937}

\bibitem[\protect\citeauthoryear{{Cucchiara} \& {Levan}}{{Cucchiara} \&
  {Levan}}{2016}]{Cucchiara2016}
{Cucchiara} A.,  {Levan} A.~J.,  2016, GRB Coordinates Network, \href
  {https://ui.adsabs.harvard.edu/abs/2016GCN.19565....1C} {19565, 1}

\bibitem[\protect\citeauthoryear{{Cucchiara}, {Fox}, {Berger}  \&
  P.A.}{{Cucchiara} et~al.}{2006}]{Cucchiara2006GCN}
{Cucchiara} A.,  {Fox} D.,  {Berger} E.,   P.A. P.,  2006, GRB Coordinates
  Network, 5470, 1

\bibitem[\protect\citeauthoryear{{Cucchiara}, {Cenko}, {Perley}, {Capone}  \&
  {Toy}}{{Cucchiara} et~al.}{2014}]{Cucchiara2014GCN}
{Cucchiara} A.,  {Cenko} S.~B.,  {Perley} D.~A.,  {Capone} J.,   {Toy} V.,
  2014, GRB Coordinates Network, \href
  {https://ui.adsabs.harvard.edu/abs/2014GCN.16774....1C} {16774, 1}

\bibitem[\protect\citeauthoryear{{D'Avanzo} et~al.,}{{D'Avanzo}
  et~al.}{2009}]{Davanzo2009}
{D'Avanzo} P.,  et~al., 2009, \mn@doi [\aap] {10.1051/0004-6361/200811294},
  \href {https://ui.adsabs.harvard.edu/abs/2009A&A...498..711D} {498, 711}

\bibitem[\protect\citeauthoryear{{Dichiara}, {Troja}, {O'Connor}, {Marshall},
  {Beniamini}, {Cannizzo}, {Lien}  \& {Sakamoto}}{{Dichiara}
  et~al.}{2020}]{dichiara2019short}
{Dichiara} S.,  {Troja} E.,  {O'Connor} B.,  {Marshall} F.~E.,  {Beniamini} P.,
   {Cannizzo} J.~K.,  {Lien} A.~Y.,   {Sakamoto} T.,  2020, \mn@doi [\mnras]
  {10.1093/mnras/staa124}, \href
  {https://ui.adsabs.harvard.edu/abs/2020MNRAS.492.5011D} {492, 5011}

\bibitem[\protect\citeauthoryear{{Eldridge}, {Stanway}, {Xiao}, {McClelland },
  {Taylor}, {Ng}, {Greis}  \& {Bray}}{{Eldridge} et~al.}{2017}]{Eldridge2017}
{Eldridge} J.~J.,  {Stanway} E.~R.,  {Xiao} L.,  {McClelland } L.~A.~S.,
  {Taylor} G.,  {Ng} M.,  {Greis} S.~M.~L.,   {Bray} J.~C.,  2017, \mn@doi
  [\pasa] {10.1017/pasa.2017.51}, \href
  {https://ui.adsabs.harvard.edu/abs/2017PASA...34...58E} {34, e058}

\bibitem[\protect\citeauthoryear{{Evans} et~al.,}{{Evans}
  et~al.}{2016}]{evans2016}
{Evans} P.~A.,  et~al., 2016, \mn@doi [\mnras] {10.1093/mnras/stv2213}, \href
  {https://ui.adsabs.harvard.edu/abs/2016MNRAS.455.1522E} {455, 1522}

\bibitem[\protect\citeauthoryear{{Evans} et~al.,}{{Evans}
  et~al.}{2017}]{Evans2017}
{Evans} P.~A.,  et~al., 2017, \mn@doi [Science] {10.1126/science.aap9580},
  \href {https://ui.adsabs.harvard.edu/abs/2017Sci...358.1565E} {358, 1565}

\bibitem[\protect\citeauthoryear{{Fang}, {Bullock}  \& {Boylan-Kolchin}}{{Fang}
  et~al.}{2013}]{Fang2013}
{Fang} T.,  {Bullock} J.,   {Boylan-Kolchin} M.,  2013, \mn@doi [\apj]
  {10.1088/0004-637X/762/1/20}, \href
  {https://ui.adsabs.harvard.edu/abs/2013ApJ...762...20F} {762, 20}

\bibitem[\protect\citeauthoryear{{Fern{\'a}ndez} \&
  {Kobayashi}}{{Fern{\'a}ndez} \& {Kobayashi}}{2019}]{Fernandez_2019}
{Fern{\'a}ndez} J.~J.,  {Kobayashi} S.,  2019, \mn@doi [\mnras]
  {10.1093/mnras/stz1353}, \href
  {https://ui.adsabs.harvard.edu/abs/2019MNRAS.487.1200F} {487, 1200}

\bibitem[\protect\citeauthoryear{{Foley}, {Bloom}  \& {Chen}}{{Foley}
  et~al.}{2005}]{Foley2005GCN}
{Foley} R.~J.,  {Bloom} J.~S.,   {Chen} H.~W.,  2005, GRB Coordinates Network,
  \href {https://ui.adsabs.harvard.edu/abs/2005GCN..3808....1F} {3808, 1}

\bibitem[\protect\citeauthoryear{{Fong} \& {Berger}}{{Fong} \&
  {Berger}}{2013}]{fong2013locations}
{Fong} W.,  {Berger} E.,  2013, \mn@doi [\apj] {10.1088/0004-637X/776/1/18},
  \href {https://ui.adsabs.harvard.edu/abs/2013ApJ...776...18F} {776, 18}

\bibitem[\protect\citeauthoryear{{Fong} et~al.,}{{Fong}
  et~al.}{2013}]{fong2013demographics}
{Fong} W.,  et~al., 2013, \mn@doi [\apj] {10.1088/0004-637X/769/1/56}, \href
  {https://ui.adsabs.harvard.edu/abs/2013ApJ...769...56F} {769, 56}

\bibitem[\protect\citeauthoryear{{Fong}, {Berger}, {Margutti}  \&
  {Zauderer}}{{Fong} et~al.}{2015a}]{Fong_2015}
{Fong} W.,  {Berger} E.,  {Margutti} R.,   {Zauderer} B.~A.,  2015a, \mn@doi
  [\apj] {10.1088/0004-637X/815/2/102}, \href
  {https://ui.adsabs.harvard.edu/abs/2015ApJ...815..102F} {815, 102}

\bibitem[\protect\citeauthoryear{{Fong}, {Berger}, {Fox}  \& {Shappee}}{{Fong}
  et~al.}{2015b}]{Fong2015GCN}
{Fong} W.,  {Berger} E.,  {Fox} D.,   {Shappee} B.~J.,  2015b, GRB Coordinates
  Network, \href {https://ui.adsabs.harvard.edu/abs/2015GCN.17333....1F}
  {17333, 1}

\bibitem[\protect\citeauthoryear{{Fong} et~al.,}{{Fong}
  et~al.}{2016}]{Fong2016}
{Fong} W.,  et~al., 2016, \mn@doi [\apj] {10.3847/1538-4357/833/2/151}, \href
  {https://ui.adsabs.harvard.edu/abs/2016ApJ...833..151F} {833, 151}

\bibitem[\protect\citeauthoryear{{Fong} et~al.,}{{Fong}
  et~al.}{2017}]{fong2017electromagnetic}
{Fong} W.,  et~al., 2017, \mn@doi [\apjl] {10.3847/2041-8213/aa9018}, \href
  {https://ui.adsabs.harvard.edu/abs/2017ApJ...848L..23F} {848, L23}

\bibitem[\protect\citeauthoryear{{Fong} et~al.,}{{Fong}
  et~al.}{2020}]{Fong2020}
{Fong} W.,  et~al., 2020, arXiv e-prints, \href
  {https://ui.adsabs.harvard.edu/abs/2020arXiv200808593F} {p. arXiv:2008.08593}

\bibitem[\protect\citeauthoryear{{Foucart}, {Duez}, {Kidder}, {Nissanke},
  {Pfeiffer}  \& {Scheel}}{{Foucart} et~al.}{2019}]{Foucart_2019}
{Foucart} F.,  {Duez} M.~D.,  {Kidder} L.~E.,  {Nissanke} S.~M.,  {Pfeiffer}
  H.~P.,   {Scheel} M.~A.,  2019, \mn@doi [\prd] {10.1103/PhysRevD.99.103025},
  \href {https://ui.adsabs.harvard.edu/abs/2019PhRvD..99j3025F} {99, 103025}

\bibitem[\protect\citeauthoryear{{Fox} et~al.,}{{Fox} et~al.}{2005}]{Fox2005}
{Fox} D.~B.,  et~al., 2005, \mn@doi [\nat] {10.1038/nature04189}, \href
  {https://ui.adsabs.harvard.edu/abs/2005Natur.437..845F} {437, 845}

\bibitem[\protect\citeauthoryear{{Fragione} \& {Kocsis}}{{Fragione} \&
  {Kocsis}}{2018}]{Fragione_2018}
{Fragione} G.,  {Kocsis} B.,  2018, \mn@doi [\prl]
  {10.1103/PhysRevLett.121.161103}, \href
  {https://ui.adsabs.harvard.edu/abs/2018PhRvL.121p1103F} {121, 161103}

\bibitem[\protect\citeauthoryear{{Fragos}, {Willems}, {Kalogera}, {Ivanova},
  {Rockefeller}, {Fryer}  \& {Young}}{{Fragos}
  et~al.}{2009}]{fragos2009understanding}
{Fragos} T.,  {Willems} B.,  {Kalogera} V.,  {Ivanova} N.,  {Rockefeller} G.,
  {Fryer} C.~L.,   {Young} P.~A.,  2009, \mn@doi [\apj]
  {10.1088/0004-637X/697/2/1057}, \href
  {https://ui.adsabs.harvard.edu/abs/2009ApJ...697.1057F} {697, 1057}

\bibitem[\protect\citeauthoryear{{Frith}, {Metcalfe}  \& {Shanks}}{{Frith}
  et~al.}{2006}]{Frith2006}
{Frith} W.~J.,  {Metcalfe} N.,   {Shanks} T.,  2006, \mn@doi [\mnras]
  {10.1111/j.1365-2966.2006.10736.x}, \href
  {https://ui.adsabs.harvard.edu/abs/2006MNRAS.371.1601F} {371, 1601}

\bibitem[\protect\citeauthoryear{{Fryer}, {Belczynski}, {Wiktorowicz},
  {Dominik}, {Kalogera}  \& {Holz}}{{Fryer} et~al.}{2012}]{Fryer2012}
{Fryer} C.~L.,  {Belczynski} K.,  {Wiktorowicz} G.,  {Dominik} M.,  {Kalogera}
  V.,   {Holz} D.~E.,  2012, \mn@doi [\apj] {10.1088/0004-637X/749/1/91}, \href
  {https://ui.adsabs.harvard.edu/abs/2012ApJ...749...91F} {749, 91}

\bibitem[\protect\citeauthoryear{{Furlong} et~al.,}{{Furlong}
  et~al.}{2015}]{furlong2015evolution}
{Furlong} M.,  et~al., 2015, \mn@doi [\mnras] {10.1093/mnras/stv852}, \href
  {https://ui.adsabs.harvard.edu/abs/2015MNRAS.450.4486F} {450, 4486}

\bibitem[\protect\citeauthoryear{{Gehrels} et~al.,}{{Gehrels}
  et~al.}{2005}]{Gehrels2005}
{Gehrels} N.,  et~al., 2005, \mn@doi [\nat] {10.1038/nature04142}, \href
  {https://ui.adsabs.harvard.edu/abs/2005Natur.437..851G} {437, 851}

\bibitem[\protect\citeauthoryear{{Gehrels}, {Cannizzo}, {Kanner}, {Kasliwal},
  {Nissanke}  \& {Singer}}{{Gehrels} et~al.}{2016}]{gehrels2016}
{Gehrels} N.,  {Cannizzo} J.~K.,  {Kanner} J.,  {Kasliwal} M.~M.,  {Nissanke}
  S.,   {Singer} L.~P.,  2016, \mn@doi [\apj] {10.3847/0004-637X/820/2/136},
  \href {https://ui.adsabs.harvard.edu/abs/2016ApJ...820..136G} {820, 136}

\bibitem[\protect\citeauthoryear{{Ghirlanda} et~al.,}{{Ghirlanda}
  et~al.}{2016}]{ghirlanda2016}
{Ghirlanda} G.,  et~al., 2016, \mn@doi [\aap] {10.1051/0004-6361/201628993},
  \href {https://ui.adsabs.harvard.edu/abs/2016A&A...594A..84G} {594, A84}

\bibitem[\protect\citeauthoryear{{Giacobbo} \& {Mapelli}}{{Giacobbo} \&
  {Mapelli}}{2018}]{giacobbo2018progenitors}
{Giacobbo} N.,  {Mapelli} M.,  2018, \mn@doi [\mnras] {10.1093/mnras/sty1999},
  \href {https://ui.adsabs.harvard.edu/abs/2018MNRAS.480.2011G} {480, 2011}

\bibitem[\protect\citeauthoryear{{Gompertz} et~al.,}{{Gompertz}
  et~al.}{2018}]{Gompertz2018}
{Gompertz} B.~P.,  et~al., 2018, \mn@doi [\apj] {10.3847/1538-4357/aac206},
  \href {https://ui.adsabs.harvard.edu/abs/2018ApJ...860...62G} {860, 62}

\bibitem[\protect\citeauthoryear{{Gompertz}, {Levan}  \& {Tanvir}}{{Gompertz}
  et~al.}{2020}]{gompertz2020search}
{Gompertz} B.~P.,  {Levan} A.~J.,   {Tanvir} N.~R.,  2020, \mn@doi [\apj]
  {10.3847/1538-4357/ab8d24}, \href
  {https://ui.adsabs.harvard.edu/abs/2020ApJ...895...58G} {895, 58}

\bibitem[\protect\citeauthoryear{{Graham} et~al.,}{{Graham}
  et~al.}{2009}]{Graham09}
{Graham} J.~F.,  et~al., 2009, \mn@doi [\apj] {10.1088/0004-637X/698/2/1620},
  \href {https://ui.adsabs.harvard.edu/abs/2009ApJ...698.1620G} {698, 1620}

\bibitem[\protect\citeauthoryear{{Haggard}, {Nynka}, {Ruan}, {Kalogera},
  {Cenko}, {Evans}  \& {Kennea}}{{Haggard} et~al.}{2017}]{Haggard2017}
{Haggard} D.,  {Nynka} M.,  {Ruan} J.~J.,  {Kalogera} V.,  {Cenko} S.~B.,
  {Evans} P.,   {Kennea} J.~A.,  2017, \mn@doi [\apjl]
  {10.3847/2041-8213/aa8ede}, \href
  {https://ui.adsabs.harvard.edu/abs/2017ApJ...848L..25H} {848, L25}

\bibitem[\protect\citeauthoryear{{Hajela} et~al.,}{{Hajela}
  et~al.}{2019}]{Hajela2019}
{Hajela} A.,  et~al., 2019, \mn@doi [\apjl] {10.3847/2041-8213/ab5226}, \href
  {https://ui.adsabs.harvard.edu/abs/2019ApJ...886L..17H} {886, L17}

\bibitem[\protect\citeauthoryear{{Hallinan} et~al.,}{{Hallinan}
  et~al.}{2017}]{Hallinan2017}
{Hallinan} G.,  et~al., 2017, \mn@doi [Science] {10.1126/science.aap9855},
  \href {https://ui.adsabs.harvard.edu/abs/2017Sci...358.1579H} {358, 1579}

\bibitem[\protect\citeauthoryear{{Hayashi}, {Kawaguchi}, {Kiuchi}, {Kyutoku}
  \& {Shibata}}{{Hayashi} et~al.}{2020}]{Hayashi2020}
{Hayashi} K.,  {Kawaguchi} K.,  {Kiuchi} K.,  {Kyutoku} K.,   {Shibata} M.,
  2020, arXiv e-prints, \href
  {https://ui.adsabs.harvard.edu/abs/2020arXiv201002563H} {p. arXiv:2010.02563}

\bibitem[\protect\citeauthoryear{{Hinderer} et~al.,}{{Hinderer}
  et~al.}{2019}]{Hinderer_2019}
{Hinderer} T.,  et~al., 2019, \mn@doi [\prd] {10.1103/PhysRevD.100.063021},
  \href {https://ui.adsabs.harvard.edu/abs/2019PhRvD.100f3021H} {100, 063021}

\bibitem[\protect\citeauthoryear{{Hjorth} et~al.,}{{Hjorth}
  et~al.}{2005}]{Hjorth2005}
{Hjorth} J.,  et~al., 2005, \mn@doi [\apjl] {10.1086/491733}, \href
  {https://ui.adsabs.harvard.edu/abs/2005ApJ...630L.117H} {630, L117}

\bibitem[\protect\citeauthoryear{{Hobbs}, {Lorimer}, {Lyne}  \&
  {Kramer}}{{Hobbs} et~al.}{2005}]{hobbs2005statistical}
{Hobbs} G.,  {Lorimer} D.~R.,  {Lyne} A.~G.,   {Kramer} M.,  2005, \mn@doi
  [\mnras] {10.1111/j.1365-2966.2005.09087.x}, \href
  {https://ui.adsabs.harvard.edu/abs/2005MNRAS.360..974H} {360, 974}

\bibitem[\protect\citeauthoryear{{Holland}, {de Pasquale}  \&
  {Markwardt}}{{Holland} et~al.}{2007}]{Holland2007GCN}
{Holland} S.~T.,  {de Pasquale} M.,   {Markwardt} C.~B.,  2007, GRB Coordinates
  Network, \href {https://ui.adsabs.harvard.edu/abs/2007GCN..7145....1H} {7145,
  1}

\bibitem[\protect\citeauthoryear{{Huerta} et~al.,}{{Huerta}
  et~al.}{2019}]{Huerta2019}
{Huerta} E.~A.,  et~al., 2019, \mn@doi [Nature Reviews Physics]
  {10.1038/s42254-019-0097-4}, \href
  {https://ui.adsabs.harvard.edu/abs/2019NatRP...1..600H} {1, 600}

\bibitem[\protect\citeauthoryear{Hunter}{Hunter}{2007}]{Hunter:2007}
Hunter J.~D.,  2007, \mn@doi [Computing in Science \& Engineering]
  {10.1109/MCSE.2007.55}, 9, 90

\bibitem[\protect\citeauthoryear{{Hurley}, {Tout}  \& {Pols}}{{Hurley}
  et~al.}{2002}]{hurley2002}
{Hurley} J.~R.,  {Tout} C.~A.,   {Pols} O.~R.,  2002, \mn@doi [\mnras]
  {10.1046/j.1365-8711.2002.05038.x}, \href
  {https://ui.adsabs.harvard.edu/abs/2002MNRAS.329..897H} {329, 897}

\bibitem[\protect\citeauthoryear{{Im}, {Park}, {Pak}, {Jeong}, {Kim}  \&
  {Kim}}{{Im} et~al.}{2010}]{ImGCN2010}
{Im} M.,  {Park} W.~K.,  {Pak} S.,  {Jeong} H.,  {Kim} E.,   {Kim} J.,  2010,
  GRB Coordinates Network, \href
  {https://ui.adsabs.harvard.edu/abs/2010GCN.11108....1I} {11108, 1}

\bibitem[\protect\citeauthoryear{{Im} et~al.,}{{Im}
  et~al.}{2017}]{Myungshin2017}
{Im} M.,  et~al., 2017, \mn@doi [\apjl] {10.3847/2041-8213/aa9367}, \href
  {https://ui.adsabs.harvard.edu/abs/2017ApJ...849L..16I} {849, L16}

\bibitem[\protect\citeauthoryear{{Izzo}, {Cano}, { de Ugarte Postig}, {Kann},
  {Thoene}  \& {Geier}}{{Izzo} et~al.}{2017}]{Izzo2017}
{Izzo} L.,  {Cano} Z.,  { de Ugarte Postig} A.,  {Kann} D.~A.,  {Thoene} C.,
  {Geier} S.,  2017, GRB Coordinates Network, 21059, 1

\bibitem[\protect\citeauthoryear{{Jiang} et~al.,}{{Jiang}
  et~al.}{2020}]{jiang2020simulating}
{Jiang} Z.,  et~al., 2020, \mn@doi [\mnras] {10.1093/mnras/staa1989}, \href
  {https://ui.adsabs.harvard.edu/abs/2020MNRAS.498..926J} {498, 926}

\bibitem[\protect\citeauthoryear{{Jin} et~al.,}{{Jin} et~al.}{2016}]{Jin2016}
{Jin} Z.-P.,  et~al., 2016, \mn@doi [Nature Communications]
  {10.1038/ncomms12898}, \href
  {https://ui.adsabs.harvard.edu/abs/2016NatCo...712898J} {7, 12898}

\bibitem[\protect\citeauthoryear{{Jin}, {Covino}, {Liao}, {Li}, {D'Avanzo},
  {Fan}  \& {Wei}}{{Jin} et~al.}{2020}]{Jin2020}
{Jin} Z.-P.,  {Covino} S.,  {Liao} N.-H.,  {Li} X.,  {D'Avanzo} P.,  {Fan}
  Y.-Z.,   {Wei} D.-M.,  2020, \mn@doi [Nature Astronomy]
  {10.1038/s41550-019-0892-y}, \href
  {https://ui.adsabs.harvard.edu/abs/2020NatAs...4...77J} {4, 77}

\bibitem[\protect\citeauthoryear{{Kann} et~al.,}{{Kann}
  et~al.}{2011}]{Kann2011}
{Kann} D.~A.,  et~al., 2011, \mn@doi [\apj] {10.1088/0004-637X/734/2/96}, \href
  {https://ui.adsabs.harvard.edu/abs/2011ApJ...734...96K} {734, 96}

\bibitem[\protect\citeauthoryear{{Kasliwal} et~al.,}{{Kasliwal}
  et~al.}{2017}]{Kasliwal2017}
{Kasliwal} M.~M.,  et~al., 2017, \mn@doi [Science] {10.1126/science.aap9455},
  \href {https://ui.adsabs.harvard.edu/abs/2017Sci...358.1559K} {358, 1559}

\bibitem[\protect\citeauthoryear{{Katz} \& {Canel}}{{Katz} \&
  {Canel}}{1996}]{katz1996long}
{Katz} J.~I.,  {Canel} L.~M.,  1996, \mn@doi [\apj] {10.1086/178018}, \href
  {https://ui.adsabs.harvard.edu/abs/1996ApJ...471..915K} {471, 915}

\bibitem[\protect\citeauthoryear{{Kawaguchi}, {Kyutoku}, {Shibata}  \&
  {Tanaka}}{{Kawaguchi} et~al.}{2016}]{Kawaguchi2016}
{Kawaguchi} K.,  {Kyutoku} K.,  {Shibata} M.,   {Tanaka} M.,  2016, \mn@doi
  [\apj] {10.3847/0004-637X/825/1/52}, \href
  {https://ui.adsabs.harvard.edu/abs/2016ApJ...825...52K} {825, 52}

\bibitem[\protect\citeauthoryear{{Kocevski} et~al.,}{{Kocevski}
  et~al.}{2010}]{Kocevski2010}
{Kocevski} D.,  et~al., 2010, \mn@doi [\mnras]
  {10.1111/j.1365-2966.2010.16327.x}, \href
  {https://ui.adsabs.harvard.edu/abs/2010MNRAS.404..963K} {404, 963}

\bibitem[\protect\citeauthoryear{{Kouveliotou}, {Meegan}, {Fishman}, {Bhat},
  {Briggs}, {Koshut}, {Paciesas}  \& {Pendleton}}{{Kouveliotou}
  et~al.}{1993}]{Kouveliotou93}
{Kouveliotou} C.,  {Meegan} C.~A.,  {Fishman} G.~J.,  {Bhat} N.~P.,  {Briggs}
  M.~S.,  {Koshut} T.~M.,  {Paciesas} W.~S.,   {Pendleton} G.~N.,  1993,
  \mn@doi [\apjl] {10.1086/186969}, \href
  {https://ui.adsabs.harvard.edu/abs/1993ApJ...413L.101K} {413, L101}

\bibitem[\protect\citeauthoryear{{Kroupa}}{{Kroupa}}{2001}]{Kroupa2001}
{Kroupa} P.,  2001, \mn@doi [\mnras] {10.1046/j.1365-8711.2001.04022.x}, \href
  {https://ui.adsabs.harvard.edu/abs/2001MNRAS.322..231K} {322, 231}

\bibitem[\protect\citeauthoryear{{Kuin} \& {Gelbord}}{{Kuin} \&
  {Gelbord}}{2011}]{Kuin2011GCN}
{Kuin} N.~P.~M.,  {Gelbord} J.~M.,  2011, GRB Coordinates Network, \href
  {https://ui.adsabs.harvard.edu/abs/2011GCN.11575....1K} {11575, 1}

\bibitem[\protect\citeauthoryear{{Lamb} et~al.,}{{Lamb}
  et~al.}{2019a}]{Lamb2019}
{Lamb} G.~P.,  et~al., 2019a, \mn@doi [\apjl] {10.3847/2041-8213/aaf96b}, \href
  {https://ui.adsabs.harvard.edu/abs/2019ApJ...870L..15L} {870, L15}

\bibitem[\protect\citeauthoryear{{Lamb} et~al.,}{{Lamb}
  et~al.}{2019b}]{lamb2019short}
{Lamb} G.~P.,  et~al., 2019b, \mn@doi [\apj] {10.3847/1538-4357/ab38bb}, \href
  {https://ui.adsabs.harvard.edu/abs/2019ApJ...883...48L} {883, 48}

\bibitem[\protect\citeauthoryear{{Lee} \& {Ramirez-Ruiz}}{{Lee} \&
  {Ramirez-Ruiz}}{2007}]{lee2007progenitors}
{Lee} W.~H.,  {Ramirez-Ruiz} E.,  2007, \mn@doi [New Journal of Physics]
  {10.1088/1367-2630/9/1/017}, \href
  {https://ui.adsabs.harvard.edu/abs/2007NJPh....9...17L} {9, 17}

\bibitem[\protect\citeauthoryear{{Leibler} \& {Berger}}{{Leibler} \&
  {Berger}}{2010}]{Leibler2010}
{Leibler} C.~N.,  {Berger} E.,  2010, \mn@doi [\apj]
  {10.1088/0004-637X/725/1/1202}, \href
  {https://ui.adsabs.harvard.edu/abs/2010ApJ...725.1202L} {725, 1202}

\bibitem[\protect\citeauthoryear{{Levan} et~al.,}{{Levan}
  et~al.}{2006}]{levan06b}
{Levan} A.~J.,  et~al., 2006, \mn@doi [\apjl] {10.1086/507625}, \href
  {https://ui.adsabs.harvard.edu/abs/2006ApJ...648L...9L} {648, L9}

\bibitem[\protect\citeauthoryear{{Levan} et~al.,}{{Levan}
  et~al.}{2007}]{levan07}
{Levan} A.~J.,  et~al., 2007, \mn@doi [\mnras]
  {10.1111/j.1365-2966.2007.11879.x}, \href
  {https://ui.adsabs.harvard.edu/abs/2007MNRAS.378.1439L} {378, 1439}

\bibitem[\protect\citeauthoryear{{Levan}, {Tanvir}, {Wiersema}, {Hartoog},
  {Kolle}, {Mendez}  \& {Kupfer}}{{Levan} et~al.}{2013}]{Levan2013GCN}
{Levan} A.~J.,  {Tanvir} N.~R.,  {Wiersema} K.,  {Hartoog} O.,  {Kolle} K.,
  {Mendez} J.,   {Kupfer} T.,  2013, GRB Coordinates Network, \href
  {https://ui.adsabs.harvard.edu/abs/2013GCN.14742....1L} {14742, 1}

\bibitem[\protect\citeauthoryear{{Levan}, {Cenko}, {Cucchiara}  \&
  {Perley}}{{Levan} et~al.}{2014}]{Levan2014GCN}
{Levan} A.~J.,  {Cenko} S.~B.,  {Cucchiara} A.,   {Perley} D.~A.,  2014, GRB
  Coordinates Network, \href
  {https://ui.adsabs.harvard.edu/abs/2014GCN.16784....1L} {16784, 1}

\bibitem[\protect\citeauthoryear{{Levan}, {Hjorth}, {Wiersema}  \&
  {Tanvir}}{{Levan} et~al.}{2015}]{Levan2015}
{Levan} A.~J.,  {Hjorth} J.,  {Wiersema} K.,   {Tanvir} N.~R.,  2015, GRB
  Coordinates Network, \href
  {https://ui.adsabs.harvard.edu/abs/2015GCN.17281....1L} {17281, 1}

\bibitem[\protect\citeauthoryear{{Levan} et~al.,}{{Levan}
  et~al.}{2017}]{Levan2017}
{Levan} A.~J.,  et~al., 2017, \mn@doi [\apjl] {10.3847/2041-8213/aa905f}, \href
  {https://ui.adsabs.harvard.edu/abs/2017ApJ...848L..28L} {848, L28}

\bibitem[\protect\citeauthoryear{{Licquia} \& {Newman}}{{Licquia} \&
  {Newman}}{2015}]{Licquia_2015}
{Licquia} T.~C.,  {Newman} J.~A.,  2015, \mn@doi [\apj]
  {10.1088/0004-637X/806/1/96}, \href
  {https://ui.adsabs.harvard.edu/abs/2015ApJ...806...96L} {806, 96}

\bibitem[\protect\citeauthoryear{{Lien} et~al.,}{{Lien}
  et~al.}{2016}]{lien2016}
{Lien} A.,  et~al., 2016, \mn@doi [\apj] {10.3847/0004-637X/829/1/7}, \href
  {https://ui.adsabs.harvard.edu/abs/2016ApJ...829....7L} {829, 7}

\bibitem[\protect\citeauthoryear{{Lyman} et~al.,}{{Lyman}
  et~al.}{2018}]{Lyman2018}
{Lyman} J.~D.,  et~al., 2018, \mn@doi [Nature Astronomy]
  {10.1038/s41550-018-0511-3}, \href
  {https://ui.adsabs.harvard.edu/abs/2018NatAs...2..751L} {2, 751}

\bibitem[\protect\citeauthoryear{Madau \& Dickinson}{Madau \&
  Dickinson}{2014}]{Madau_2014}
Madau P.,  Dickinson M.,  2014, \mn@doi [Annual Review of Astronomy and
  Astrophysics] {10.1146/annurev-astro-081811-125615}, 52, 415–486

\bibitem[\protect\citeauthoryear{{Mainetti}, {Lupi}, {Campana}  \&
  {Colpi}}{{Mainetti} et~al.}{2016}]{Mainetti2016}
{Mainetti} D.,  {Lupi} A.,  {Campana} S.,   {Colpi} M.,  2016, \mn@doi [\mnras]
  {10.1093/mnras/stw197}, \href
  {https://ui.adsabs.harvard.edu/abs/2016MNRAS.457.2516M} {457, 2516}

\bibitem[\protect\citeauthoryear{{Maller} \& {Bullock}}{{Maller} \&
  {Bullock}}{2004}]{Maller2004}
{Maller} A.~H.,  {Bullock} J.~S.,  2004, \mn@doi [\mnras]
  {10.1111/j.1365-2966.2004.08349.x}, \href
  {https://ui.adsabs.harvard.edu/abs/2004MNRAS.355..694M} {355, 694}

\bibitem[\protect\citeauthoryear{{Mandhai}, {Tanvir}, {Lamb}, {Levan}  \&
  {Tsang}}{{Mandhai} et~al.}{2018}]{mandhai2018rate}
{Mandhai} S.,  {Tanvir} N.,  {Lamb} G.,  {Levan} A.,   {Tsang} D.,  2018,
  \mn@doi [Galaxies] {10.3390/galaxies6040130}, \href
  {https://ui.adsabs.harvard.edu/abs/2018Galax...6..130M} {6, 130}

\bibitem[\protect\citeauthoryear{{Mapelli}, {Giacobbo}, {Toffano}, {Ripamonti},
  {Bressan}, {Spera}  \& {Branchesi}}{{Mapelli} et~al.}{2018}]{Mapelli_2018}
{Mapelli} M.,  {Giacobbo} N.,  {Toffano} M.,  {Ripamonti} E.,  {Bressan} A.,
  {Spera} M.,   {Branchesi} M.,  2018, \mn@doi [\mnras]
  {10.1093/mnras/sty2663}, \href
  {https://ui.adsabs.harvard.edu/abs/2018MNRAS.481.5324M} {481, 5324}

\bibitem[\protect\citeauthoryear{{Margutti} et~al.,}{{Margutti}
  et~al.}{2012}]{Margutti12}
{Margutti} R.,  et~al., 2012, \mn@doi [\apj] {10.1088/0004-637X/756/1/63},
  \href {https://ui.adsabs.harvard.edu/abs/2012ApJ...756...63M} {756, 63}

\bibitem[\protect\citeauthoryear{{Margutti} et~al.,}{{Margutti}
  et~al.}{2017}]{Margutti2017}
{Margutti} R.,  et~al., 2017, \mn@doi [\apjl] {10.3847/2041-8213/aa9057}, \href
  {https://ui.adsabs.harvard.edu/abs/2017ApJ...848L..20M} {848, L20}

\bibitem[\protect\citeauthoryear{{Mazets} \& {Golenetskii}}{{Mazets} \&
  {Golenetskii}}{1981}]{Mazets81}
{Mazets} E.~P.,  {Golenetskii} S.~V.,  1981, \mn@doi [\apss]
  {10.1007/BF00651384}, \href
  {https://ui.adsabs.harvard.edu/abs/1981Ap&SS..75...47M} {75, 47}

\bibitem[\protect\citeauthoryear{{McAlpine} et~al.,}{{McAlpine}
  et~al.}{2016}]{McAlpine2015}
{McAlpine} S.,  et~al., 2016, \mn@doi [Astronomy and Computing]
  {10.1016/j.ascom.2016.02.004}, \href
  {https://ui.adsabs.harvard.edu/abs/2016A&C....15...72M} {15, 72}

\bibitem[\protect\citeauthoryear{{McCarthy}, {Zheng}  \&
  {Ramirez-Ruiz}}{{McCarthy} et~al.}{2020}]{McCarthy2020}
{McCarthy} K.~S.,  {Zheng} Z.,   {Ramirez-Ruiz} E.,  2020, arXiv e-prints,
  \href {https://ui.adsabs.harvard.edu/abs/2020arXiv200715024M} {p.
  arXiv:2007.15024}

\bibitem[\protect\citeauthoryear{{McCully} et~al.,}{{McCully}
  et~al.}{2017}]{McCully2017}
{McCully} C.,  et~al., 2017, \mn@doi [\apjl] {10.3847/2041-8213/aa9111}, \href
  {https://ui.adsabs.harvard.edu/abs/2017ApJ...848L..32M} {848, L32}

\bibitem[\protect\citeauthoryear{McKinney}{McKinney}{2010}]{mckinney-proc-scipy-2010}
McKinney W.,  2010, in van~der Walt S.,  Millman J.,  eds, Proceedings of the
  9th Python in Science Conference. pp 51 -- 56

\bibitem[\protect\citeauthoryear{{M{\'e}sz{\'a}ros} \&
  {Rees}}{{M{\'e}sz{\'a}ros} \& {Rees}}{1997}]{Meszaros1997}
{M{\'e}sz{\'a}ros} P.,  {Rees} M.~J.,  1997, \mn@doi [\apj] {10.1086/303625},
  \href {https://ui.adsabs.harvard.edu/abs/1997ApJ...476..232M} {476, 232}

\bibitem[\protect\citeauthoryear{{M{\'e}sz{\'a}ros}, {Fox}, {Hanna}  \&
  {Murase}}{{M{\'e}sz{\'a}ros} et~al.}{2019}]{Meszaros2019}
{M{\'e}sz{\'a}ros} P.,  {Fox} D.~B.,  {Hanna} C.,   {Murase} K.,  2019, \mn@doi
  [Nature Reviews Physics] {10.1038/s42254-019-0101-z}, \href
  {https://ui.adsabs.harvard.edu/abs/2019NatRP...1..585M} {1, 585}

\bibitem[\protect\citeauthoryear{{Metcalfe}, {Shanks}, {Weilbacher},
  {McCracken}, {Fong}  \& {Thompson}}{{Metcalfe} et~al.}{2006}]{Metcalf2006}
{Metcalfe} N.,  {Shanks} T.,  {Weilbacher} P.~M.,  {McCracken} H.~J.,  {Fong}
  R.,   {Thompson} D.,  2006, \mn@doi [\mnras]
  {10.1111/j.1365-2966.2006.10534.x}, \href
  {https://ui.adsabs.harvard.edu/abs/2006MNRAS.370.1257M} {370, 1257}

\bibitem[\protect\citeauthoryear{{Miyamoto} \& {Nagai}}{{Miyamoto} \&
  {Nagai}}{1975}]{MiyamotoNagai95}
{Miyamoto} M.,  {Nagai} R.,  1975, \pasj, \href
  {https://ui.adsabs.harvard.edu/abs/1975PASJ...27..533M} {27, 533}

\bibitem[\protect\citeauthoryear{{Mogushi}, {Cavagli{\`a}}  \&
  {Siellez}}{{Mogushi} et~al.}{2019}]{Mogushi2019}
{Mogushi} K.,  {Cavagli{\`a}} M.,   {Siellez} K.,  2019, \mn@doi [\apj]
  {10.3847/1538-4357/ab1f76}, \href
  {https://ui.adsabs.harvard.edu/abs/2019ApJ...880...55M} {880, 55}

\bibitem[\protect\citeauthoryear{{Mooley} et~al.,}{{Mooley}
  et~al.}{2018}]{Mooley2018}
{Mooley} K.~P.,  et~al., 2018, \mn@doi [\nat] {10.1038/s41586-018-0486-3},
  \href {https://ui.adsabs.harvard.edu/abs/2018Natur.561..355M} {561, 355}

\bibitem[\protect\citeauthoryear{{Morgan} et~al.,}{{Morgan}
  et~al.}{2020}]{DES2020GW190814}
{Morgan} R.,  et~al., 2020, \mn@doi [\apj] {10.3847/1538-4357/abafaa}, \href
  {https://ui.adsabs.harvard.edu/abs/2020ApJ...901...83M} {901, 83}

\bibitem[\protect\citeauthoryear{{Murase} \& {Bartos}}{{Murase} \&
  {Bartos}}{2019}]{Murase2019}
{Murase} K.,  {Bartos} I.,  2019, \mn@doi [Annual Review of Nuclear and
  Particle Science] {10.1146/annurev-nucl-101918-023510}, \href
  {https://ui.adsabs.harvard.edu/abs/2019ARNPS..69..477M} {69, 477}

\bibitem[\protect\citeauthoryear{{Nakar}}{{Nakar}}{2007}]{Nakar2007Review}
{Nakar} E.,  2007, \mn@doi [\physrep] {10.1016/j.physrep.2007.02.005}, \href
  {https://ui.adsabs.harvard.edu/abs/2007PhR...442..166N} {442, 166}

\bibitem[\protect\citeauthoryear{{Narayan}, {Paczynski}  \& {Piran}}{{Narayan}
  et~al.}{1992}]{narayan1992gamma}
{Narayan} R.,  {Paczynski} B.,   {Piran} T.,  1992, \mn@doi [\apjl]
  {10.1086/186493}, \href
  {https://ui.adsabs.harvard.edu/abs/1992ApJ...395L..83N} {395, L83}

\bibitem[\protect\citeauthoryear{{Narayan}, {Piran}  \& {Kumar}}{{Narayan}
  et~al.}{2001}]{narayan2001accretion}
{Narayan} R.,  {Piran} T.,   {Kumar} P.,  2001, \mn@doi [\apj]
  {10.1086/322267}, \href
  {https://ui.adsabs.harvard.edu/abs/2001ApJ...557..949N} {557, 949}

\bibitem[\protect\citeauthoryear{{Navarro}, {Frenk}  \& {White}}{{Navarro}
  et~al.}{1996}]{Navarro96}
{Navarro} J.~F.,  {Frenk} C.~S.,   {White} S. D.~M.,  1996, \mn@doi [\apj]
  {10.1086/177173}, \href
  {https://ui.adsabs.harvard.edu/abs/1996ApJ...462..563N} {462, 563}

\bibitem[\protect\citeauthoryear{{Nicholl} et~al.,}{{Nicholl}
  et~al.}{2017}]{Nicholl2017}
{Nicholl} M.,  et~al., 2017, \mn@doi [\apjl] {10.3847/2041-8213/aa9029}, \href
  {https://ui.adsabs.harvard.edu/abs/2017ApJ...848L..18N} {848, L18}

\bibitem[\protect\citeauthoryear{{Nynka}, {Ruan}, {Haggard}  \&
  {Evans}}{{Nynka} et~al.}{2018}]{Nynka2018}
{Nynka} M.,  {Ruan} J.~J.,  {Haggard} D.,   {Evans} P.~A.,  2018, \mn@doi
  [\apjl] {10.3847/2041-8213/aad32d}, \href
  {https://ui.adsabs.harvard.edu/abs/2018ApJ...862L..19N} {862, L19}

\bibitem[\protect\citeauthoryear{{O'Connor}, {Beniamini}  \&
  {Kouveliotou}}{{O'Connor} et~al.}{2020}]{oconnor2020constraints}
{O'Connor} B.,  {Beniamini} P.,   {Kouveliotou} C.,  2020, \mn@doi [\mnras]
  {10.1093/mnras/staa1433}, \href
  {https://ui.adsabs.harvard.edu/abs/2020MNRAS.495.4782O} {495, 4782}

\bibitem[\protect\citeauthoryear{{O'Connor} et~al.,}{{O'Connor}
  et~al.}{2021}]{Oconnor21}
{O'Connor} B.,  et~al., 2021, \mn@doi [\mnras] {10.1093/mnras/stab132}, \href
  {https://ui.adsabs.harvard.edu/abs/2021MNRAS.502.1279O} {502, 1279}

\bibitem[\protect\citeauthoryear{Oliphant}{Oliphant}{2006}]{oliphant2006guide}
Oliphant T.~E.,  2006, A guide to NumPy.
 Vol. 1, Trelgol Publishing USA

\bibitem[\protect\citeauthoryear{{Palmese} et~al.,}{{Palmese}
  et~al.}{2017}]{Palmese_2017}
{Palmese} A.,  et~al., 2017, \mn@doi [\apjl] {10.3847/2041-8213/aa9660}, \href
  {https://ui.adsabs.harvard.edu/abs/2017ApJ...849L..34P} {849, L34}

\bibitem[\protect\citeauthoryear{{Paterson} et~al.,}{{Paterson}
  et~al.}{2020}]{Paterson2020}
{Paterson} K.,  et~al., 2020, \mn@doi [\apjl] {10.3847/2041-8213/aba4b0}, \href
  {https://ui.adsabs.harvard.edu/abs/2020ApJ...898L..32P} {898, L32}

\bibitem[\protect\citeauthoryear{{Paul}}{{Paul}}{2018}]{paul2018}
{Paul} D.,  2018, \mn@doi [\mnras] {10.1093/mnras/sty840}, \href
  {https://ui.adsabs.harvard.edu/abs/2018MNRAS.477.4275P} {477, 4275}

\bibitem[\protect\citeauthoryear{{Perego}, {Radice}  \& {Bernuzzi}}{{Perego}
  et~al.}{2017}]{Perego_2017}
{Perego} A.,  {Radice} D.,   {Bernuzzi} S.,  2017, \mn@doi [\apjl]
  {10.3847/2041-8213/aa9ab9}, \href
  {https://ui.adsabs.harvard.edu/abs/2017ApJ...850L..37P} {850, L37}

\bibitem[\protect\citeauthoryear{{Perley}, {Bloom}, {Modjaz}, {Poznanski}  \&
  {Thoene}}{{Perley} et~al.}{2007}]{Perley2007GCN}
{Perley} D.~A.,  {Bloom} J.~S.,  {Modjaz} M.,  {Poznanski} D.,   {Thoene}
  C.~C.,  2007, GRB Coordinates Network, \href
  {https://ui.adsabs.harvard.edu/abs/2007GCN..7140....1P} {7140, 1}

\bibitem[\protect\citeauthoryear{{Perna} \& {Belczynski}}{{Perna} \&
  {Belczynski}}{2002}]{Perna_2002}
{Perna} R.,  {Belczynski} K.,  2002, \mn@doi [\apj] {10.1086/339571}, \href
  {https://ui.adsabs.harvard.edu/abs/2002ApJ...570..252P} {570, 252}

\bibitem[\protect\citeauthoryear{{Pian} et~al.,}{{Pian}
  et~al.}{2017}]{Pian2017}
{Pian} E.,  et~al., 2017, \mn@doi [\nat] {10.1038/nature24298}, \href
  {https://ui.adsabs.harvard.edu/abs/2017Natur.551...67P} {551, 67}

\bibitem[\protect\citeauthoryear{{Piranomonte} et~al.,}{{Piranomonte}
  et~al.}{2008}]{Piramonte2008}
{Piranomonte} S.,  et~al., 2008, \mn@doi [\aap] {10.1051/0004-6361:200810547},
  \href {https://ui.adsabs.harvard.edu/abs/2008A&A...491..183P} {491, 183}

\bibitem[\protect\citeauthoryear{{Piro} et~al.,}{{Piro}
  et~al.}{2019}]{Piro2019}
{Piro} L.,  et~al., 2019, \mn@doi [\mnras] {10.1093/mnras/sty3047}, \href
  {https://ui.adsabs.harvard.edu/abs/2019MNRAS.483.1912P} {483, 1912}

\bibitem[\protect\citeauthoryear{{Planck Collaboration} et~al.,}{{Planck
  Collaboration} et~al.}{2014}]{ade2014planck}
{Planck Collaboration} et~al., 2014, \mn@doi [\aap]
  {10.1051/0004-6361/201321591}, \href
  {https://ui.adsabs.harvard.edu/abs/2014A&A...571A..16P} {571, A16}

\bibitem[\protect\citeauthoryear{{Price-Whelan} et~al.,}{{Price-Whelan}
  et~al.}{2018}]{astropy:2018}
{Price-Whelan} A.~M.,  et~al., 2018, \mn@doi [\aj] {10.3847/1538-3881/aabc4f},
  \href {https://ui.adsabs.harvard.edu/#abs/2018AJ....156..123T} {156, 123}

\bibitem[\protect\citeauthoryear{{Prochaska}, {Bloom}, {Chen}, {Hansen},
  {Kalirai}, {Rich}  \& {Richer}}{{Prochaska} et~al.}{2005}]{Prochaska2005GCN}
{Prochaska} J.~X.,  {Bloom} J.~S.,  {Chen} H.~W.,  {Hansen} B.,  {Kalirai} J.,
  {Rich} M.,   {Richer} H.,  2005, GRB Coordinates Network, \href
  {https://ui.adsabs.harvard.edu/abs/2005GCN..3700....1P} {3700, 1}

\bibitem[\protect\citeauthoryear{{Qu} et~al.,}{{Qu} et~al.}{2017}]{Qu_2016}
{Qu} Y.,  et~al., 2017, \mn@doi [\mnras] {10.1093/mnras/stw2437}, \href
  {https://ui.adsabs.harvard.edu/abs/2017MNRAS.464.1659Q} {464, 1659}

\bibitem[\protect\citeauthoryear{{Racusin}, {Burrows}  \& {Gehrels}}{{Racusin}
  et~al.}{2006}]{Racusin2006GCN}
{Racusin} J.~L.,  {Burrows} D.~N.,   {Gehrels} N.,  2006, GRB Coordinates
  Network, \href {https://ui.adsabs.harvard.edu/abs/2006GCN..5773....1R} {5773,
  1}

\bibitem[\protect\citeauthoryear{{Rau}, {McBreen}  \& {Kruehler}}{{Rau}
  et~al.}{2009}]{Rau2009GCN}
{Rau} A.,  {McBreen} S.,   {Kruehler} T.,  2009, GRB Coordinates Network, \href
  {https://ui.adsabs.harvard.edu/abs/2009GCN..9353....1R} {9353, 1}

\bibitem[\protect\citeauthoryear{{Resmi} et~al.,}{{Resmi}
  et~al.}{2018}]{Resmi2018}
{Resmi} L.,  et~al., 2018, \mn@doi [\apj] {10.3847/1538-4357/aae1a6}, \href
  {https://ui.adsabs.harvard.edu/abs/2018ApJ...867...57R} {867, 57}

\bibitem[\protect\citeauthoryear{{Rodriguez} \& {Loeb}}{{Rodriguez} \&
  {Loeb}}{2018}]{Rodriguez_2018}
{Rodriguez} C.~L.,  {Loeb} A.,  2018, \mn@doi [\apjl]
  {10.3847/2041-8213/aae377}, \href
  {https://ui.adsabs.harvard.edu/abs/2018ApJ...866L...5R} {866, L5}

\bibitem[\protect\citeauthoryear{{Rodriguez}, {Morscher}, {Pattabiraman},
  {Chatterjee}, {Haster}  \& {Rasio}}{{Rodriguez}
  et~al.}{2015}]{Rodriguez_2015}
{Rodriguez} C.~L.,  {Morscher} M.,  {Pattabiraman} B.,  {Chatterjee} S.,
  {Haster} C.-J.,   {Rasio} F.~A.,  2015, \mn@doi [\prl]
  {10.1103/PhysRevLett.115.051101}, \href
  {https://ui.adsabs.harvard.edu/abs/2015PhRvL.115e1101R} {115, 051101}

\bibitem[\protect\citeauthoryear{{Rodriguez}, {Chatterjee}  \&
  {Rasio}}{{Rodriguez} et~al.}{2016}]{Rodriguez_2016}
{Rodriguez} C.~L.,  {Chatterjee} S.,   {Rasio} F.~A.,  2016, \mn@doi [\prd]
  {10.1103/PhysRevD.93.084029}, \href
  {https://ui.adsabs.harvard.edu/abs/2016PhRvD..93h4029R} {93, 084029}

\bibitem[\protect\citeauthoryear{{Rothberg}, {Kuhn}, {Veillet}  \&
  {Allanson}}{{Rothberg} et~al.}{2020}]{Rothberg2020}
{Rothberg} B.,  {Kuhn} O.,  {Veillet} C.,   {Allanson} S.,  2020, GRB
  Coordinates Network, \href
  {https://ui.adsabs.harvard.edu/abs/2020GCN.28319....1R} {28319, 1}

\bibitem[\protect\citeauthoryear{{Rowlinson} et~al.,}{{Rowlinson}
  et~al.}{2010}]{rowlinson2010discovery}
{Rowlinson} A.,  et~al., 2010, \mn@doi [\mnras]
  {10.1111/j.1365-2966.2010.17115.x}, \href
  {https://ui.adsabs.harvard.edu/abs/2010MNRAS.408..383R} {408, 383}

\bibitem[\protect\citeauthoryear{{Ruiz}, {Paschalidis}, {Tsokaros}  \&
  {Shapiro}}{{Ruiz} et~al.}{2020}]{Ruiz2020}
{Ruiz} M.,  {Paschalidis} V.,  {Tsokaros} A.,   {Shapiro} S.~L.,  2020, \mn@doi
  [\prd] {10.1103/PhysRevD.102.124077}, \href
  {https://ui.adsabs.harvard.edu/abs/2020PhRvD.102l4077R} {102, 124077}

\bibitem[\protect\citeauthoryear{{Samsing} \& {D'Orazio}}{{Samsing} \&
  {D'Orazio}}{2018}]{Samsing_2018}
{Samsing} J.,  {D'Orazio} D.~J.,  2018, \mn@doi [\mnras]
  {10.1093/mnras/sty2334}, \href
  {https://ui.adsabs.harvard.edu/abs/2018MNRAS.481.5445S} {481, 5445}

\bibitem[\protect\citeauthoryear{{Santoliquido}, {Mapelli}, {Bouffanais},
  {Giacobbo}, {Di Carlo}, {Rastello}, {Artale}  \& {Ballone}}{{Santoliquido}
  et~al.}{2020}]{Santoliquido2020}
{Santoliquido} F.,  {Mapelli} M.,  {Bouffanais} Y.,  {Giacobbo} N.,  {Di Carlo}
  U.~N.,  {Rastello} S.,  {Artale} M.~C.,   {Ballone} A.,  2020, arXiv
  e-prints, \href {https://ui.adsabs.harvard.edu/abs/2020arXiv200409533S} {p.
  arXiv:2004.09533}

\bibitem[\protect\citeauthoryear{{Sari}, {Piran}  \& {Narayan}}{{Sari}
  et~al.}{1998}]{Sari1998}
{Sari} R.,  {Piran} T.,   {Narayan} R.,  1998, \mn@doi [\apjl]
  {10.1086/311269}, \href
  {https://ui.adsabs.harvard.edu/abs/1998ApJ...497L..17S} {497, L17}

\bibitem[\protect\citeauthoryear{{Schaye} et~al.,}{{Schaye}
  et~al.}{2015}]{Schaye_2014}
{Schaye} J.,  et~al., 2015, \mn@doi [\mnras] {10.1093/mnras/stu2058}, \href
  {https://ui.adsabs.harvard.edu/abs/2015MNRAS.446..521S} {446, 521}

\bibitem[\protect\citeauthoryear{{Selsing} et~al.,}{{Selsing}
  et~al.}{2016}]{Selsing2016GCN}
{Selsing} J.,  et~al., 2016, GRB Coordinates Network, \href
  {https://ui.adsabs.harvard.edu/abs/2016GCN.19274....1S} {19274, 1}

\bibitem[\protect\citeauthoryear{{Selsing} et~al.,}{{Selsing}
  et~al.}{2018}]{selsing18}
{Selsing} J.,  et~al., 2018, \mn@doi [\aap] {10.1051/0004-6361/201731475},
  \href {https://ui.adsabs.harvard.edu/abs/2018A&A...616A..48S} {616, A48}

\bibitem[\protect\citeauthoryear{{Selsing} et~al.,}{{Selsing}
  et~al.}{2019}]{Selsing19}
{Selsing} J.,  et~al., 2019, \mn@doi [\aap] {10.1051/0004-6361/201832835},
  \href {https://ui.adsabs.harvard.edu/abs/2019A&A...623A..92S} {623, A92}

\bibitem[\protect\citeauthoryear{{Shappee} et~al.,}{{Shappee}
  et~al.}{2017}]{Shappee2017}
{Shappee} B.~J.,  et~al., 2017, \mn@doi [Science] {10.1126/science.aaq0186},
  \href {https://ui.adsabs.harvard.edu/abs/2017Sci...358.1574S} {358, 1574}

\bibitem[\protect\citeauthoryear{{Shen}, {Cooke}, {Ramirez-Ruiz}, {Madau},
  {Mayer}  \& {Guedes}}{{Shen} et~al.}{2015}]{Shen2015}
{Shen} S.,  {Cooke} R.~J.,  {Ramirez-Ruiz} E.,  {Madau} P.,  {Mayer} L.,
  {Guedes} J.,  2015, \mn@doi [\apj] {10.1088/0004-637X/807/2/115}, \href
  {https://ui.adsabs.harvard.edu/abs/2015ApJ...807..115S} {807, 115}

\bibitem[\protect\citeauthoryear{{Shibata} \& {Taniguchi}}{{Shibata} \&
  {Taniguchi}}{2011}]{Shibata_2011}
{Shibata} M.,  {Taniguchi} K.,  2011, \mn@doi [Living Reviews in Relativity]
  {10.12942/lrr-2011-6}, \href
  {https://ui.adsabs.harvard.edu/abs/2011LRR....14....6S} {14, 6}

\bibitem[\protect\citeauthoryear{{Smartt} et~al.,}{{Smartt}
  et~al.}{2017}]{Smartt2017}
{Smartt} S.~J.,  et~al., 2017, \mn@doi [\nat] {10.1038/nature24303}, \href
  {https://ui.adsabs.harvard.edu/abs/2017Natur.551...75S} {551, 75}

\bibitem[\protect\citeauthoryear{{Soares-Santos} et~al.,}{{Soares-Santos}
  et~al.}{2017}]{Soares-Santos2017}
{Soares-Santos} M.,  et~al., 2017, \mn@doi [\apjl] {10.3847/2041-8213/aa9059},
  \href {https://ui.adsabs.harvard.edu/abs/2017ApJ...848L..16S} {848, L16}

\bibitem[\protect\citeauthoryear{{Soderberg} et~al.,}{{Soderberg}
  et~al.}{2006}]{Soderberg2006}
{Soderberg} A.~M.,  et~al., 2006, \mn@doi [\apj] {10.1086/506429}, \href
  {https://ui.adsabs.harvard.edu/abs/2006ApJ...650..261S} {650, 261}

\bibitem[\protect\citeauthoryear{{Stanway} \& {Eldridge}}{{Stanway} \&
  {Eldridge}}{2018}]{StanwayEldridgeBPASS2017}
{Stanway} E.~R.,  {Eldridge} J.~J.,  2018, \mn@doi [\mnras]
  {10.1093/mnras/sty1353}, \href
  {https://ui.adsabs.harvard.edu/abs/2018MNRAS.479...75S} {479, 75}

\bibitem[\protect\citeauthoryear{{Sun}, {Zhang}  \& {Li}}{{Sun}
  et~al.}{2015}]{sun2015}
{Sun} H.,  {Zhang} B.,   {Li} Z.,  2015, \mn@doi [\apj]
  {10.1088/0004-637X/812/1/33}, \href
  {https://ui.adsabs.harvard.edu/abs/2015ApJ...812...33S} {812, 33}

\bibitem[\protect\citeauthoryear{{Tan} \& {Yu}}{{Tan} \& {Yu}}{2020}]{Tan2020}
{Tan} W.-W.,  {Yu} Y.-W.,  2020, arXiv e-prints, \href
  {https://ui.adsabs.harvard.edu/abs/2020arXiv200602060T} {p. arXiv:2006.02060}

\bibitem[\protect\citeauthoryear{{Tanaka}, {Hotokezaka}, {Kyutoku}, {Wanajo},
  {Kiuchi}, {Sekiguchi}  \& {Shibata}}{{Tanaka} et~al.}{2014}]{Tanaka2014}
{Tanaka} M.,  {Hotokezaka} K.,  {Kyutoku} K.,  {Wanajo} S.,  {Kiuchi} K.,
  {Sekiguchi} Y.,   {Shibata} M.,  2014, \mn@doi [\apj]
  {10.1088/0004-637X/780/1/31}, \href
  {https://ui.adsabs.harvard.edu/abs/2014ApJ...780...31T} {780, 31}

\bibitem[\protect\citeauthoryear{{Tanaka} et~al.,}{{Tanaka}
  et~al.}{2017}]{Tanaka2017}
{Tanaka} M.,  et~al., 2017, \mn@doi [\pasj] {10.1093/pasj/psx121}, \href
  {https://ui.adsabs.harvard.edu/abs/2017PASJ...69..102T} {69, 102}

\bibitem[\protect\citeauthoryear{{Tanvir} et~al.,}{{Tanvir}
  et~al.}{2010}]{Tanvir2010}
{Tanvir} N.~R.,  et~al., 2010, GRB Coordinates Network, 11123, 1

\bibitem[\protect\citeauthoryear{{Tanvir}, {Levan}, {Fruchter}, {Hjorth},
  {Hounsell}, {Wiersema}  \& {Tunnicliffe}}{{Tanvir} et~al.}{2013}]{Tanvir2013}
{Tanvir} N.~R.,  {Levan} A.~J.,  {Fruchter} A.~S.,  {Hjorth} J.,  {Hounsell}
  R.~A.,  {Wiersema} K.,   {Tunnicliffe} R.~L.,  2013, \mn@doi [\nat]
  {10.1038/nature12505}, \href
  {https://ui.adsabs.harvard.edu/abs/2013Natur.500..547T} {500, 547}

\bibitem[\protect\citeauthoryear{{Tanvir} et~al.,}{{Tanvir}
  et~al.}{2017}]{Tanvir2017}
{Tanvir} N.~R.,  et~al., 2017, \mn@doi [\apjl] {10.3847/2041-8213/aa90b6},
  \href {https://ui.adsabs.harvard.edu/abs/2017ApJ...848L..27T} {848, L27}

\bibitem[\protect\citeauthoryear{{Thob} et~al.,}{{Thob}
  et~al.}{2019}]{Thob2019EAGLE}
{Thob} A. C.~R.,  et~al., 2019, \mn@doi [\mnras] {10.1093/mnras/stz448}, \href
  {https://ui.adsabs.harvard.edu/abs/2019MNRAS.485..972T} {485, 972}

\bibitem[\protect\citeauthoryear{{Thone}, {de Ugarte Postigo}, {Gorosabel},
  {Tanvir}  \& {Fynbo}}{{Thone} et~al.}{2013}]{Thone2013GCN}
{Thone} C.~C.,  {de Ugarte Postigo} A.,  {Gorosabel} J.,  {Tanvir} N.,
  {Fynbo} J.~P.~U.,  2013, GRB Coordinates Network, \href
  {https://ui.adsabs.harvard.edu/abs/2013GCN.14744....1T} {14744, 1}

\bibitem[\protect\citeauthoryear{{Troja} et~al.,}{{Troja}
  et~al.}{2016}]{Troja2016}
{Troja} E.,  et~al., 2016, \mn@doi [\apj] {10.3847/0004-637X/827/2/102}, \href
  {https://ui.adsabs.harvard.edu/abs/2016ApJ...827..102T} {827, 102}

\bibitem[\protect\citeauthoryear{{Troja} et~al.,}{{Troja}
  et~al.}{2017}]{Troja2017}
{Troja} E.,  et~al., 2017, \mn@doi [\nat] {10.1038/nature24290}, \href
  {https://ui.adsabs.harvard.edu/abs/2017Natur.551...71T} {551, 71}

\bibitem[\protect\citeauthoryear{{Troja} et~al.,}{{Troja}
  et~al.}{2019a}]{Troja2019}
{Troja} E.,  et~al., 2019a, \mn@doi [\mnras] {10.1093/mnras/stz2248}, \href
  {https://ui.adsabs.harvard.edu/abs/2019MNRAS.489.1919T} {489, 1919}

\bibitem[\protect\citeauthoryear{{Troja} et~al.,}{{Troja}
  et~al.}{2019b}]{Troja2019GRB160821B}
{Troja} E.,  et~al., 2019b, \mn@doi [\mnras] {10.1093/mnras/stz2255}, \href
  {https://ui.adsabs.harvard.edu/abs/2019MNRAS.489.2104T} {489, 2104}

\bibitem[\protect\citeauthoryear{{Troja} et~al.,}{{Troja}
  et~al.}{2020}]{Troja2020}
{Troja} E.,  et~al., 2020, \mn@doi [\mnras] {10.1093/mnras/staa2626}, \href
  {https://ui.adsabs.harvard.edu/abs/2020MNRAS.498.5643T} {498, 5643}

\bibitem[\protect\citeauthoryear{{Tsutsui}, {Yonetoku}, {Nakamura}, {Takahashi}
   \& {Morihara}}{{Tsutsui} et~al.}{2013}]{Tsutsui2013}
{Tsutsui} R.,  {Yonetoku} D.,  {Nakamura} T.,  {Takahashi} K.,   {Morihara} Y.,
   2013, \mn@doi [\mnras] {10.1093/mnras/stt262}, \href
  {https://ui.adsabs.harvard.edu/abs/2013MNRAS.431.1398T} {431, 1398}

\bibitem[\protect\citeauthoryear{{Tunnicliffe} et~al.,}{{Tunnicliffe}
  et~al.}{2014}]{Tunnicliffe_2014}
{Tunnicliffe} R.~L.,  et~al., 2014, \mn@doi [\mnras] {10.1093/mnras/stt1975},
  \href {https://ui.adsabs.harvard.edu/abs/2014MNRAS.437.1495T} {437, 1495}

\bibitem[\protect\citeauthoryear{{Villar} et~al.,}{{Villar}
  et~al.}{2017}]{Villar2017}
{Villar} V.~A.,  et~al., 2017, \mn@doi [\apjl] {10.3847/2041-8213/aa9c84},
  \href {https://ui.adsabs.harvard.edu/abs/2017ApJ...851L..21V} {851, L21}

\bibitem[\protect\citeauthoryear{{Virtanen} et~al.,}{{Virtanen}
  et~al.}{2020}]{2020SciPy-NMeth}
{Virtanen} P.,  et~al., 2020, \mn@doi [Nature Methods]
  {https://doi.org/10.1038/s41592-019-0686-2}, \href {https://rdcu.be/b08Wh}
  {17, 261}

\bibitem[\protect\citeauthoryear{{Wanderman} \& {Piran}}{{Wanderman} \&
  {Piran}}{2015}]{wanderman2015}
{Wanderman} D.,  {Piran} T.,  2015, \mn@doi [\mnras] {10.1093/mnras/stv123},
  \href {https://ui.adsabs.harvard.edu/abs/2015MNRAS.448.3026W} {448, 3026}

\bibitem[\protect\citeauthoryear{{Wang} et~al.,}{{Wang}
  et~al.}{2017}]{Wang2017}
{Wang} H.,  et~al., 2017, \mn@doi [\apjl] {10.3847/2041-8213/aa9e08}, \href
  {https://ui.adsabs.harvard.edu/abs/2017ApJ...851L..18W} {851, L18}

\bibitem[\protect\citeauthoryear{{Ye}, {Fong}, {Kremer}, {Rodriguez},
  {Chatterjee}, {Fragione}  \& {Rasio}}{{Ye} et~al.}{2019}]{Ye_2019}
{Ye} C.~S.,  {Fong} W.-f.,  {Kremer} K.,  {Rodriguez} C.~L.,  {Chatterjee} S.,
  {Fragione} G.,   {Rasio} F.~A.,  2019, arXiv e-prints, \href
  {https://ui.adsabs.harvard.edu/abs/2019arXiv191010740Y} {p. arXiv:1910.10740}

\bibitem[\protect\citeauthoryear{{Yonetoku}, {Nakamura}, {Sawano}, {Takahashi}
  \& {Toyanago}}{{Yonetoku} et~al.}{2014}]{yonetoku2014}
{Yonetoku} D.,  {Nakamura} T.,  {Sawano} T.,  {Takahashi} K.,   {Toyanago} A.,
  2014, \mn@doi [\apj] {10.1088/0004-637X/789/1/65}, \href
  {https://ui.adsabs.harvard.edu/abs/2014ApJ...789...65Y} {789, 65}

\bibitem[\protect\citeauthoryear{{Zevin}, {Kelley}, {Nugent}, {Fong}, {Berry}
  \& {Kalogera}}{{Zevin} et~al.}{2019}]{zevin2019forward}
{Zevin} M.,  {Kelley} L.~Z.,  {Nugent} A.,  {Fong} W.-f.,  {Berry} C. P.~L.,
  {Kalogera} V.,  2019, arXiv e-prints, \href
  {https://ui.adsabs.harvard.edu/abs/2019arXiv191003598Z} {p. arXiv:1910.03598}

\bibitem[\protect\citeauthoryear{{Zhang}, {Fan}, {Dyks}, {Kobayashi},
  {M{\'e}sz{\'a}ros}, {Burrows}, {Nousek}  \& {Gehrels}}{{Zhang}
  et~al.}{2006}]{Zhang2006}
{Zhang} B.,  {Fan} Y.~Z.,  {Dyks} J.,  {Kobayashi} S.,  {M{\'e}sz{\'a}ros} P.,
  {Burrows} D.~N.,  {Nousek} J.~A.,   {Gehrels} N.,  2006, \mn@doi [\apj]
  {10.1086/500723}, \href
  {https://ui.adsabs.harvard.edu/abs/2006ApJ...642..354Z} {642, 354}

\bibitem[\protect\citeauthoryear{{Zhang} et~al.,}{{Zhang}
  et~al.}{2007}]{Zhang2007}
{Zhang} B.,  et~al., 2007, \mn@doi [\apj] {10.1086/510110}, \href
  {https://ui.adsabs.harvard.edu/abs/2007ApJ...655..989Z} {655, 989}

\bibitem[\protect\citeauthoryear{{Ziaeepour} et~al.,}{{Ziaeepour}
  et~al.}{2006}]{Ziaeepour2006GCN}
{Ziaeepour} H.,  et~al., 2006, GRB Coordinates Network, \href
  {https://ui.adsabs.harvard.edu/abs/2006GCN..5948....1Z} {5948, 1}

\bibitem[\protect\citeauthoryear{{de Ugarte Postigo}, {Kann}, {Izzo}, {Thoene},
  {Blazek}, {Agui Fernandez}  \& {Lombardi}}{{de Ugarte Postigo}
  et~al.}{2020}]{deUgartePostigo2020}
{de Ugarte Postigo} A.,  {Kann} D.~A.,  {Izzo} L.,  {Thoene} C.~C.,  {Blazek}
  M.,  {Agui Fernandez} J.~F.,   {Lombardi} G.,  2020, GRB Coordinates Network,
  \href {https://ui.adsabs.harvard.edu/abs/2020GCN.29132....1D} {29132, 1}

\bibitem[\protect\citeauthoryear{{van de Voort}, {Pakmor}, {Grand },
  {Springel}, {G{\'o}mez}  \& {Marinacci}}{{van de Voort}
  et~al.}{2020}]{vandeVoort2020}
{van de Voort} F.,  {Pakmor} R.,  {Grand } R. J.~J.,  {Springel} V.,
  {G{\'o}mez} F.~A.,   {Marinacci} F.,  2020, \mn@doi [\mnras]
  {10.1093/mnras/staa754}, \href
  {https://ui.adsabs.harvard.edu/abs/2020MNRAS.494.4867V} {494, 4867}

\makeatother
\end{thebibliography}

\clearpage
\newpage
\appendix

\section{BPASS/Hobbs Binaries}\label{app:Hobbs}
Here we show the equivalent figures and tables from the main text, for BPASS/Hobbs simulated binaries.

\begin{figure*}

\centering
\vbox{
\hbox{
	\includegraphics[width=0.50\textwidth]{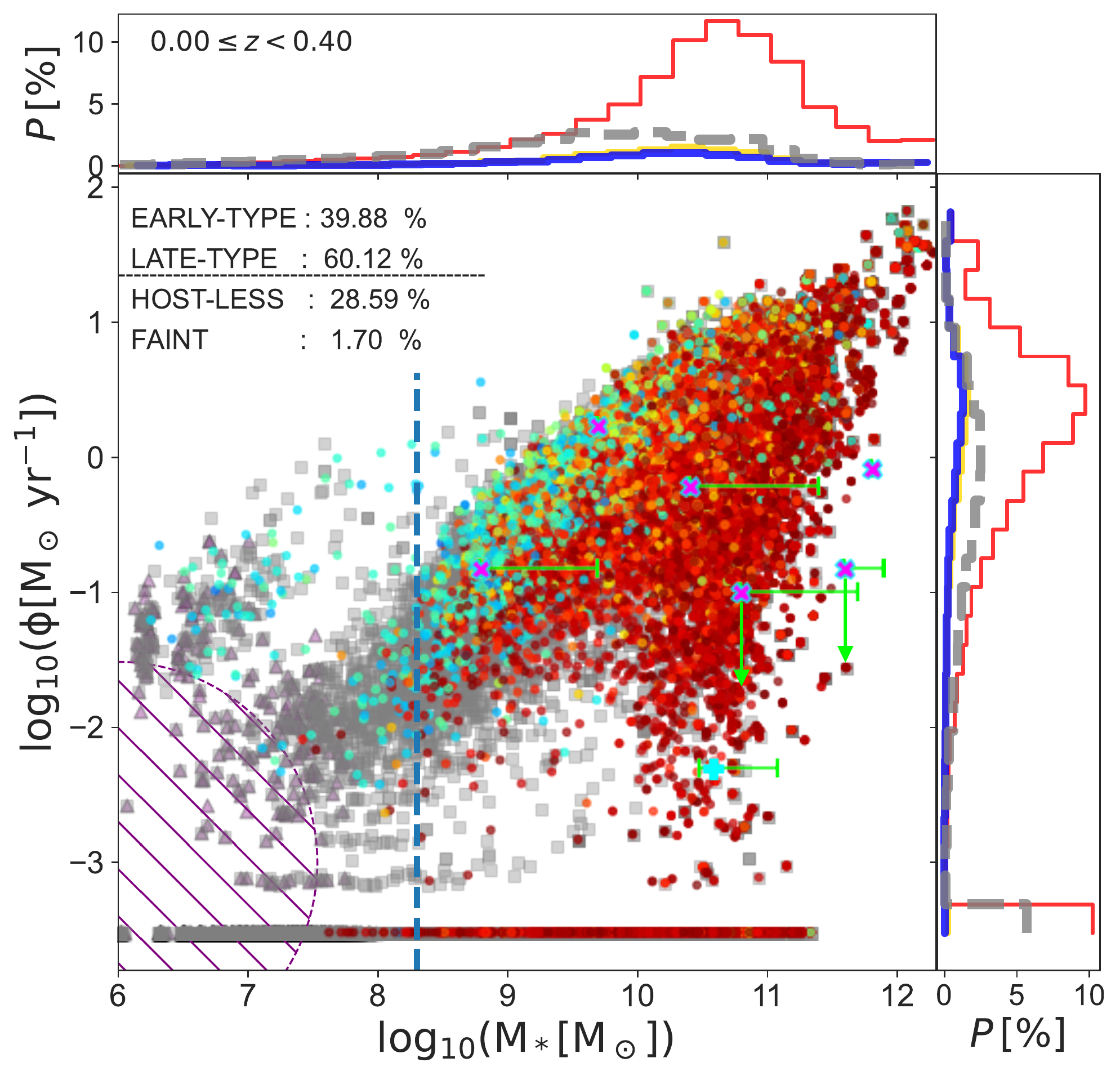}	
	\includegraphics[width=0.50\textwidth]{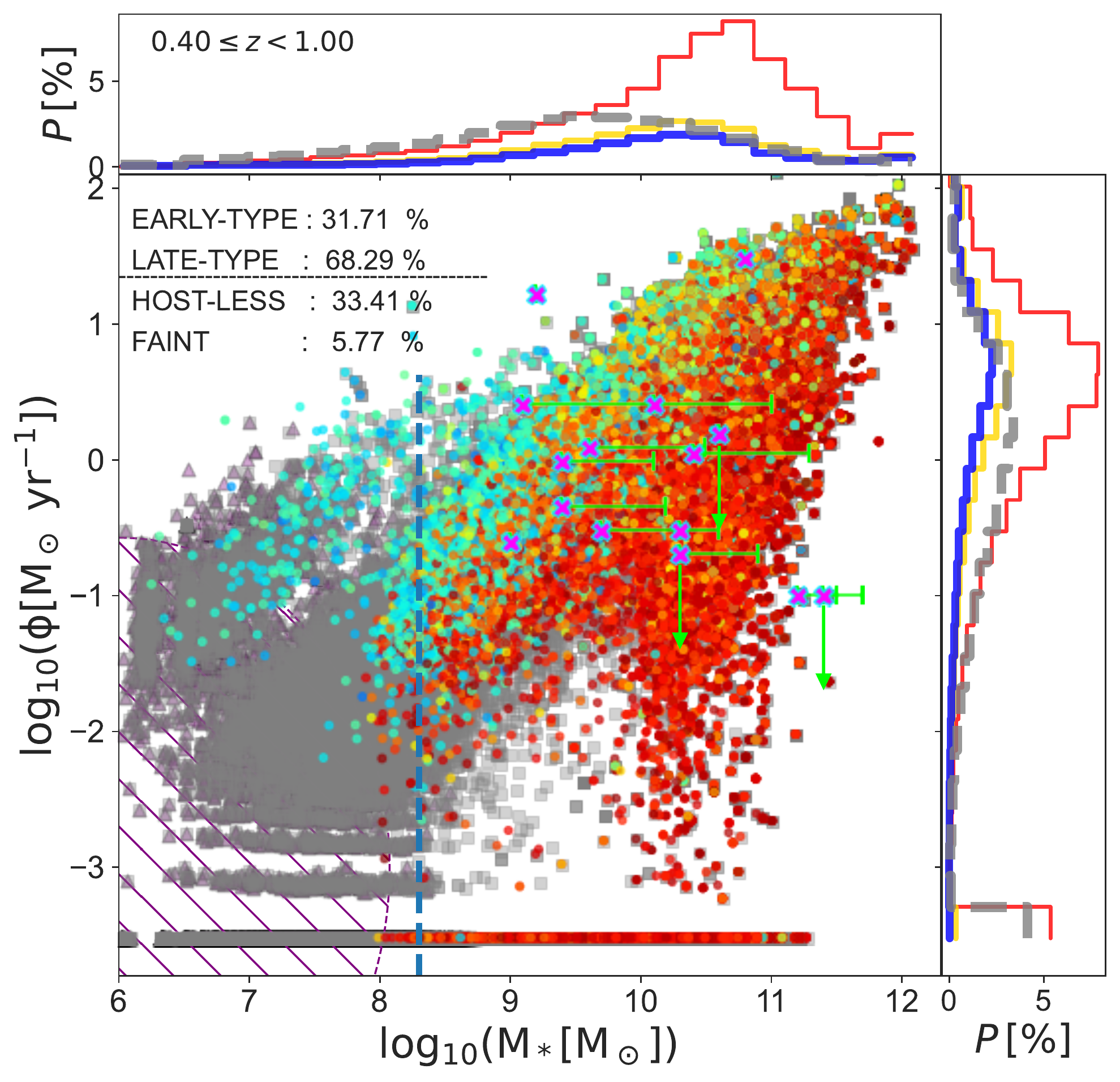}	
		}
\hbox{
	\includegraphics[width=0.50\textwidth]{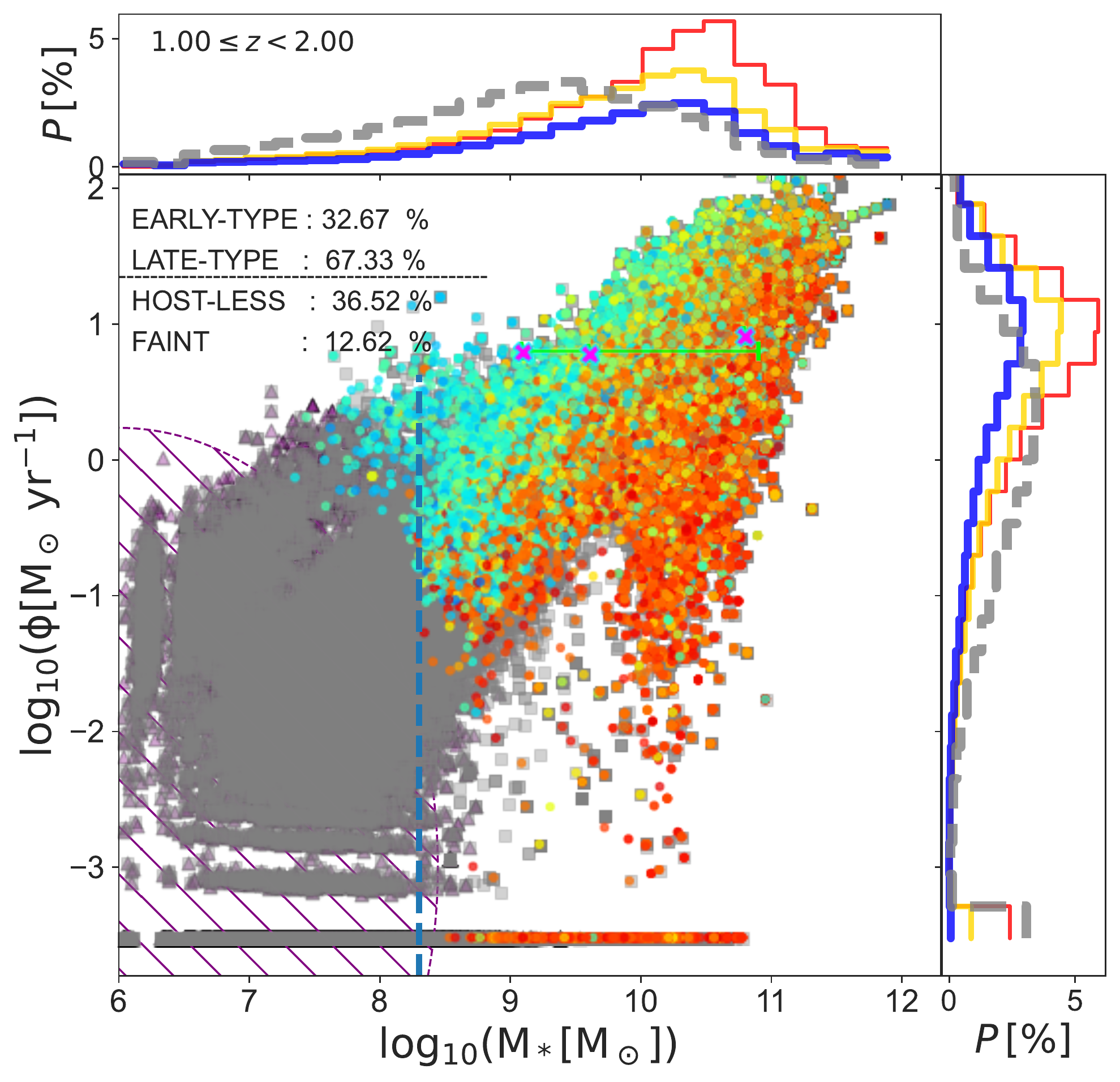}
	\includegraphics[width=0.50\textwidth]{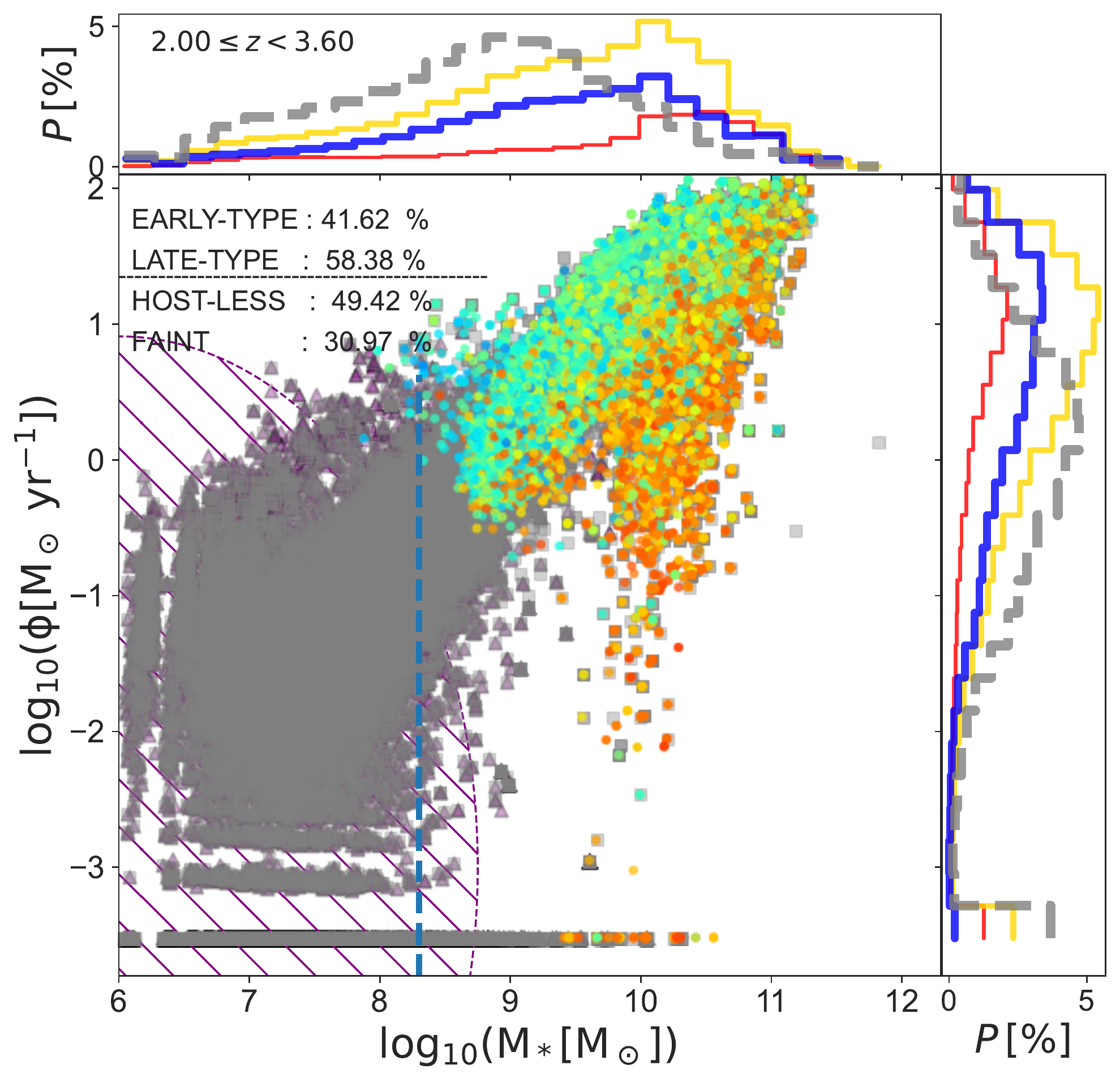}	
		}
\hbox{

\includegraphics[width=0.7\textwidth,trim={0 0 0.2cm 0.4cm},clip]{plots/colorbar.pdf}
\hsize=.2\linewidth
\vbox{
\includegraphics[scale=0.6,trim={0 0.3cm 0.4cm 0.2cm},clip]{plots/m-sfr-legend.pdf}\\
\includegraphics[scale=0.6,trim={0 0.4cm 0.4cm 0.4cm},clip]{plots/main_legend.pdf}
}

}
		}
 \caption{[BPASS/Hobbs - NSNS] Similar to Figure \ref{fig:nsns-ev}.
  }
 \label{fig:hobbs-nsns-ev}
\end{figure*}

\begin{figure*}

\centering
\vbox{
\hbox{
	\includegraphics[width=0.50\textwidth]{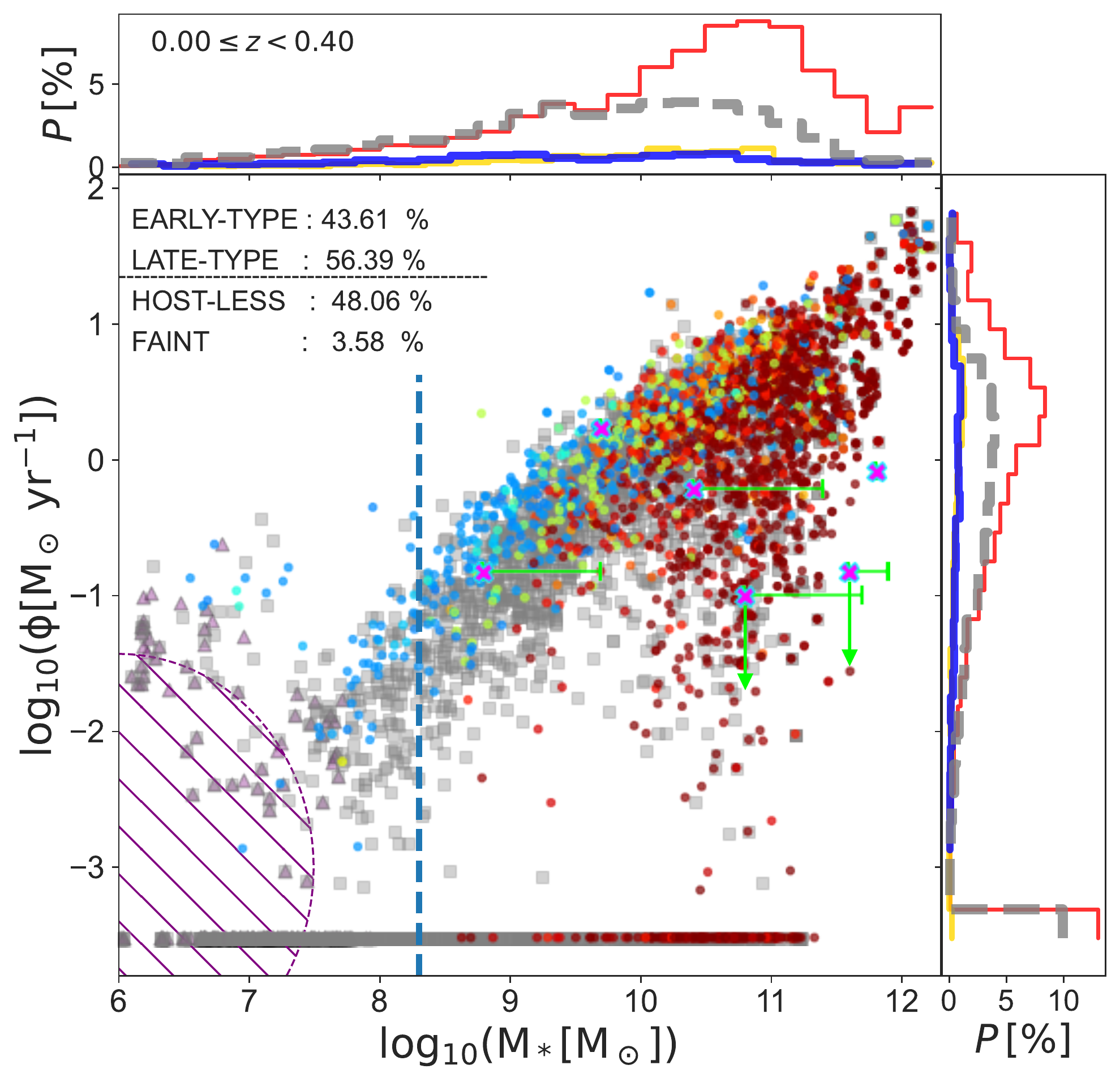}	
	\includegraphics[width=0.50\textwidth]{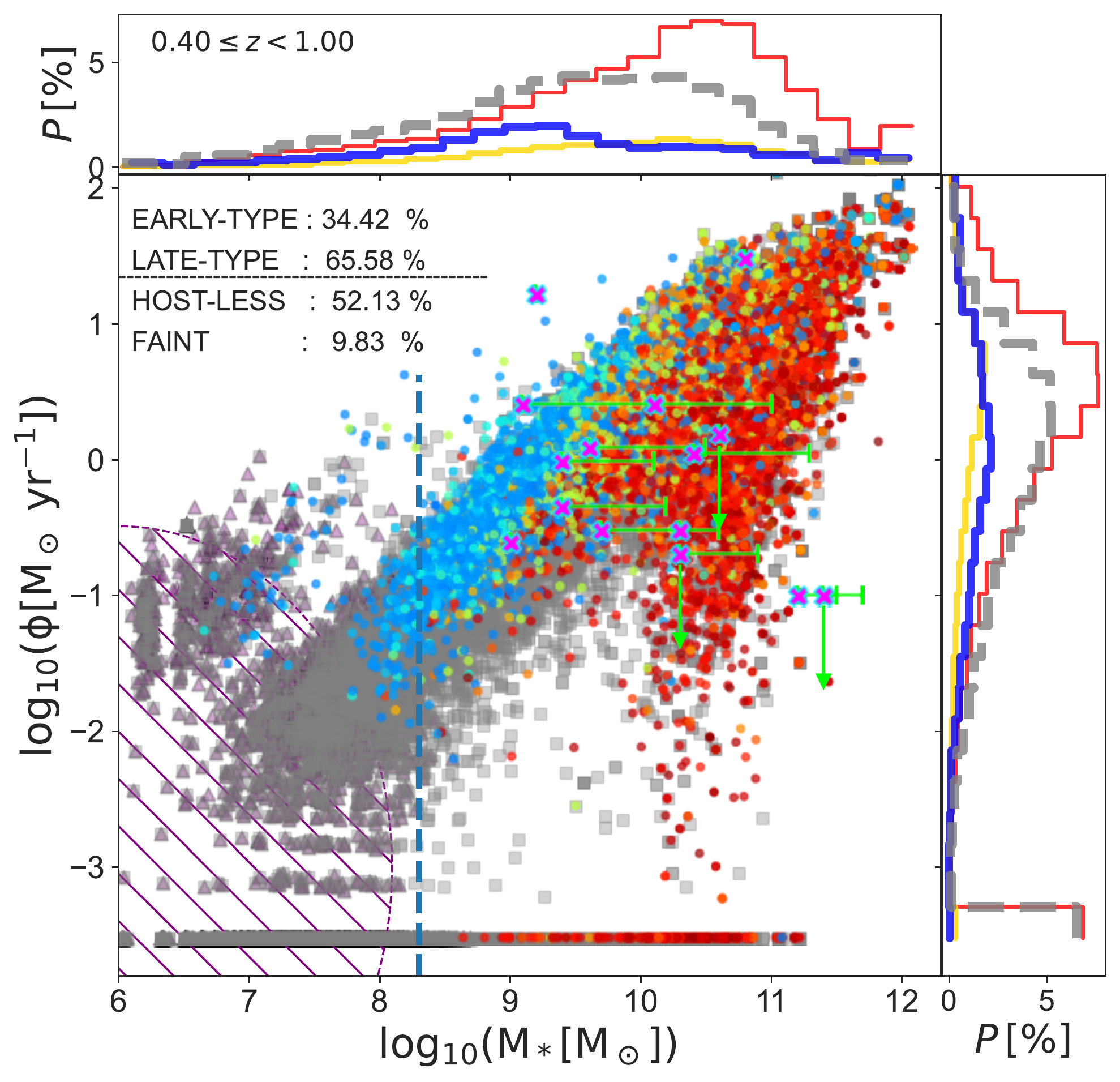}	
		}
\hbox{
	\includegraphics[width=0.50\textwidth]{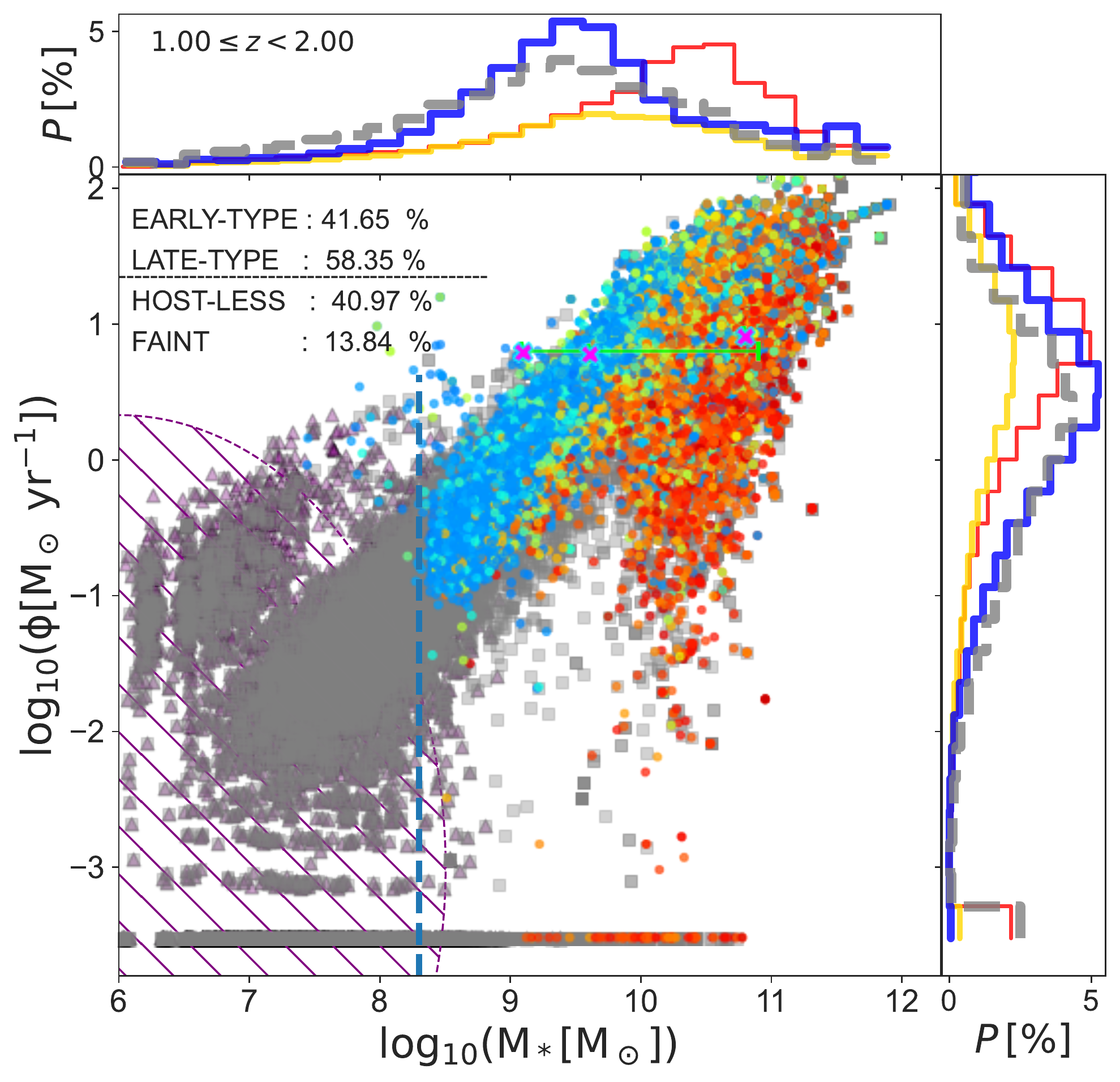}
	\includegraphics[width=0.50\textwidth]{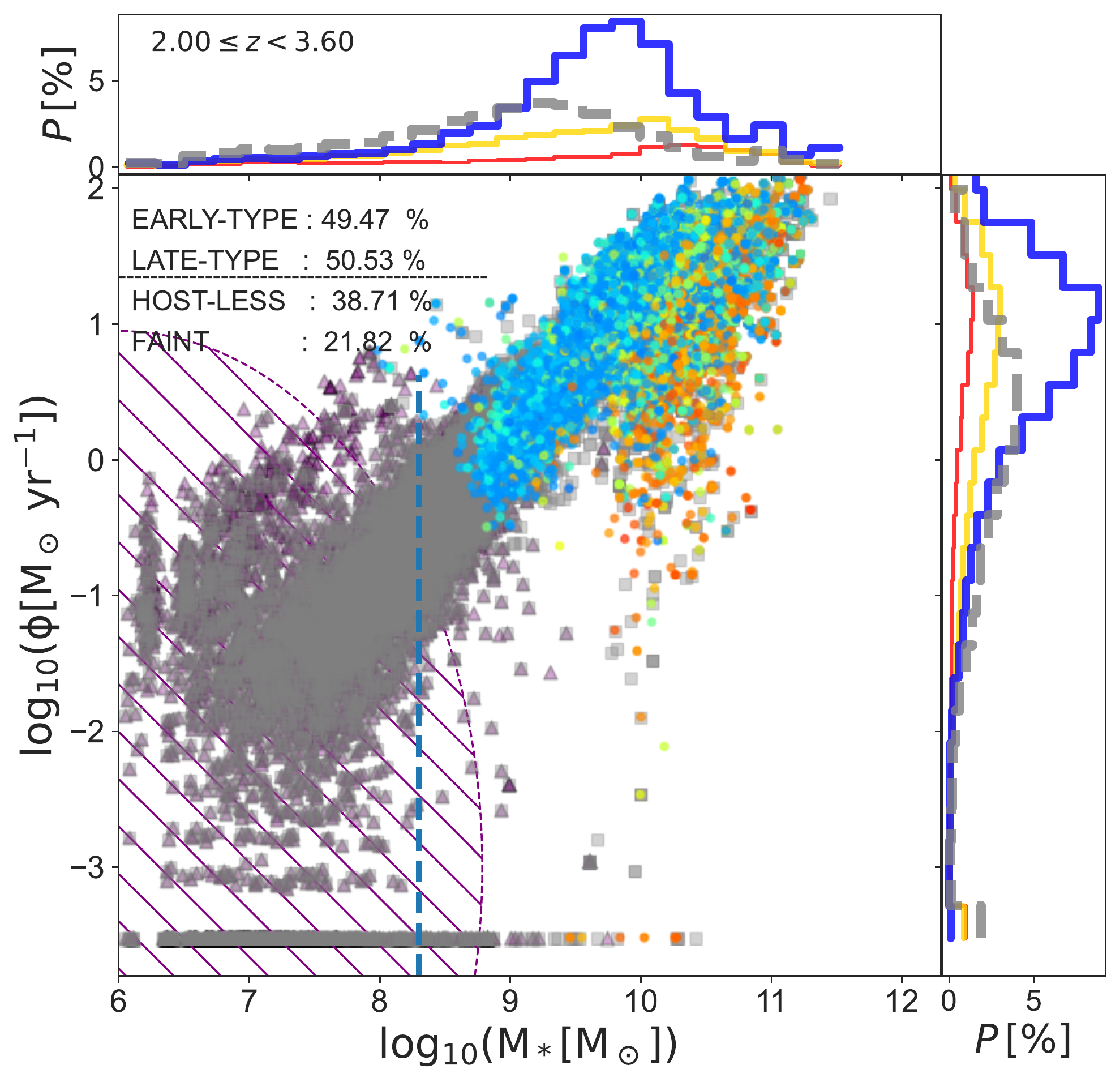}	
		}

\hbox{

\includegraphics[width=0.7\textwidth,trim={0 0 0.2cm 0.4cm},clip]{plots/colorbar.pdf}
\hsize=.2\linewidth
\vbox{
\includegraphics[scale=0.6,trim={0 0.3cm 0.4cm 0.2cm},clip]{plots/m-sfr-legend.pdf}\\
\includegraphics[scale=0.6,trim={0 0.4cm 0.4cm 0.4cm},clip]{plots/main_legend.pdf}
}

}

		}
 \caption{[BPASS/Hobbs - eBHNS] Similar to Figure \ref{fig:bhns-ev}.
  }
  
 \label{fig:hobbs-bhns-ev}
\end{figure*}

\begin{figure*}
\includegraphics[width =  1.\textwidth]{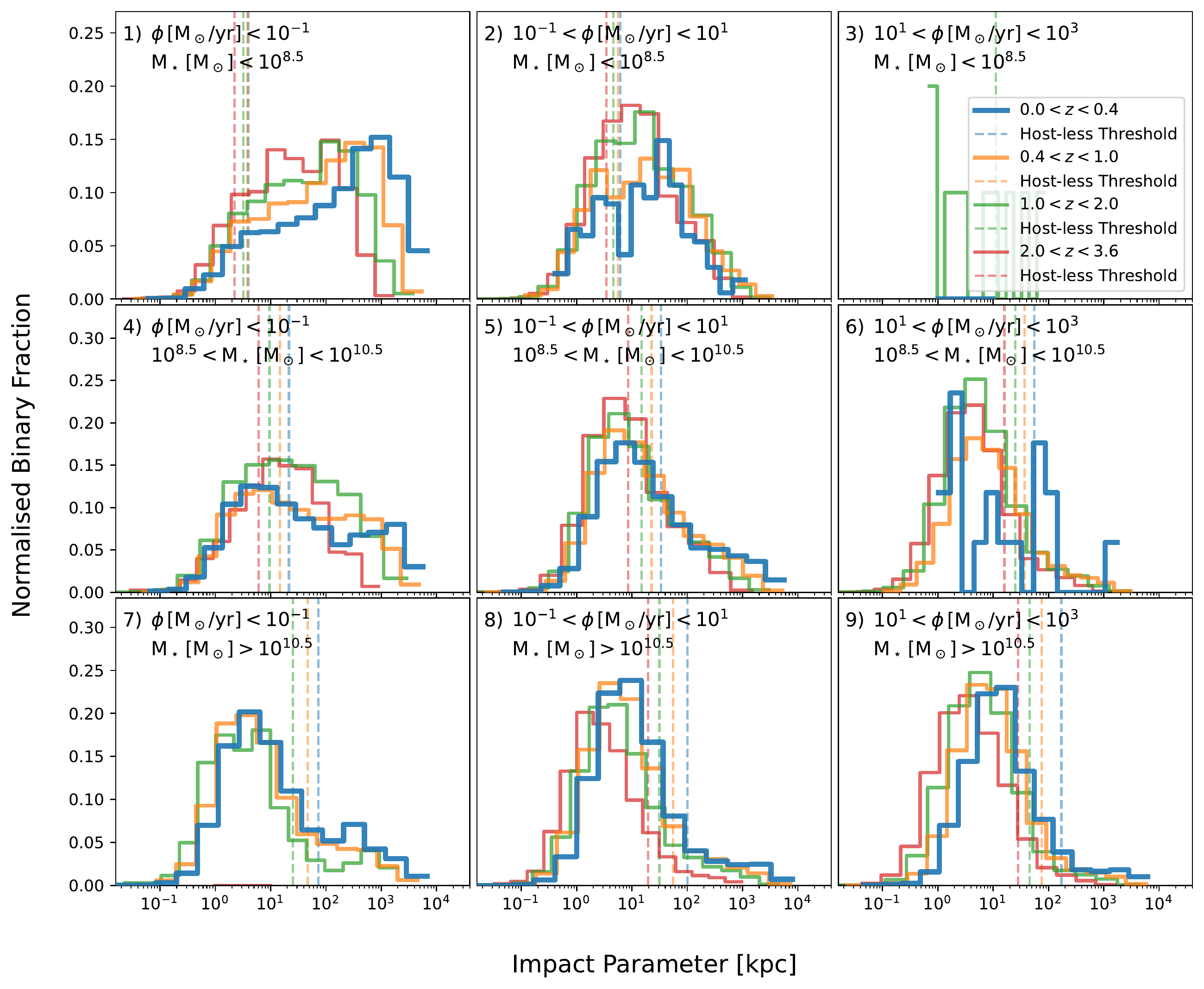}

\caption{[BPASS/Hobbs - NSNS] 
For varied redshift slices, each panel corresponds to a breakdown of the 
impact parameter distributions for different galaxy masses (row), and star formation rates (column).  
Each histogram is composed of the number of binaries with a given impact parameter normalised over the total summation of merging systems within the respective panel and redshift. 
The dashed vertical line corresponds to the galaxy averaged impact parameter corresponding to the $P_{\rm chance}$ limit used to identify systems that may be potentially classed as "host-less". 
A numerical breakdown of each panel is shown in Table \ref{tab:hobbs-nsns-ip-table}. 
}
\label{fig:hobbs-nsns-ip}
\end{figure*}

\begin{table*}

\caption{[BPASS/Hobbs - NSNS] A breakdown of the relative binary mergers occurring within each panel of Figure \ref{fig:hobbs-nsns-ip}. The top division refer to the relative fractions of binaries coalescing for a given redshift slice. The last column indicates the population of the binaries contained within each slice as a percentage of the cumulative population. The second section of the table shows the estimated host-less population per panel [overall fraction corresponding to the redshift slice] based on the criteria defined in Section \ref{ss:proj-dist}. }
\label{tab:hobbs-nsns-ip-table}
\centering
\scalebox{0.87}[0.87]{
\begin{tabular}{|c|c|c|c|c|c|c|c|c|c|c|}
\hline 
\multirow{2}{*}{z} & 1  & 2  & 3  & 4  & 5  & 6  & 7  & 8  & 9  & \multirow{2}{*}{Total Population (\%)}\tabularnewline
\cline{2-10} \cline{3-10} \cline{4-10} \cline{5-10} \cline{6-10} \cline{7-10} \cline{8-10} \cline{9-10} \cline{10-10} 
 & \multicolumn{9}{c|}{Binary Fraction (\%)} & \tabularnewline
\hline 
0.0\textless z\textless 0.4  & 4.9 & 0.3 & 0.0 & 9.0 & 34.8 & 0.0 & 5.2 & 37.4 & 8.3 & 6.9\tabularnewline
0.4\textless z\textless 1.0  & 6.6 & 1.2 & 0.0 & 4.3 & 45.0 & 1.3 & 1.6 & 23.2 & 16.8 & 31.6\tabularnewline
1.0\textless z\textless 2.0  & 7.9 & 3.5 & 0.0 & 1.5 & 47.0 & 10.4 & 0.1 & 7.4 & 22.2 & 43.4\tabularnewline
2.0\textless z\textless 3.6  & 11.6 & 9.5 & 0.0 & 0.6 & 41.1 & 21.9 & 0.0 & 1.7 & 13.5 & 18.1\tabularnewline
\hline 
 & \multicolumn{9}{c|}{Host-less Fraction/Panel {[}Relative to z-slice{]} (\%) } & Total Host-less Fraction (\%)\tabularnewline
\hline 
0.0\textless z\textless 0.4  & 92.2 {[}4.5{]} & 67.3 {[}0.2{]} & 0.0 {[}0.0{]} & 53.6 {[}4.8{]} & 36.1 {[}12.6{]} & 35.3 {[}0.0{]} & 22.8 {[}1.2{]} & 12.6 {[}4.7{]} & 6.2 {[}0.5{]} & 28.6\tabularnewline
0.4\textless z\textless 1.0  & 92.0 {[}6.1{]} & 72.0 {[}0.8{]} & 50.0 {[}0.0{]} & 60.7 {[}2.6{]} & 39.7 {[}17.9{]} & 16.8 {[}0.2{]} & 19.0 {[}0.3{]} & 14.9 {[}3.4{]} & 12.0 {[}2.0{]} & 33.4\tabularnewline
1.0\textless z\textless 2.0  & 93.2 {[}7.3{]} & 79.6 {[}2.8{]} & 40.0 {[}0.0{]} & 68.3 {[}1.0{]} & 42.9 {[}20.2{]} & 16.1 {[}1.7{]} & 18.7 {[}0.0{]} & 16.0 {[}1.2{]} & 10.5 {[}2.3{]} & 36.5\tabularnewline
2.0\textless z\textless 3.6  & 95.1 {[}11.0{]} & 87.5 {[}8.3{]} & 33.3 {[}0.0{]} & 73.5 {[}0.5{]} & 56.6 {[}23.3{]} & 20.2 {[}4.4{]} & 0.0 {[}0.0{]} & 12.5 {[}0.2{]} & 12.6 {[}1.7{]} & 49.4\tabularnewline
\hline 
\end{tabular}
}
\end{table*}

\begin{figure*}

\includegraphics[width = 1.\textwidth]{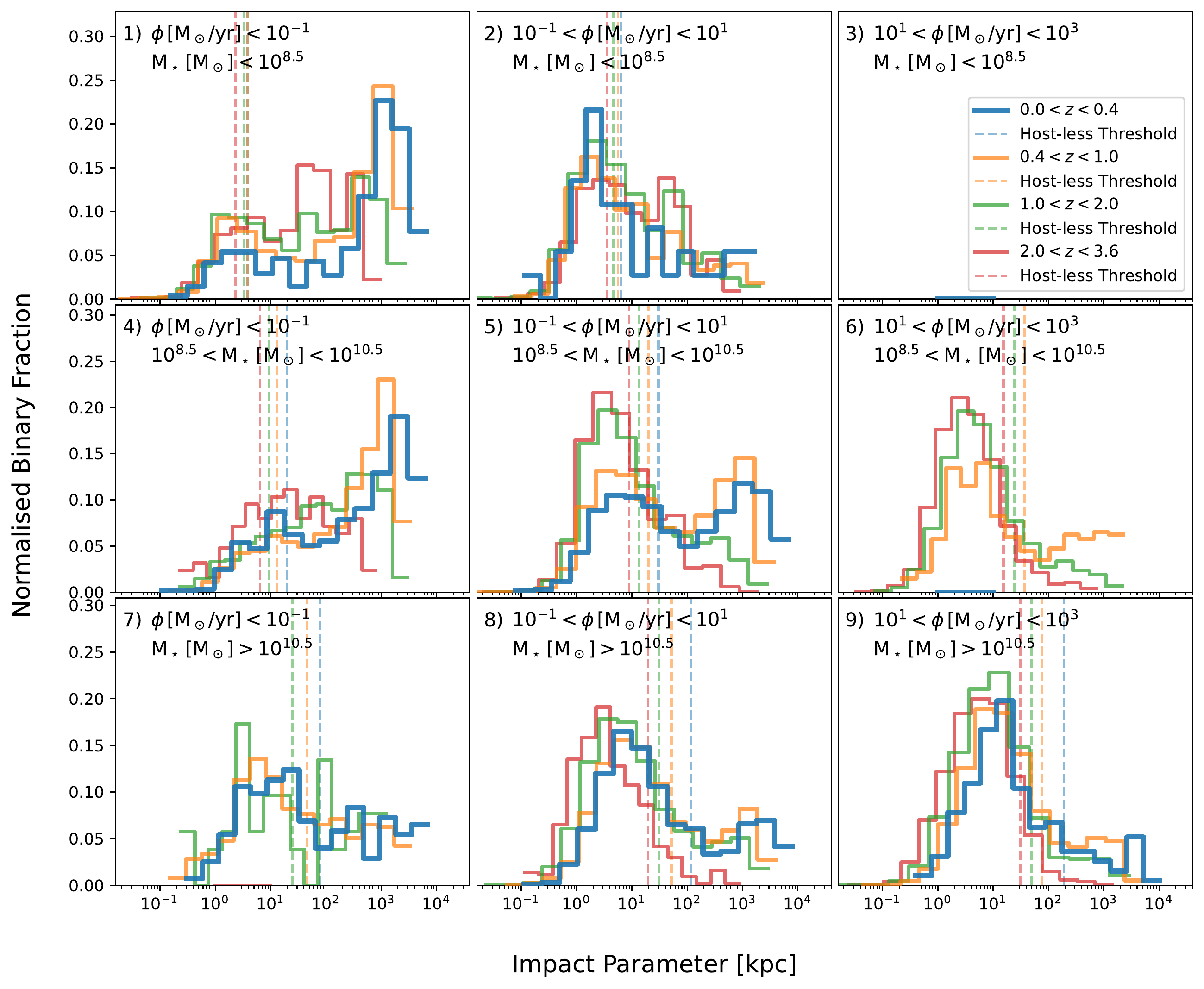}

\caption{[BPASS/Hobbs - eBHNS] For varied redshift slices, each panel corresponds to a breakdown of the
impact parameter distributions for different galaxy masses (row), and star formation rates (column).
Each histogram is composed of the number of binaries with a given impact parameter normalised over the total summation of merging systems within the respective panel and redshift. 
The dashed vertical line corresponds to the galaxy averaged impact parameter corresponding to the $P_{\rm chance}$ limit used to identify systems that may be potentially classed as "host-less".  
A numerical breakdown of each panel is shown in Table \ref{tab:hobbs-bhns-ip-table}. }
\label{fig:hobbs-bhns-ip}
\end{figure*}

\begin{table*}

\caption{[BPASS/Hobbs - eBHNS] A breakdown of the relative binary mergers occurring within each panel of Figure \ref{fig:hobbs-bhns-ip}. The top division refer to the relative fractions of binaries coalescing for a given redshift slice. The last column indicates the population of the binaries contained within each slice as a percentage of the cumulative population. The second section of the table shows the estimated host-less population per panel [overall fraction corresponding to the redshift slice] based on the criteria defined in Section \ref{ss:proj-dist}. }
\label{tab:hobbs-bhns-ip-table}
\centering
\scalebox{0.87}[0.87]{
\begin{tabular}{|c|c|c|c|c|c|c|c|c|c|c|}
\hline 
\multirow{2}{*}{z} & 1  & 2  & 3  & 4  & 5  & 6  & 7  & 8  & 9  & \multirow{2}{*}{Total Population (\%)}\tabularnewline
\cline{2-10} \cline{3-10} \cline{4-10} \cline{5-10} \cline{6-10} \cline{7-10} \cline{8-10} \cline{9-10} \cline{10-10} 
 & \multicolumn{9}{c|}{Binary Fraction (\%)} & \tabularnewline
\hline 
0.0\textless z\textless 0.4  & 9.9 & 0.7 & 0.0 & 10.2 & 32.0 & 0.1 & 4.9 & 34.3 & 8.0 & 3.4\tabularnewline
0.4\textless z\textless 1.0  & 11.2 & 2.0 & 0.0 & 4.9 & 48.2 & 1.0 & 0.9 & 17.7 & 14.0 & 24.1\tabularnewline
1.0\textless z\textless 2.0  & 7.9 & 4.2 & 0.0 & 1.2 & 54.0 & 8.4 & 0.1 & 5.7 & 18.5 & 48.0\tabularnewline
2.0\textless z\textless 3.6  & 7.1 & 7.1 & 0.0 & 0.3 & 44.6 & 27.1 & 0.0 & 1.1 & 12.6 & 24.5\tabularnewline
\hline 
 & \multicolumn{9}{c|}{Host-less Fraction/Panel {[}Relative to z-slice{]} (\%) } & Total Host-less Fraction (\%)\tabularnewline
\hline 
0.0\textless z\textless 0.4  & 88.5 {[}8.7{]} & 45.9 {[}0.3{]} & 0.0 {[}0.0{]} & 75.8 {[}7.7{]} & 57.8 {[}18.5{]} & 25.0 {[}0.0{]} & 38.5 {[}1.9{]} & 28.6 {[}9.8{]} & 13.7 {[}1.1{]} & 48.1\tabularnewline
0.4\textless z\textless 1.0  & 84.9 {[}9.5{]} & 54.8 {[}1.1{]} & 0.0 {[}0.0{]} & 83.6 {[}4.1{]} & 57.2 {[}27.6{]} & 36.9 {[}0.4{]} & 40.8 {[}0.4{]} & 34.2 {[}6.0{]} & 21.6 {[}3.0{]} & 52.1\tabularnewline
1.0\textless z\textless 2.0  & 84.3 {[}6.7{]} & 64.1 {[}2.7{]} & 0.0 {[}0.0{]} & 81.8 {[}1.0{]} & 43.7 {[}23.6{]} & 21.1 {[}1.8{]} & 46.2 {[}0.0{]} & 28.6 {[}1.6{]} & 19.5 {[}3.6{]} & 41.0\tabularnewline
2.0\textless z\textless 3.6  & 91.6 {[}6.5{]} & 78.2 {[}5.6{]} & 25.0 {[}0.0{]} & 75.4 {[}0.2{]} & 44.7 {[}19.9{]} & 15.9 {[}4.3{]} & 0.0 {[}0.0{]} & 13.5 {[}0.1{]} & 15.9 {[}2.0{]} & 38.7\tabularnewline
\hline 
\end{tabular}
}
\end{table*}

\begin{figure*}
    \centering
    \hbox{
    \includegraphics[width=\textwidth]{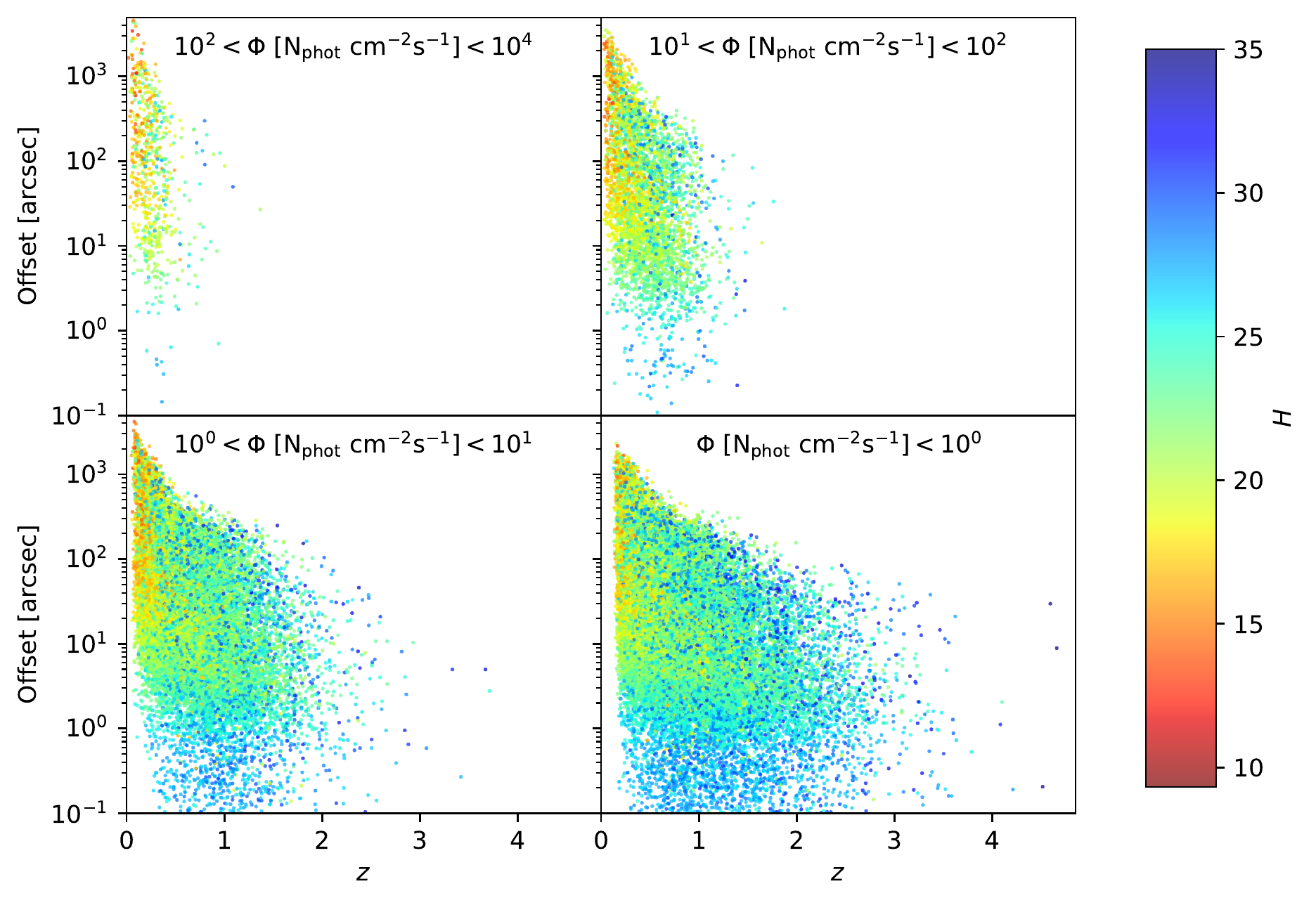}    
    }

    \caption{[BPASS/Hobbs] Similar to Figure \ref{fig:phot-z-grid}. }
    \label{fig:hobbs-phot-z-grid}
\end{figure*}

\begin{figure}
\vbox{
\includegraphics[width=\columnwidth]{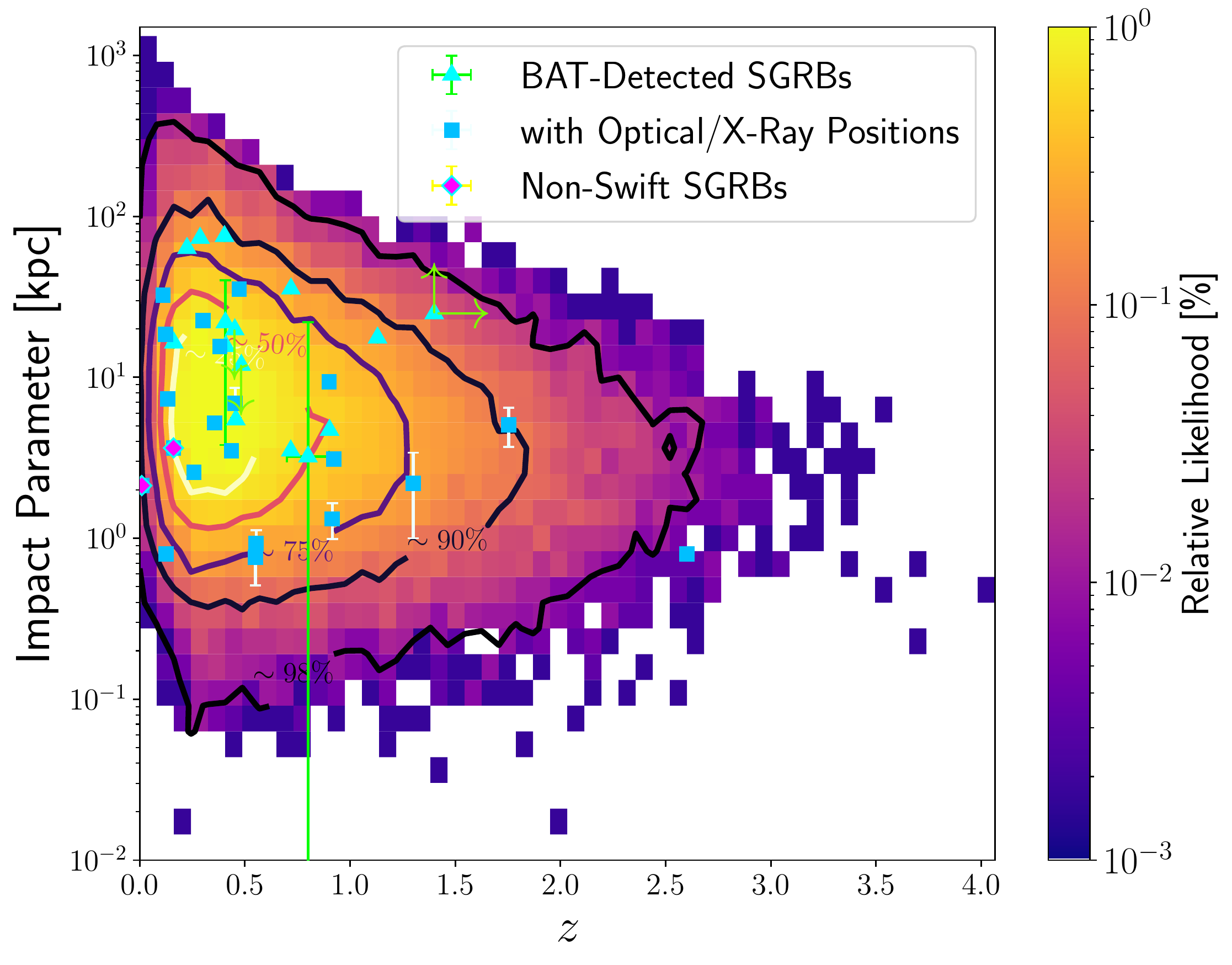}
}

\caption{[BPASS/Hobbs] Similar to Figure \ref{fig:obs-ip-z-dist}. 
}

\end{figure}

\begin{figure}
\vbox{
\includegraphics[width=\columnwidth]{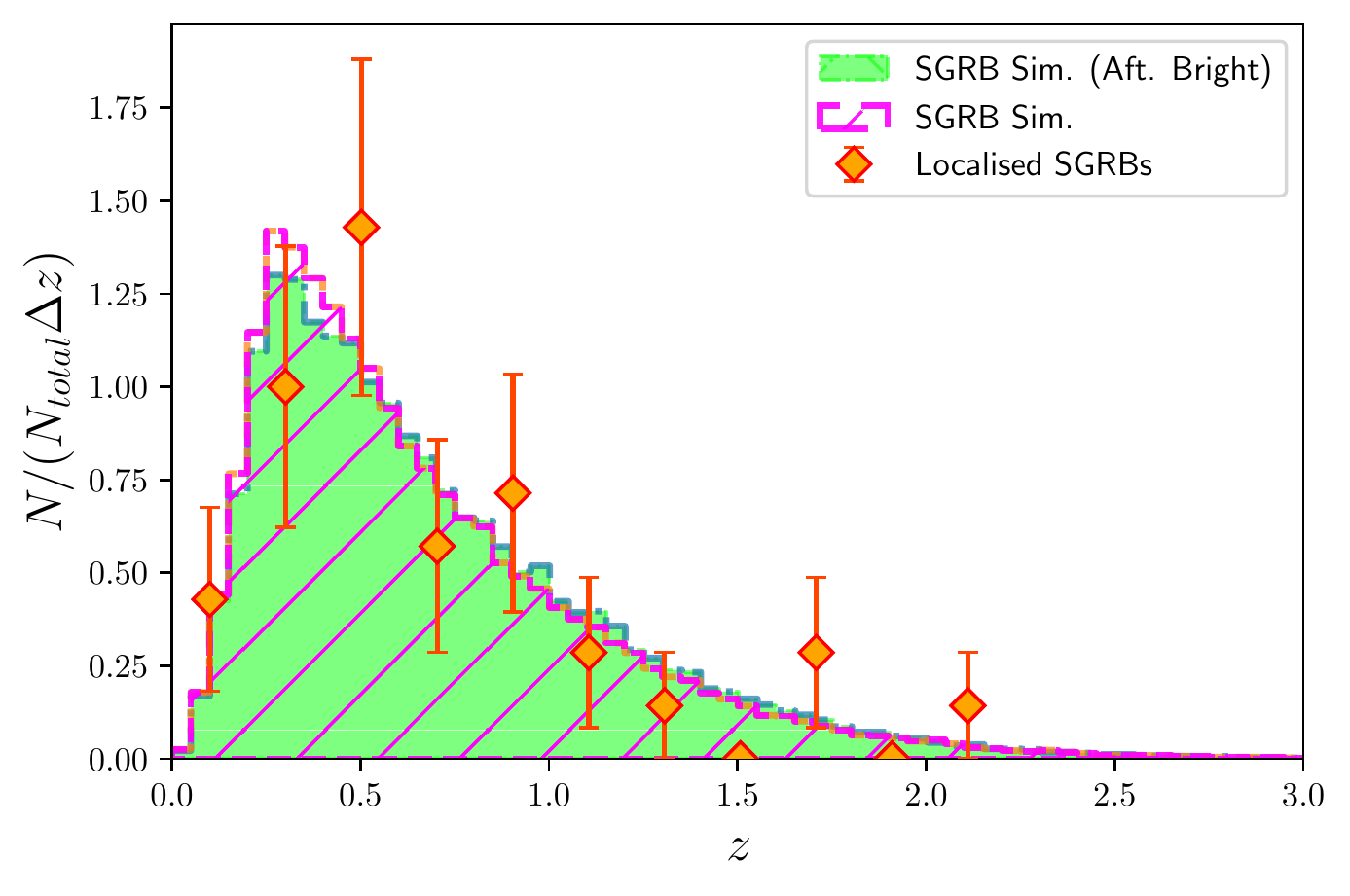}
}

\caption{ [BPASS/Hobbs] Similar to Figure \ref{fig:sgrb-obvs-dist}.
}
\label{fig:hobbs-amb-dens}
\end{figure}

\begin{figure}
    \centering
    \hbox{
    \includegraphics[width=0.5\textwidth]{{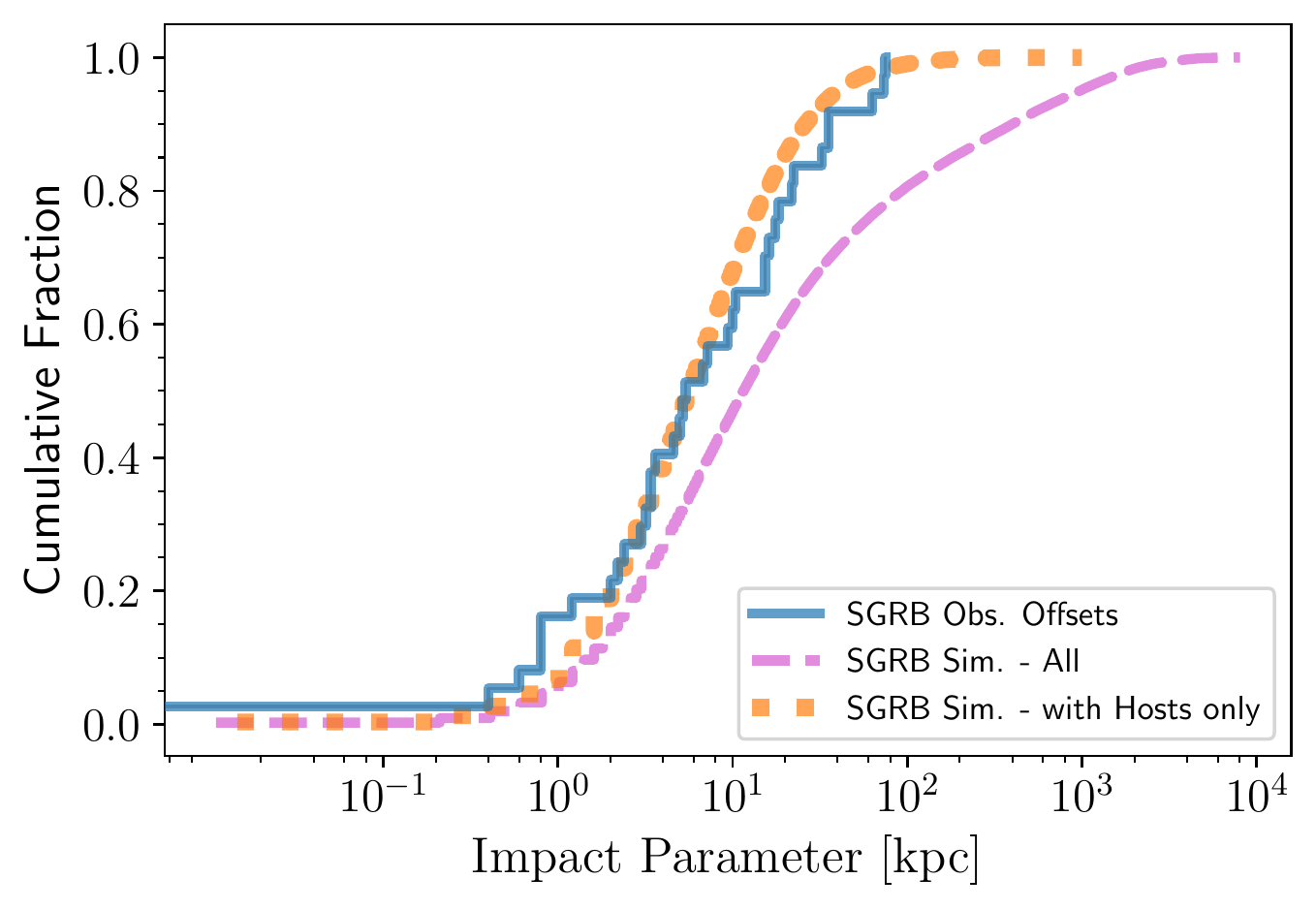}}
    }
    \caption{[BPASS/Hobbs]  Similar to Figure \ref{fig:ip_cumal}. Refer to Table \ref{tab:sgrb-tab} for sources used.  
    }
    \label{fig:hobbs-ip_cumal}
\end{figure}


\clearpage
\newpage
\section{COSMIC Binaries}\label{app:COSMIC}
Here we show the equivalent figures and tables from the main text, for COSMIC simulated binaries.

\begin{figure*}

\centering
\vbox{
\hbox{
	\includegraphics[width=0.50\textwidth]{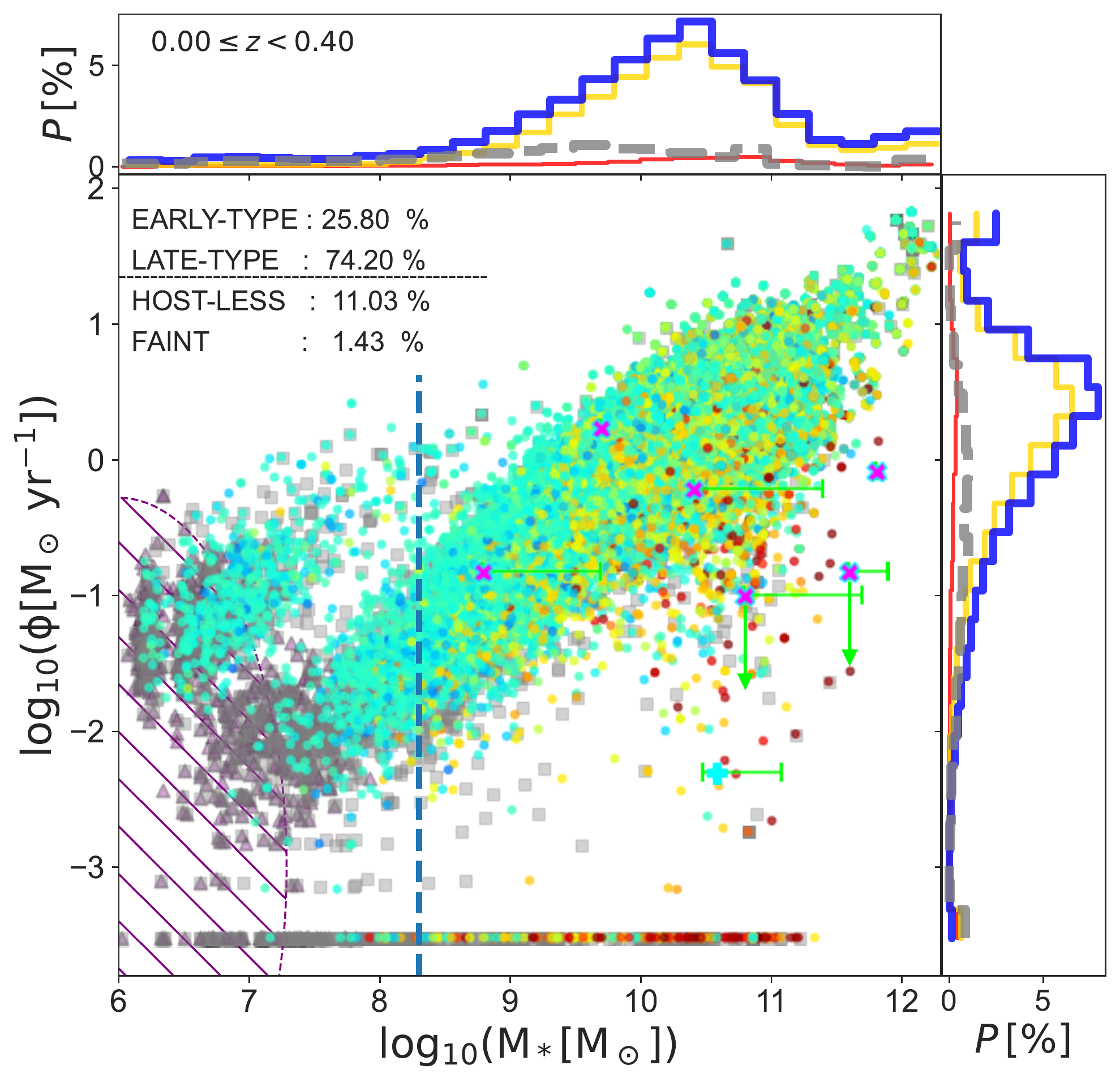}	
	\includegraphics[width=0.50\textwidth]{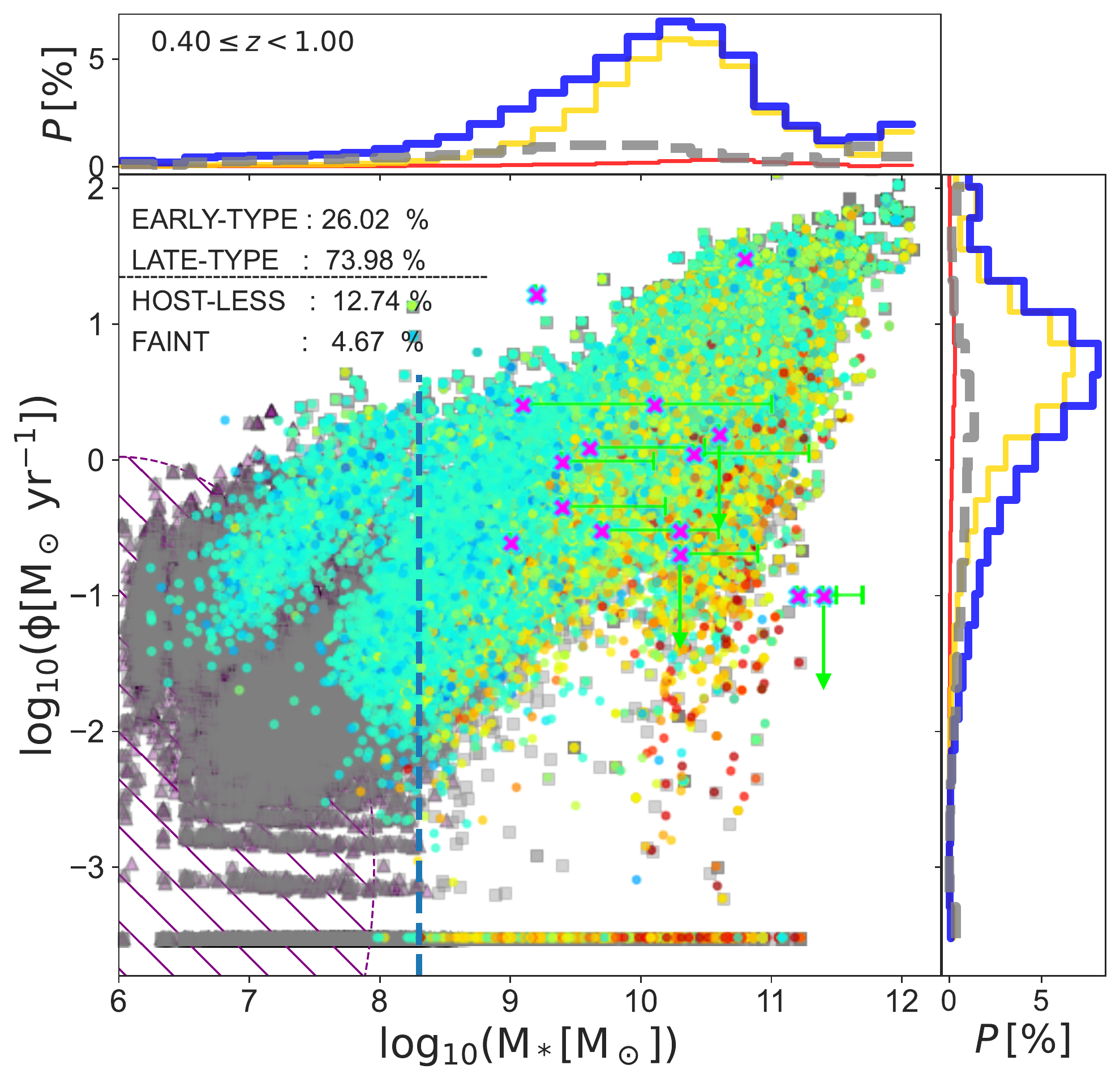}	
		}
\hbox{
	\includegraphics[width=0.50\textwidth]{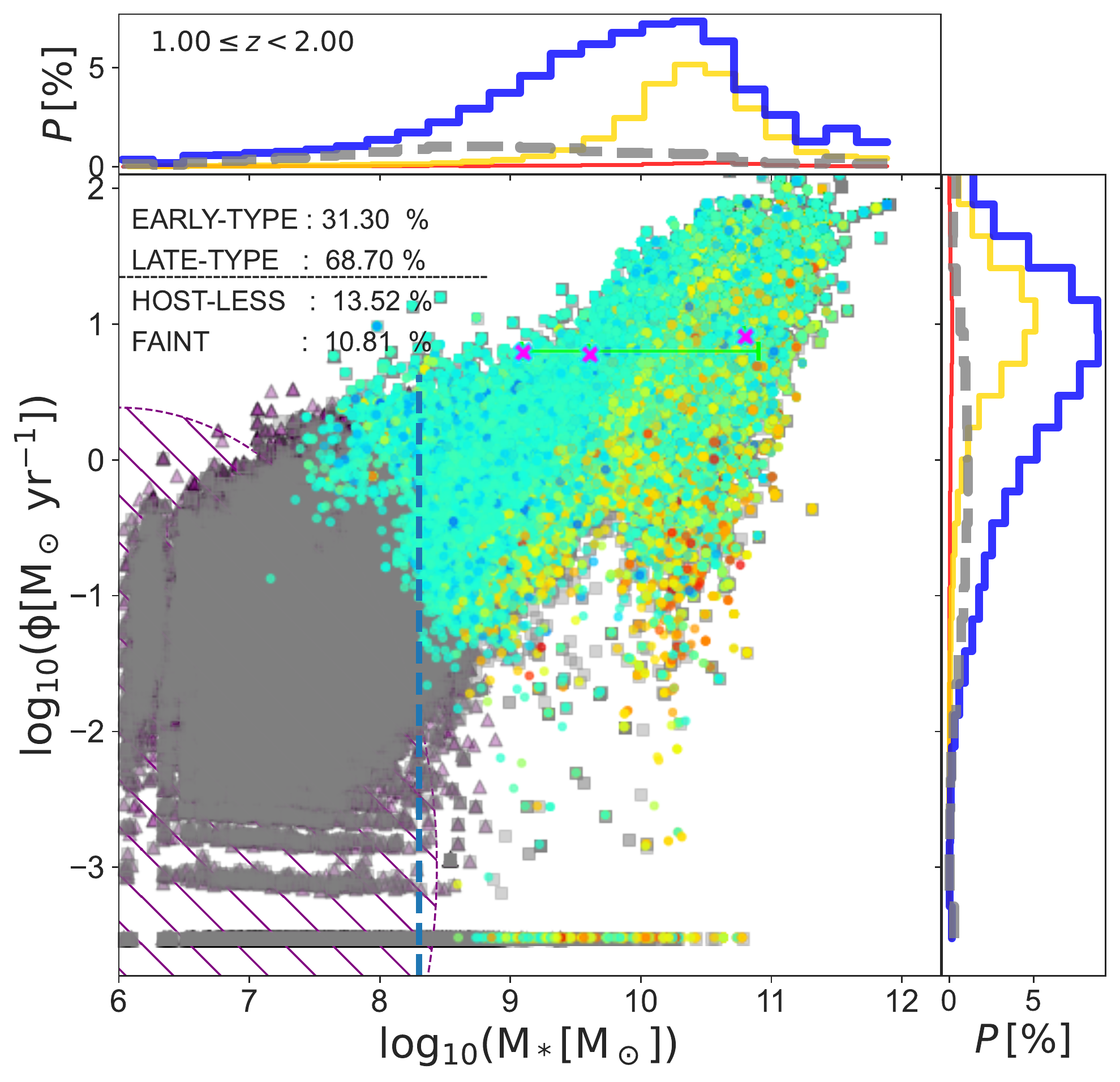}
	\includegraphics[width=0.50\textwidth]{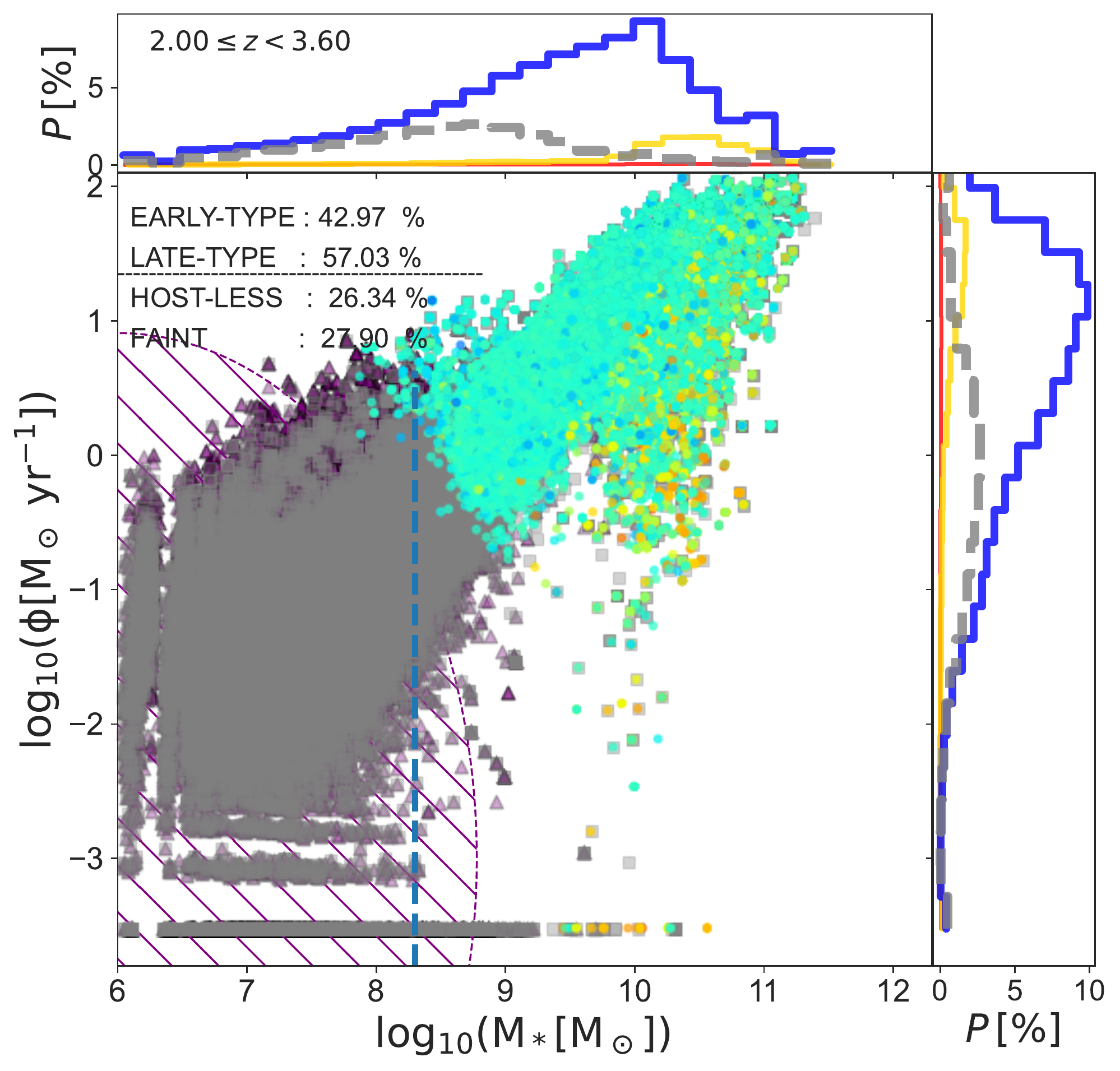}	
		}
\hbox{

\includegraphics[width=0.7\textwidth,trim={0 0 0.2cm 0.4cm},clip]{plots/colorbar.pdf}
\hsize=.2\linewidth
\vbox{
\includegraphics[scale=0.6,trim={0 0.3cm 0.4cm 0.2cm},clip]{plots/m-sfr-legend.pdf}\\
\includegraphics[scale=0.6,trim={0 0.4cm 0.4cm 0.4cm},clip]{plots/main_legend.pdf}
}
}
		}
 \caption{[COSMIC - NSNS] Similar to Figure \ref{fig:nsns-ev}.
  }
 \label{fig:cosmic-nsns-ev}
\end{figure*}

\begin{figure*}

\centering
\vbox{
\hbox{
	\includegraphics[width=0.50\textwidth]{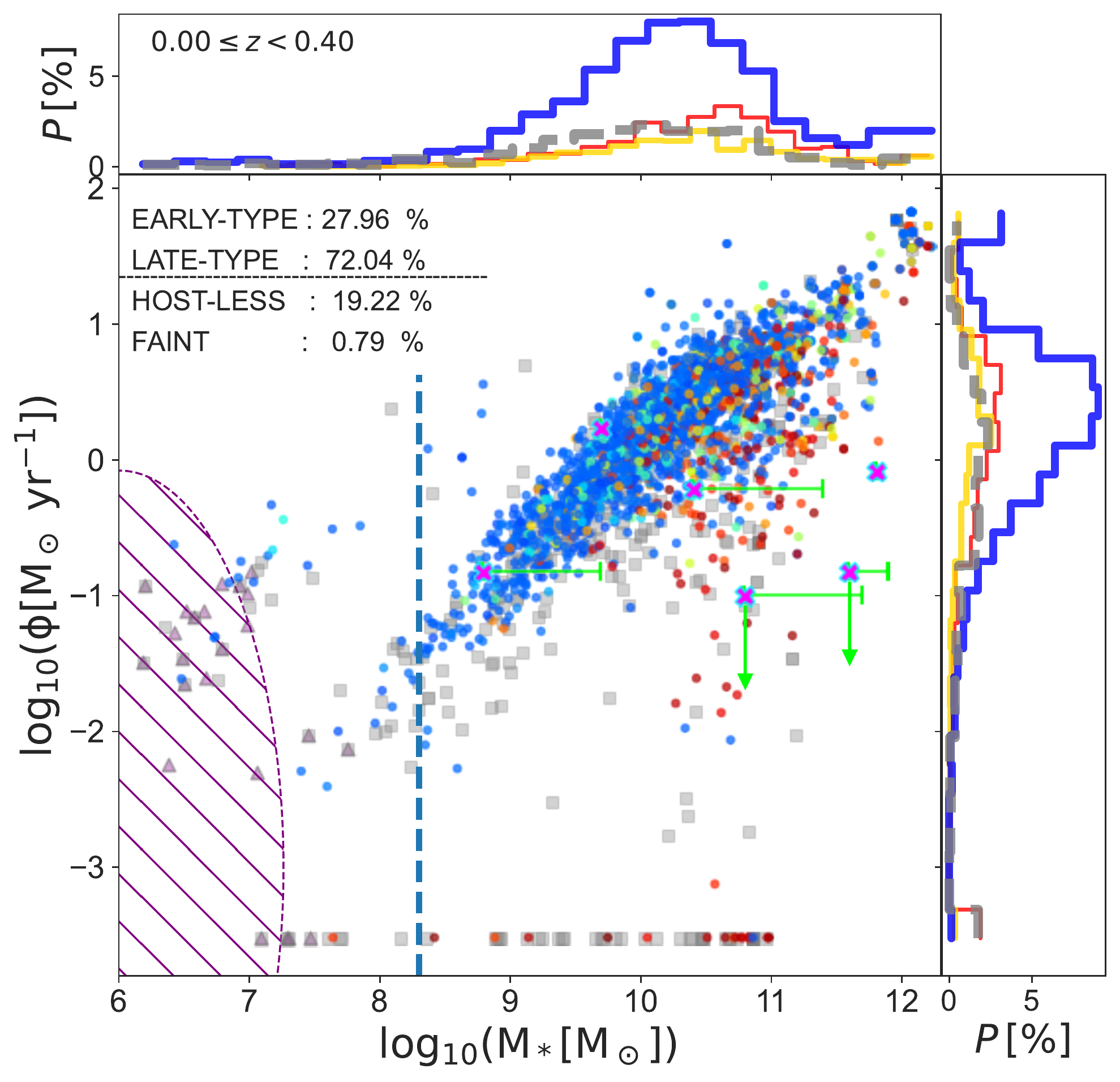}	
	\includegraphics[width=0.50\textwidth]{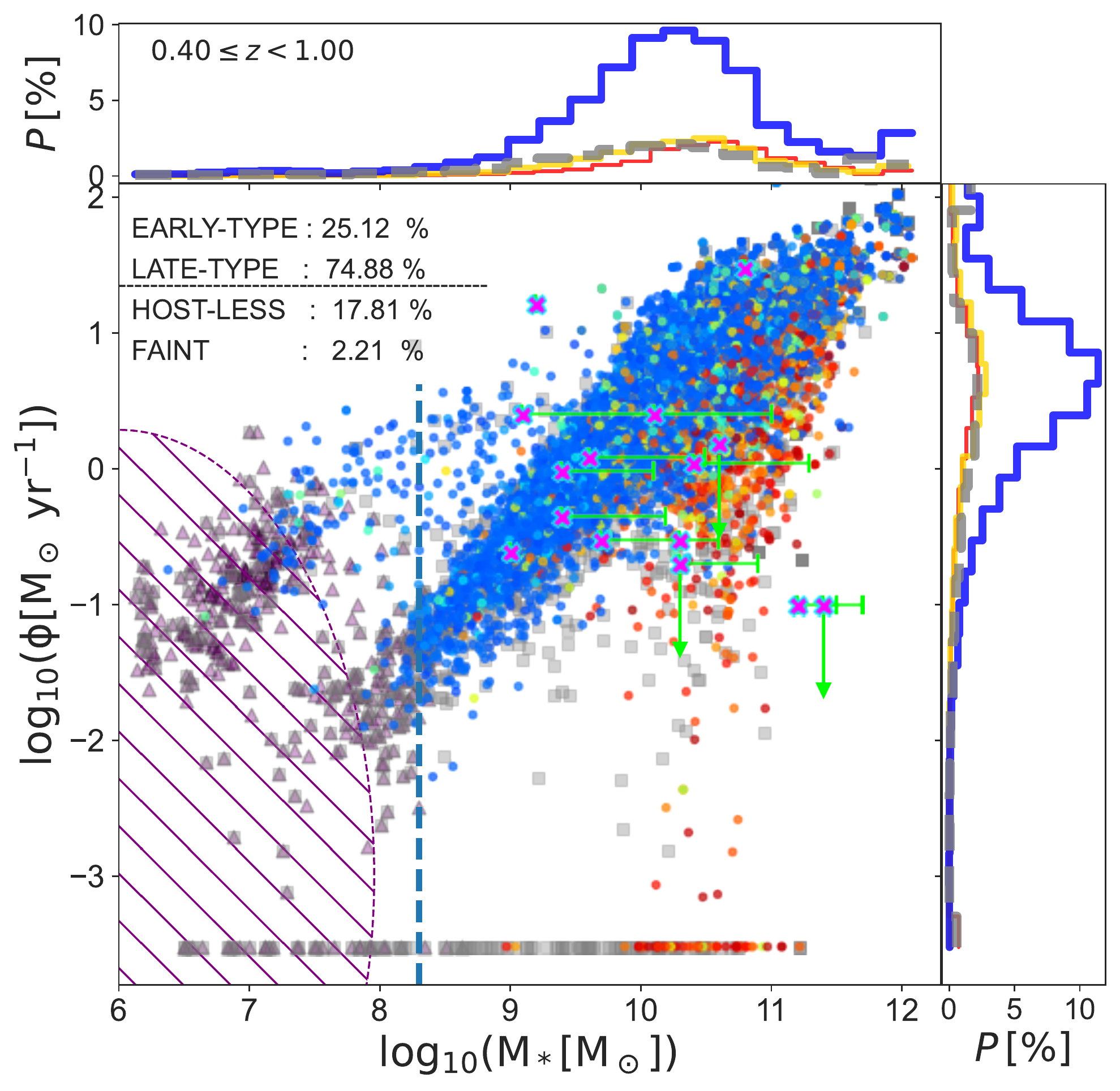}	
		}
\hbox{
	\includegraphics[width=0.50\textwidth]{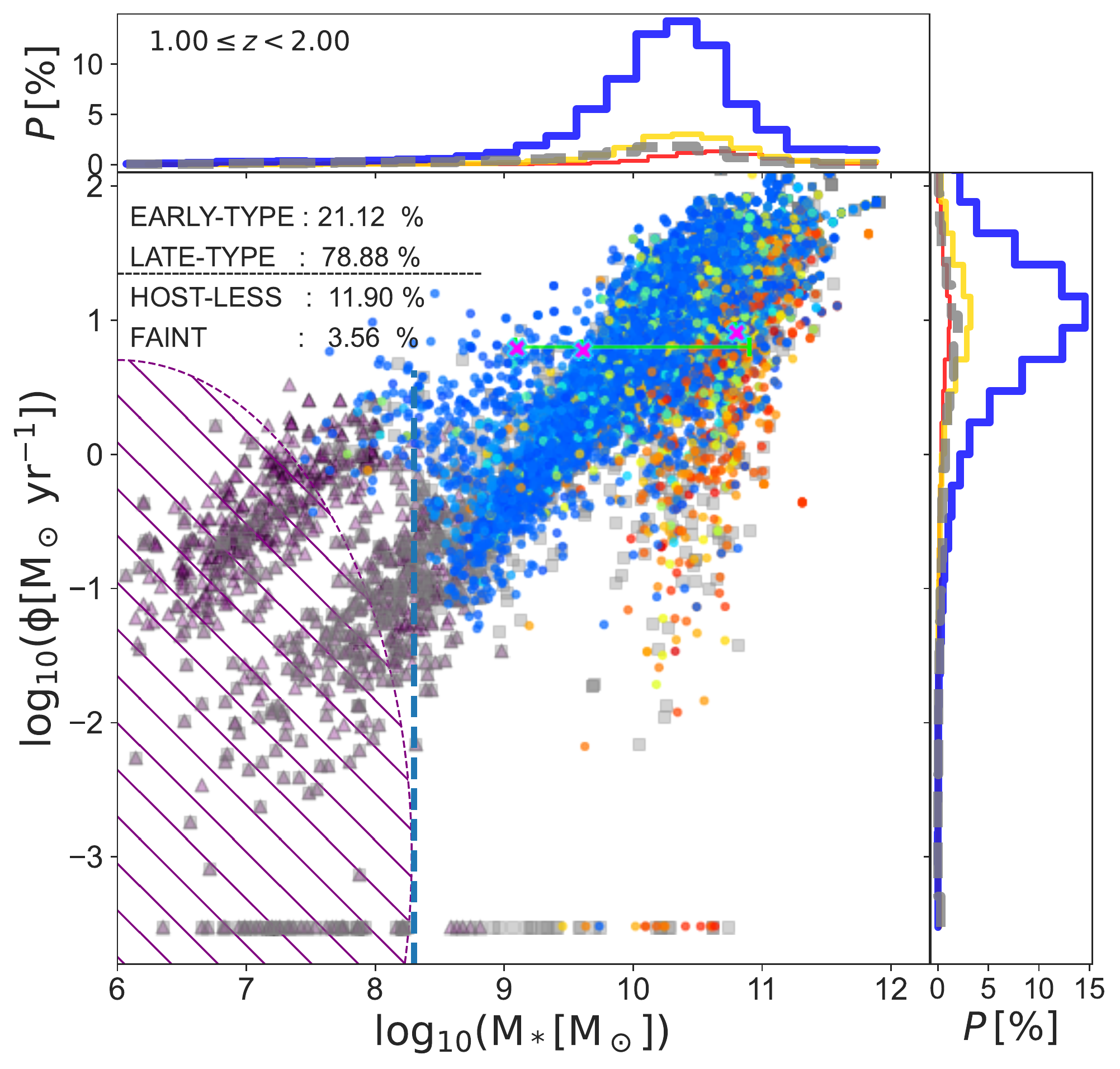}
	\includegraphics[width=0.50\textwidth]{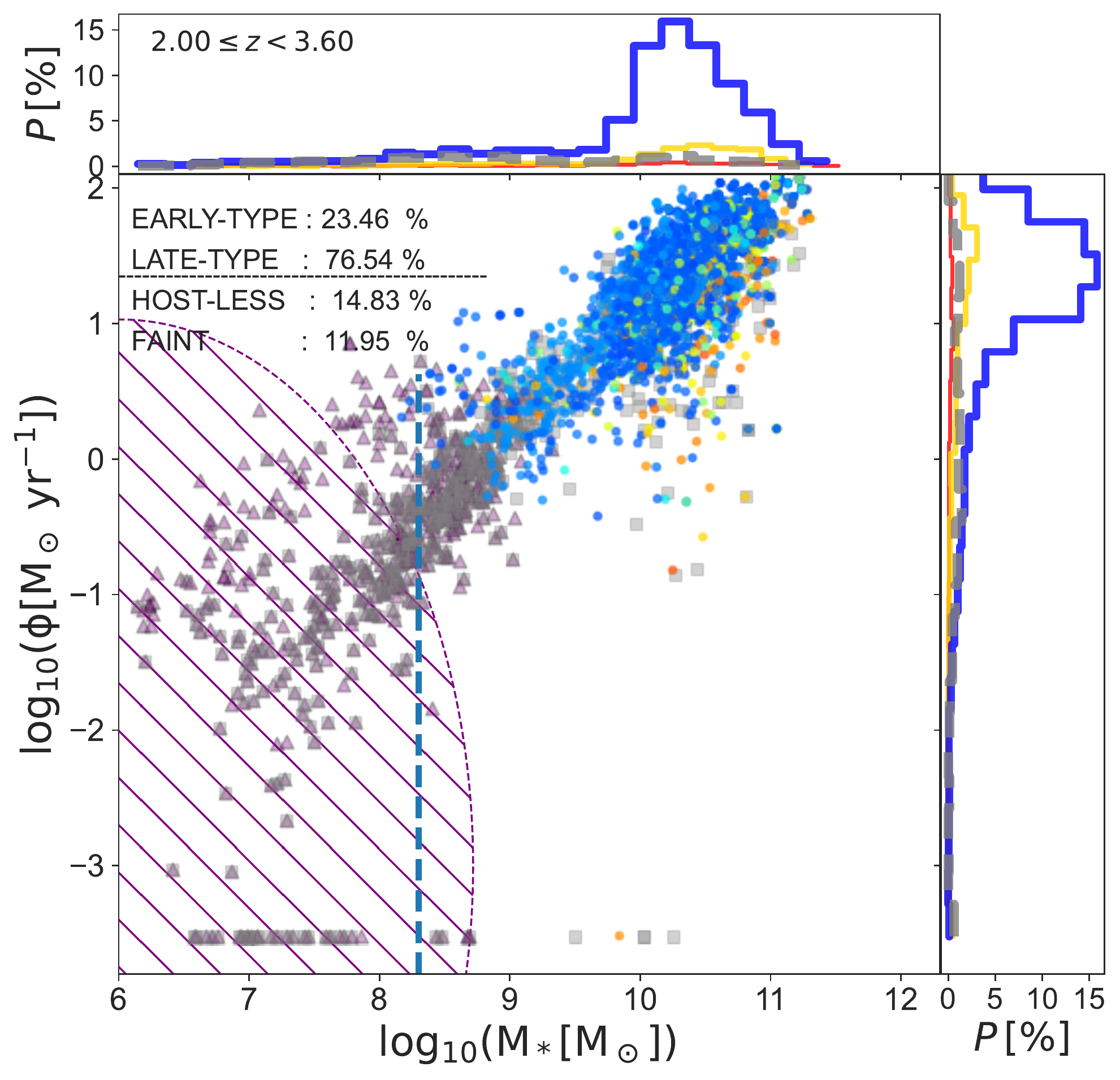}	
		}

\hbox{

\includegraphics[width=0.7\textwidth,trim={0 0 0.2cm 0.4cm},clip]{plots/colorbar.pdf}
\hsize=.2\linewidth
\vbox{
\includegraphics[scale=0.6,trim={0 0.3cm 0.4cm 0.2cm},clip]{plots/m-sfr-legend.pdf}\\
\includegraphics[scale=0.6,trim={0 0.4cm 0.4cm 0.4cm},clip]{plots/main_legend.pdf}
}

}

		}
 \caption{[COSMIC - eBHNS] Similar to Figure \ref{fig:bhns-ev}. 
  }
  
 \label{fig:cosmic-bhns-ev}
\end{figure*}

\begin{figure*}
\includegraphics[width =  1.\textwidth]{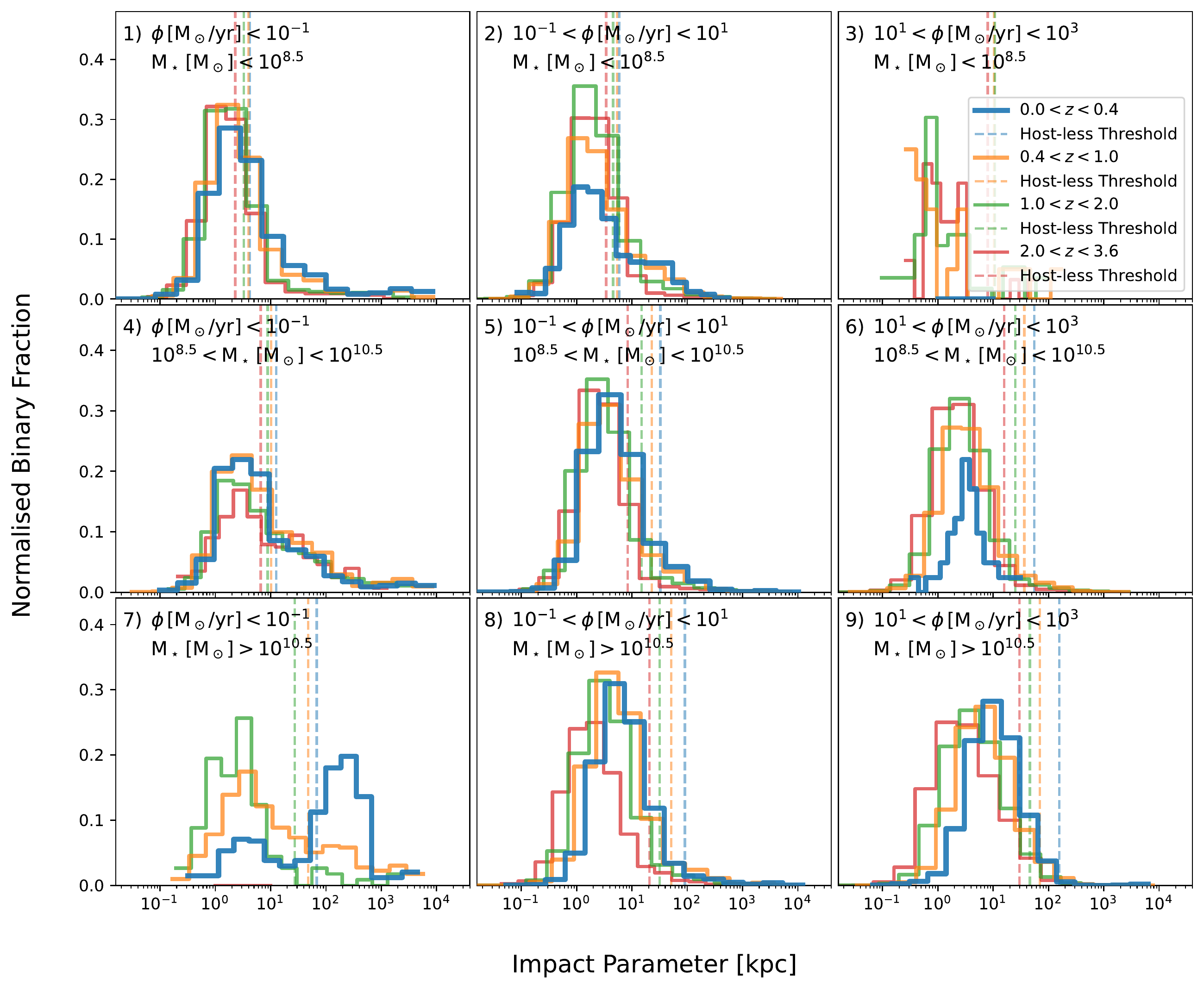}

\caption{[COSMIC - NSNS] For varied redshift slices, each panel corresponds to a breakdown of the 
impact parameter distributions for different galaxy masses (row), and star formation rates (column).
Each histogram is composed of the number of binaries with a given impact parameter normalised over the total summation of merging systems within the respective panel and redshift. 
The dashed vertical line corresponds to the galaxy averaged impact parameter corresponding to the $P_{\rm chance}$ limit used to identify systems that may be potentially classed as "host-less". 
A numerical breakdown of each panel is shown in Table \ref{tab:cosmic-nsns-ip-table}. }
\label{fig:cosmic-nsns-ip}
\end{figure*}

\begin{table*}

\caption{[COSMIC - NSNS] A breakdown of the relative binary mergers occurring within each panel of Figure \ref{fig:cosmic-nsns-ip}. The top division refer to the relative fractions of binaries coalescing for a given redshift slice. The last column indicates the population of the binaries contained within each slice as a percentage of the cumulative population. The second section of the table shows the estimated host-less population per panel [overall fraction corresponding to the redshift slice] based on the criteria defined in Section \ref{ss:proj-dist}.}
\label{tab:cosmic-nsns-ip-table}
\centering
\scalebox{0.87}[0.87]{
\begin{tabular}{|c|c|c|c|c|c|c|c|c|c|c|}
\hline 
\multirow{2}{*}{z} & 1  & 2  & 3  & 4  & 5  & 6  & 7  & 8  & 9  & \multirow{2}{*}{Total Population (\%)}\tabularnewline
\cline{2-10} \cline{3-10} \cline{4-10} \cline{5-10} \cline{6-10} \cline{7-10} \cline{8-10} \cline{9-10} \cline{10-10} 
 & \multicolumn{9}{c|}{Binary Fraction (\%)} & \tabularnewline
\hline 
0.0\textless z\textless 0.4  & 4.7 & 1.3 & 0.0 & 3.1 & 53.9 & 0.1 & 0.7 & 26.7 & 9.6 & 3.3\tabularnewline
0.4\textless z\textless 1.0  & 4.6 & 2.4 & 0.0 & 1.2 & 55.9 & 2.5 & 0.1 & 14.4 & 19.0 & 26.5\tabularnewline
1.0\textless z\textless 2.0  & 5.2 & 4.8 & 0.0 & 0.4 & 48.1 & 13.8 & 0.0 & 4.3 & 23.4 & 46.4\tabularnewline
2.0\textless z\textless 3.6  & 7.6 & 11.0 & 0.0 & 0.1 & 39.7 & 26.9 & 0.0 & 0.9 & 13.7 & 23.8\tabularnewline
\hline 
 & \multicolumn{9}{c|}{Host-less Fraction/Panel {[}Relative to z-slice{]} (\%) } & Total Host-less Fraction (\%)\tabularnewline
\hline 
0.0\textless z\textless 0.4  & 48.5 {[}2.3{]} & 31.8 {[}0.4{]} & 0.0 {[}0.0{]} & 31.4 {[}1.0{]} & 10.2 {[}5.5{]} & 0.0 {[}0.0{]} & 63.6 {[}0.4{]} & 3.9 {[}1.0{]} & 4.3 {[}0.4{]} & 11.0\tabularnewline
0.4\textless z\textless 1.0  & 48.2 {[}2.2{]} & 29.3 {[}0.7{]} & 20.0 {[}0.0{]} & 40.4 {[}0.5{]} & 11.3 {[}6.3{]} & 4.7 {[}0.1{]} & 26.8 {[}0.0{]} & 6.6 {[}0.9{]} & 10.2 {[}1.9{]} & 12.7\tabularnewline
1.0\textless z\textless 2.0  & 56.1 {[}2.9{]} & 38.9 {[}1.9{]} & 14.3 {[}0.0{]} & 49.5 {[}0.2{]} & 13.4 {[}6.4{]} & 4.0 {[}0.6{]} & 8.0 {[}0.0{]} & 5.3 {[}0.2{]} & 5.6 {[}1.3{]} & 13.5\tabularnewline
2.0\textless z\textless 3.6  & 70.6 {[}5.4{]} & 59.8 {[}6.6{]} & 6.5 {[}0.0{]} & 57.0 {[}0.1{]} & 28.6 {[}11.4{]} & 6.1 {[}1.7{]} & 0.0 {[}0.0{]} & 5.3 {[}0.0{]} & 9.0 {[}1.2{]} & 26.3\tabularnewline
\hline 
\end{tabular}
}
\end{table*}

\begin{figure*}

\includegraphics[width = 1.\textwidth]{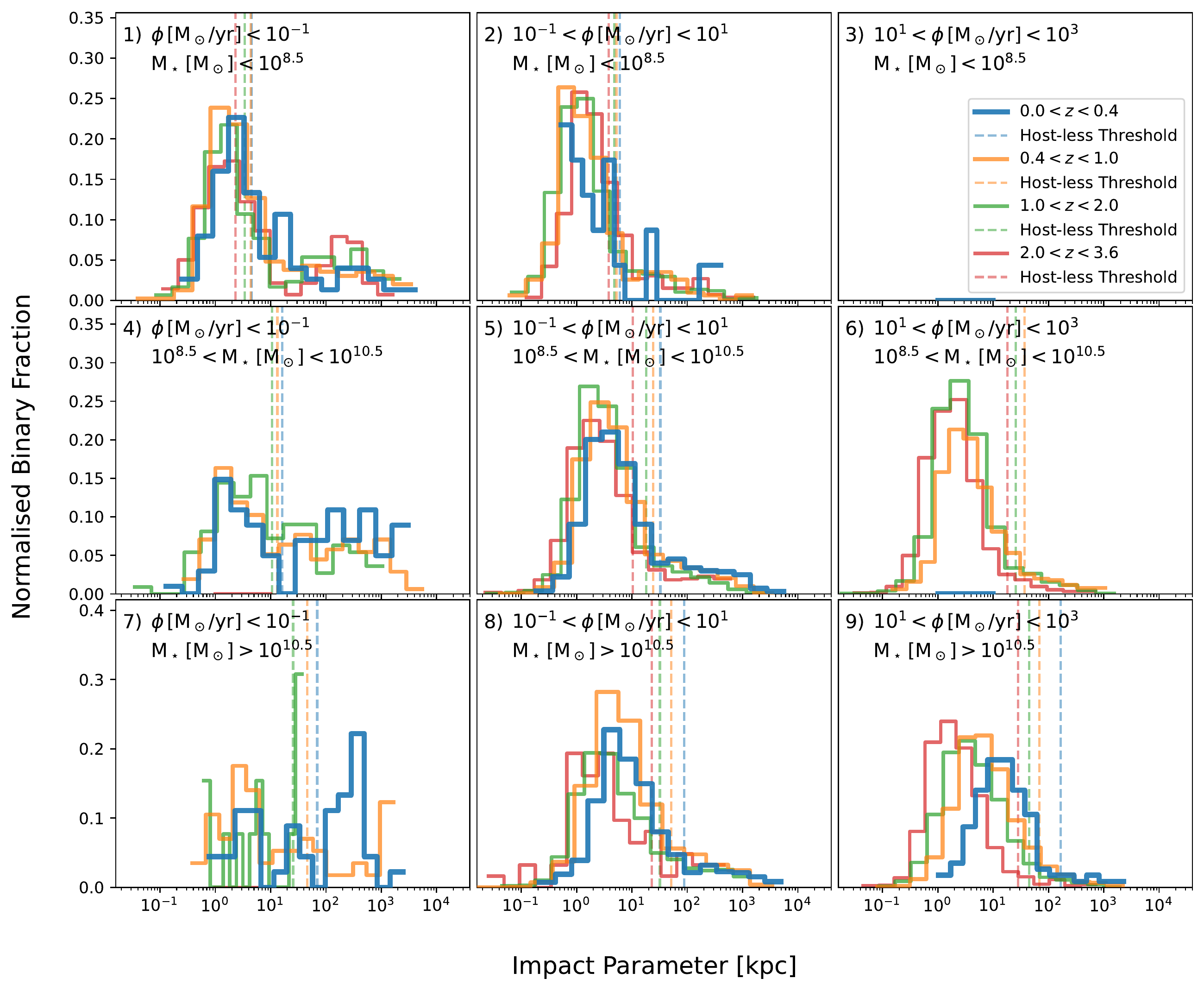}

\caption{[COSMIC - eBHNS] For varied redshift slices, each panel corresponds to a breakdown of the
impact parameter distributions for different galaxy masses (row), and star formation rates (column).
Each histogram is composed of the number of binaries with a given impact parameter normalised over the total summation of merging systems within the respective panel and redshift. 
The dashed vertical line corresponds to the galaxy averaged impact parameter corresponding to the $P_{\rm chance}$ limit used to identify systems that may be potentially classed as "host-less".  
A numerical breakdown of each panel is shown in Table \ref{tab:cosmic-bhns-ip-table}. }
\label{fig:cosmic-bhns-ip}
\end{figure*}

\begin{table*}

\caption{[COSMIC - eBHNS] A breakdown of the relative binary mergers occurring within each panel of Figure \ref{fig:cosmic-bhns-ip}. The top division refer to the relative fractions of binaries coalescing for a given redshift slice. The last column indicates the population of the binaries contained within each slice as a percentage of the cumulative population. The second section of the table shows the estimated host-less population per panel [overall fraction corresponding to the redshift slice] based on the criteria defined in Section \ref{ss:proj-dist}.}
\label{tab:cosmic-bhns-ip-table}
\centering
\scalebox{0.87}[0.87]{
\begin{tabular}{|c|c|c|c|c|c|c|c|c|c|c|}
\hline 
\multirow{2}{*}{z} & 1  & 2  & 3  & 4  & 5  & 6  & 7  & 8  & 9  & \multirow{2}{*}{Total Population (\%)}\tabularnewline
\cline{2-10} \cline{3-10} \cline{4-10} \cline{5-10} \cline{6-10} \cline{7-10} \cline{8-10} \cline{9-10} \cline{10-10} 
 & \multicolumn{9}{c|}{Binary Fraction (\%)} & \tabularnewline
\hline 
0.0\textless z\textless 0.4  & 2.6 & 0.8 & 0.0 & 3.5 & 52.5 & 0.2 & 1.5 & 29.5 & 9.4 & 5.7\tabularnewline
0.4\textless z\textless 1.0  & 2.0 & 1.6 & 0.0 & 1.6 & 55.0 & 2.5 & 0.3 & 16.8 & 20.3 & 39.4\tabularnewline
1.0\textless z\textless 2.0  & 1.3 & 2.5 & 0.0 & 0.5 & 41.5 & 18.6 & 0.1 & 5.5 & 30.1 & 45.8\tabularnewline
2.0\textless z\textless 3.6  & 3.0 & 5.7 & 0.0 & 0.2 & 20.2 & 40.7 & 0.0 & 1.4 & 28.9 & 9.0\tabularnewline
\hline 
 & \multicolumn{9}{c|}{Host-less Fraction/Panel {[}Relative to z-slice{]} (\%) } & Total Host-less Fraction (\%)\tabularnewline
\hline 
0.0\textless z\textless 0.4  & 52.0 {[}1.3{]} & 30.4 {[}0.2{]} & 0.0 {[}0.0{]} & 60.4 {[}2.1{]} & 19.8 {[}10.4{]} & 0.0 {[}0.0{]} & 55.6 {[}0.9{]} & 12.0 {[}3.5{]} & 8.1 {[}0.8{]} & 19.2\tabularnewline
0.4\textless z\textless 1.0  & 49.2 {[}1.0{]} & 19.9 {[}0.3{]} & 0.0 {[}0.0{]} & 48.1 {[}0.8{]} & 18.1 {[}10.0{]} & 8.3 {[}0.2{]} & 28.1 {[}0.1{]} & 14.8 {[}2.5{]} & 14.9 {[}3.0{]} & 17.8\tabularnewline
1.0\textless z\textless 2.0  & 64.9 {[}0.8{]} & 25.2 {[}0.6{]} & 0.0 {[}0.0{]} & 48.6 {[}0.2{]} & 15.0 {[}6.2{]} & 7.3 {[}1.4{]} & 38.5 {[}0.0{]} & 15.8 {[}0.9{]} & 5.8 {[}1.7{]} & 11.9\tabularnewline
2.0\textless z\textless 3.6  & 77.0 {[}2.3{]} & 54.6 {[}3.1{]} & 0.0 {[}0.0{]} & 62.5 {[}0.1{]} & 26.9 {[}5.4{]} & 6.0 {[}2.4{]} & 0.0 {[}0.0{]} & 19.4 {[}0.3{]} & 4.0 {[}1.2{]} & 14.8\tabularnewline
\hline 
\end{tabular}
}
\end{table*}

\begin{figure*}
    \centering
    \hbox{
    \includegraphics[width=\textwidth]{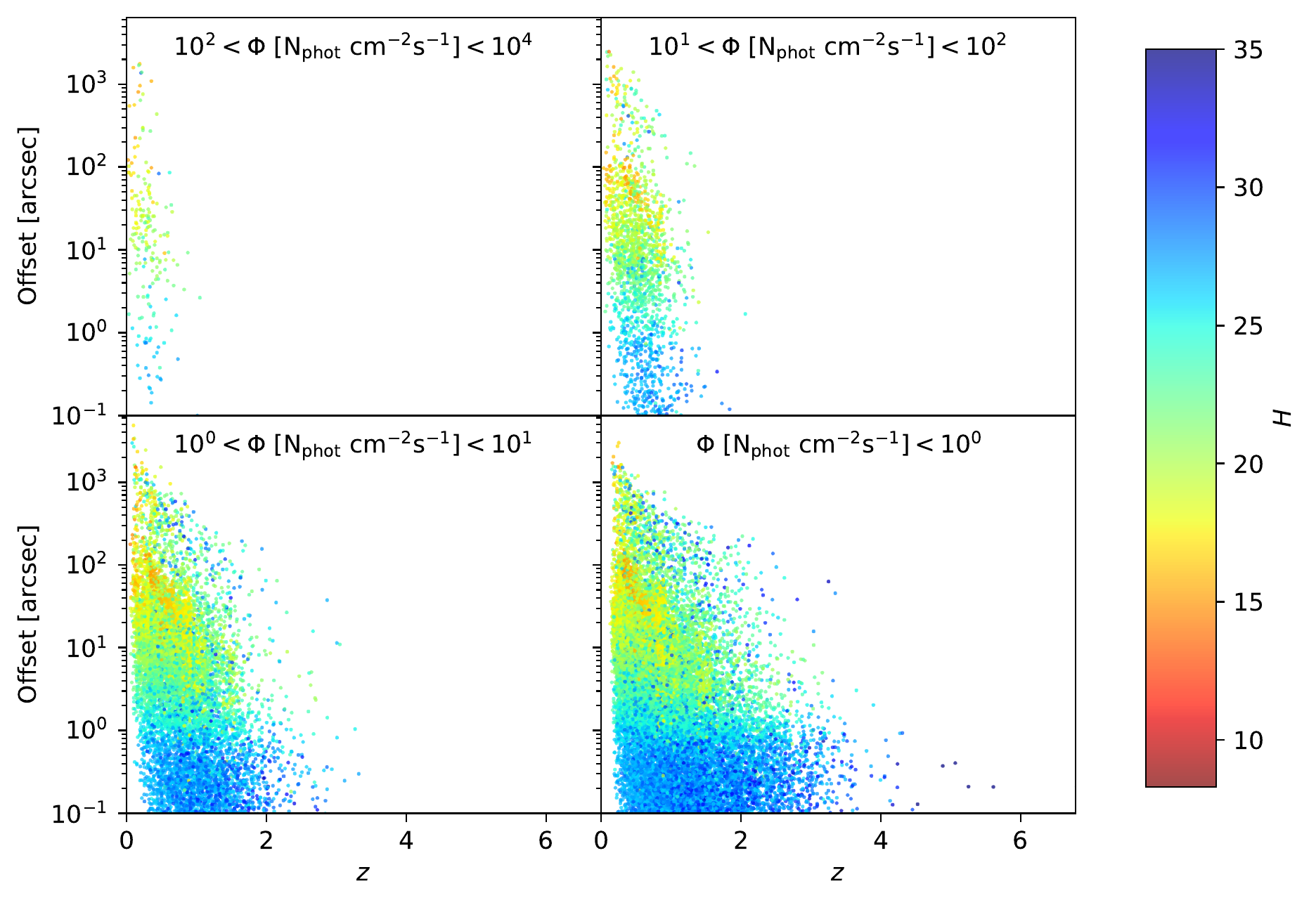}    
    }

    \caption{[COSMIC] Similar to Figure \ref{fig:phot-z-grid}.}
    \label{fig:cosmic-phot-z-grid}
\end{figure*}

\begin{figure}
\vbox{
\includegraphics[width=\columnwidth]{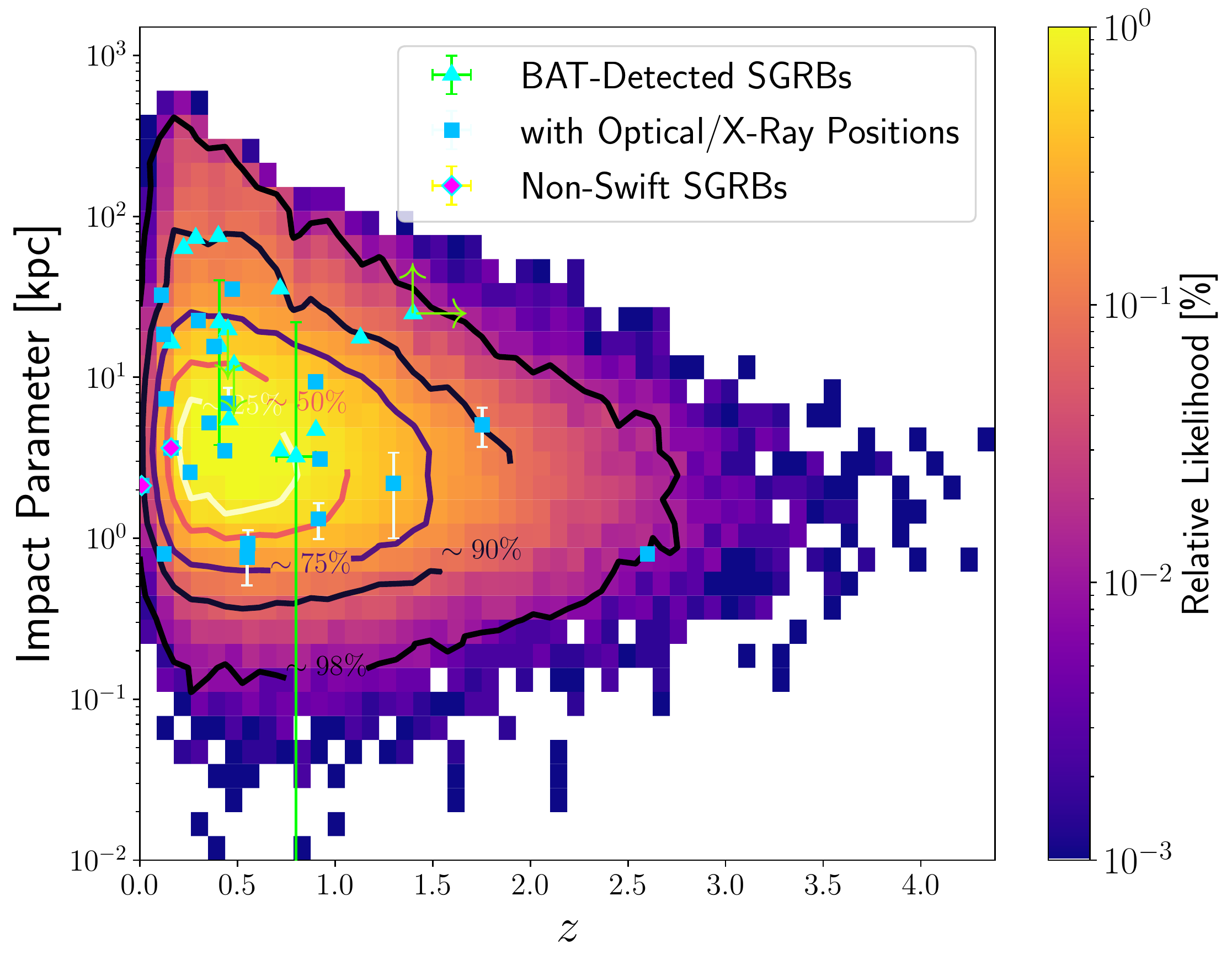}
}

\caption{[COSMIC] Similar to Figure \ref{fig:obs-ip-z-dist}. 
}

\end{figure}

\begin{figure}
\vbox{
\includegraphics[width=\columnwidth]{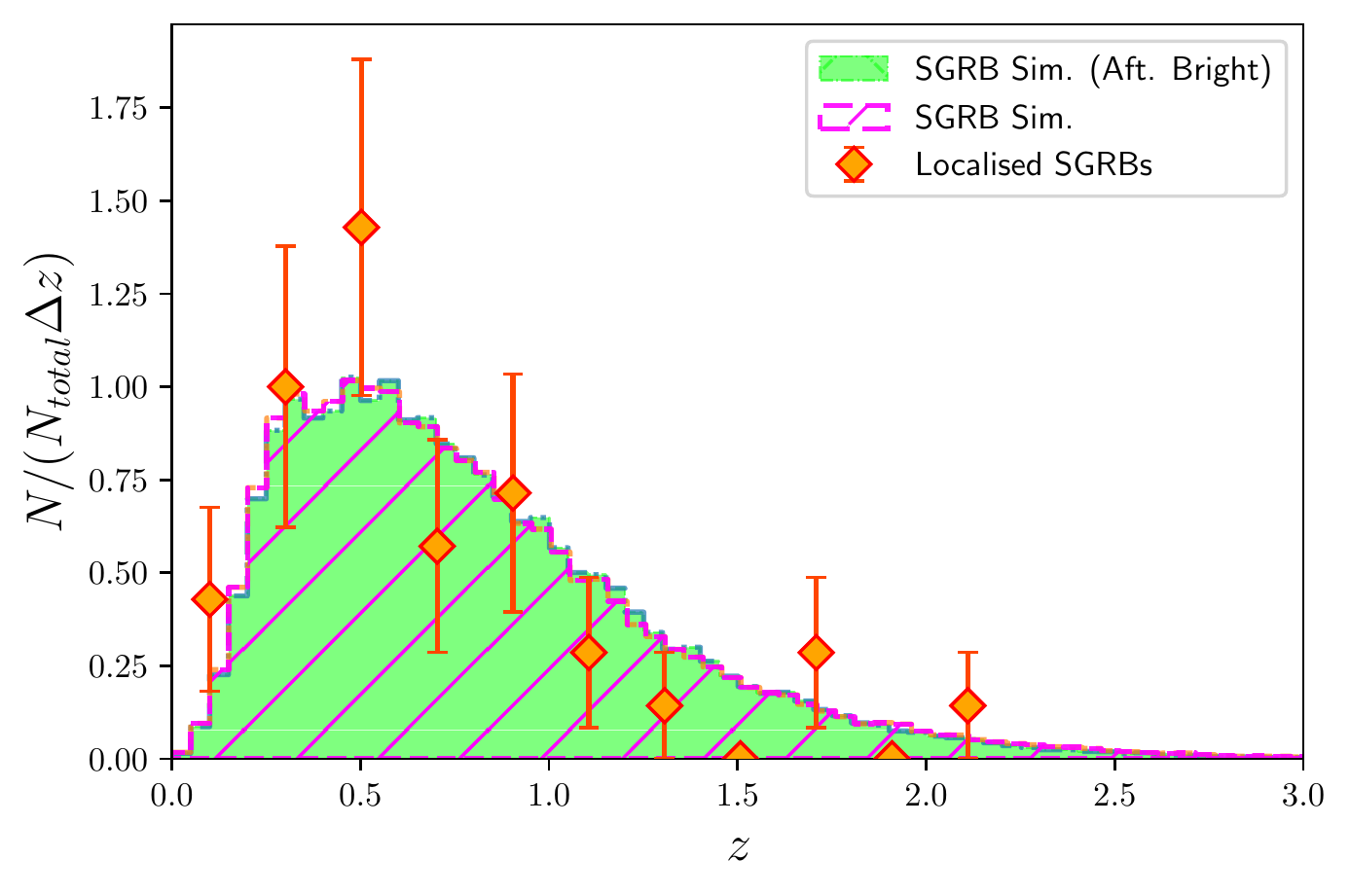}
}

\caption{[COSMIC] Similar to Figure \ref{fig:sgrb-obvs-dist}. 
}
\label{fig:cosmic-amb-dens}
\end{figure}

\begin{figure}
    \centering
    \hbox{
    \includegraphics[width=0.5\textwidth]{{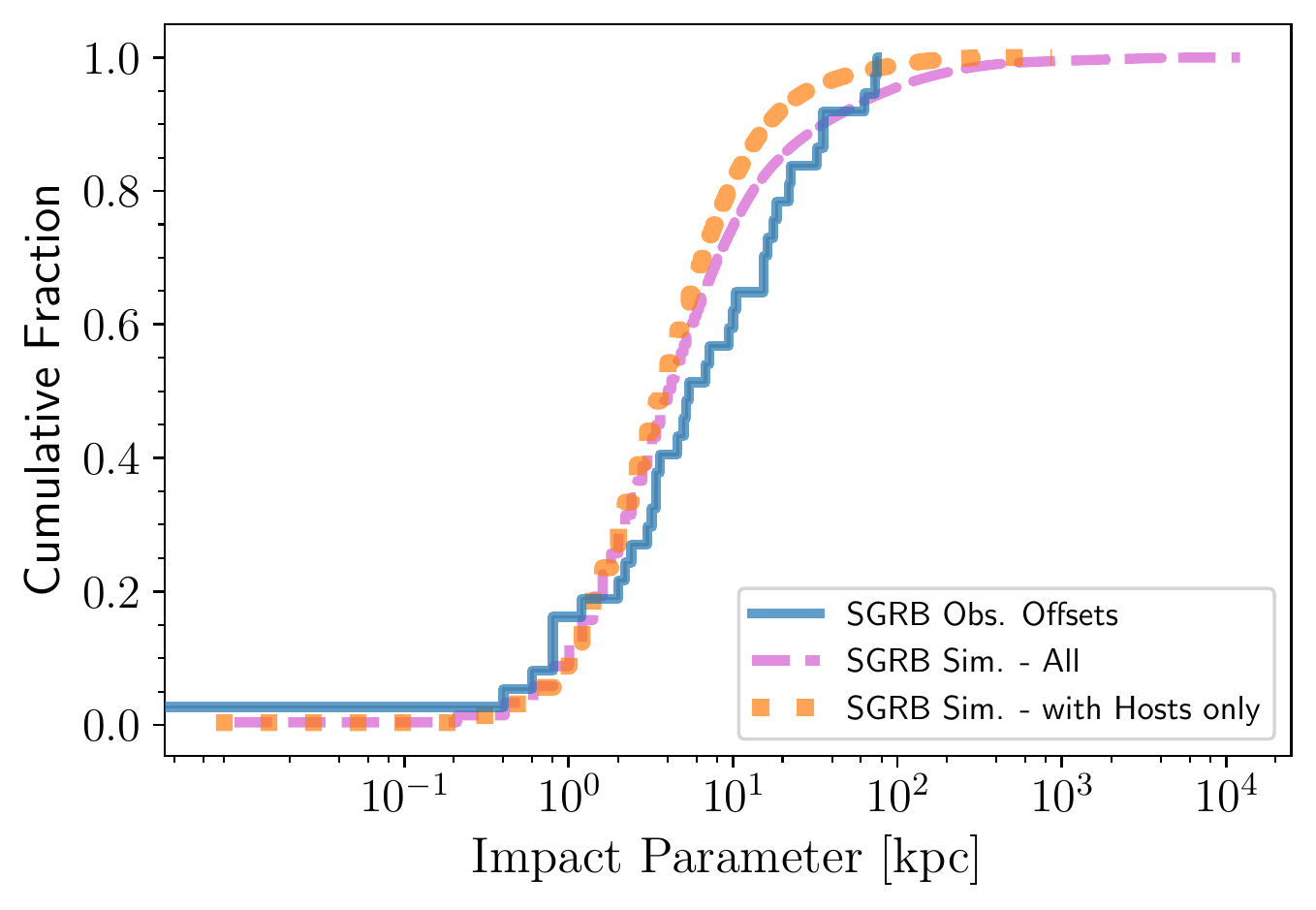}}
    }
    \caption{[COSMIC]  Similar to Figure \ref{fig:ip_cumal}. Refer to Table \ref{tab:sgrb-tab} for sources used. 
    }
    \label{fig:cosmic-ip_cumal}
\end{figure}


\end{document}